\documentclass[twocolumn]{aastex63}

\usepackage{amsmath}
\usepackage[normalem]{ulem}

\newcommand*\fgr[1]{Figure~\ref{#1}}
\newcommand*\sectn[1]{Section~\ref{#1}}
\newcommand*\eqn[1]{Equation~\ref{#1}}
\received{}
\revised{}
\accepted{}
\shorttitle{}
\shortauthors{Wang et al.}
\graphicspath{{./}{figures/}}

\begin{document}

\title{Hot Jupiter and ultra-cold Saturn formation in dense star clusters}

\correspondingauthor{Yi-Han Wang}
\email{yihan.wang.1@stonybrook.edu}

\author{Yi-Han Wang}
\affiliation{Department of Physics and Astronomy, Stony Brook University, Stony Brook, NY, 11794, USA}

\author{Nathan W. C. Leigh}
\affiliation{Departamento de Astronom\'ia, Facultad de Ciencias F\'isicas y Matem\'aticas,
Universidad de Concepci\'on, Concepci\'on, Chile}
\affiliation{Department of Astrophysics, American Museum of Natural History, Central Park West and 79th Street, New York, NY 10024}

\author{Rosalba Perna}
\affiliation{Department of Physics and Astronomy, Stony Brook University, Stony Brook, NY, 11794, USA}
\affiliation{Center for Computational Astrophysics, Flatiron Institute, 162 5th Avenue, New York, NY 10010, \
USA\\}

\author{Michael M. Shara}
\affiliation{Department of Astrophysics, American Museum of Natural History, Central Park West and 79th Street, New York, NY 10024}

\begin{abstract}
The discovery of  high incidence of hot Jupiters in dense clusters challenges the field-based hot Jupiter formation theory. In dense clusters, interactions between planetary systems and flyby stars are relatively common. This has a significant impact on planetary systems, dominating hot Jupiter formation. In this paper, we perform high precision, few-body simulations of stellar flybys and subsequent planet migration in clusters. A large parameter space exploration demonstrates that close flybys that change the architecture of the planetary system can activate high eccentricity migration mechanisms: Lidov-Kozai and planet-planet scattering, leading to high hot Jupiter formation rate in dense clusters. Our simulations predict that many of the hot Jupiters are accompanied by ``ultra-cold Saturns", expelled to apastra of thousands of AU. This increase is particularly remarkable for planetary systems originally hosting two giant planets with semi-major axis ratios $\sim$ 4 and the flyby star approaching nearly perpendicular to the planetary orbital plane. The estimated lower limit to the hot Jupiter formation rate of a virialized cluster is $\sim 1.6\times10^{-4}({\sigma}/{\rm 1kms^{-1}})^5({a_{\rm p}}/{\rm 20 AU})({M_{\rm c}}/{\rm 1000M_\odot})^{-2}$~Gyr$^{-1}$ per star, where $\sigma$ is the cluster velocity dispersion, $a_{\rm p}$ is the size of the planetary system and $M_{\rm c}$ is the mass of the cluster. Our simulations yield a hot Jupiter abundance which is $\sim$ 50 times smaller than that observed in the old open cluster M67. We expect that interactions involving binary stars, as well as a third or more giant planets, will close the discrepancy.

\end{abstract}


\section{Introduction}
The first exoplanet discovered around a main sequence star, 51 Peg b \citep{Mayor1995}, with a very tight orbit, challenged astronomers' general picture of planetary system formation. This gas giant orbits about ten times closer to its host star than similar planets in our Solar System, which are only found outwards of five Astronomical Units (AU). The discovery of the first so-called hot Jupiter not only initiated a revolution in planetary formation science, but also opened up the field of exoplanet discovery and characterization to include systems not observed in our own Solar System. Because they are among the easier extrasolar planets to observe, hot Jupiters have been rapidly discovered in the past decade, adding to the large sample of exoplanets. The increasingly large number of hot Jupiters indicates that they play a major role in the field of exoplanets. As such, they have inspired extensive theoretical work on exoplanet formation, thereby broadening our understanding of planetary system formation and evolution. 

Three main hot Jupiter formation theories have been proposed in the past decades.  These are: in situ formation, disk migration and high eccentricity tidal migration.  In the in situ formation theory, a hot Jupiter can be formed if a gravitational instability creates bound clumps from the fragments of the proto-planetary disk \citep{Boss1997,Durisen2007}, or if the proto-planetary core accretes large quantities of gas from the proto-planetary disk \citep{Perri1974, Pollack1996,Chabrier2014}. \citet{Batygin2016} have argued that hot Jupiters can form with super-Earth cores in the inner parts of disks. \citet{Bailey2018} argue that the inner boundary of the period-mass distribution of close-in giant planets conforms to the expectations of the in situ formation scenario. A recent strong argument against in situ formation is the measurement \citep{Gunther2020} of a period (and hence semimajor axis) decrease of 
the hot Jupiters in HD 189733 and six other systems.

In the gas disk migration theory, torques from the gaseous proto-planetary disk may significantly shrink a giant planet's orbit to form a hot Jupiter \citep{Goldreich1980, Lin1986, Lin1996, Ida2008, Brucalassi2014}. The efficiency of disk migration depends on the strength of the corotation torques which is in turn determined by the disk conditions. If the migration timescale is shorter than the disk lifetime, the giant may migrate all the way into the Roche radius and be tidally disrupted by its host star. If the migration timescale is comparable to or greater than the disk lifetime, the giant planet may not migrate far enough inwards to make a hot Jupiter. 

In high eccentricity tidal migration, several dynamical mechanisms have been proposed to excite the eccentricity of a cold Jupiter to high values.  This is done by removing/exchanging angular momentum with the giant planet's orbit, so that tidal migration becomes efficient enough to act on sufficiently short timescales to make a hot Jupiter. The proposed mechanisms are planet-planet scattering that converts Keplerian sheer into an angular momentum deficit and triggers high eccentricity migration \citep{Rasio1996,Weidenschilling1996, Ford2006,Chatterjee2008}, secular interactions including Lidov-Kozai oscillations from the companion star \citep{Wu2003, Fabrycky2007, Katz2011,Naoz2012,Petrovich2015, Anderson16, Storch2014, Storch2014b}, Lidov-Kozai oscillations from another giant planet \citep{Naoz2011,Teyssandier2013}, secular chaos from multiple planets \citep{Wu2011,Hamers2017,Xu2016,Li2019,Li2020} and coplanar secular effects \citep{Li2014}.    

No single mechanism mentioned above has yet been able to explain the distribution of orbital properties of the observed hot Jupiters. \citet{Ford2003} and \citet{Ford2006} proposed that multiple planet scatterings (i.e., sequential scatterings acting on the same planet but separated in time) may  be promising to explain the observations, but this ultimately fails due to a low efficiency in producing hot planets. \citet{Adams2005} demonstrated that the combination of stellar scattering and subsequent tidal migration may produce the observed hot Jupiter properties. This illustrates the possibility to tweak some of the existing theories for producing hot Jupiters by introducing more degrees of complexity. However, this is non trivial since further complications arise from the fact that a large fraction of stars are born in clusters or associations. Stellar interactions (scatterings) within a cluster/association can significantly change the orbits of the planets or the disk \citep{Hurley2002,Adams2006, Spurzem2009, Li2015, Shara2016}. These perturbed planetary systems may eventually escape into the field from their birth environment due to the dissociation of the cluster.

Indeed, planets have been detected in star clusters \citep{Lovis2007, Sato2007}. \citet{Meibom2013} detected two small planets in the open cluster NGC 6811, suggesting that planets are as common in open clusters as in the field. \citet{Quinn2012} and \citet{Brucalassi2016} reported the discovery of four giant planets orbiting M67 stars, including three hot Jupiters. \citet{Quinn2014} reported an eccentric hot Jupiter in the Hyades open cluster. All of these observations provide promising evidence that stellar interactions in clusters/associations may affect the planet formation process and/or perturb their orbital parameters post-formation on short timescales (i.e., shorter than the timescale for cluster disruption), and thus may play a non-trivial role in hot Jupiter formation.

\citet{ Hurley2002, Spurzem2009} and  \citet{Parker2012} showed that the orbits of single planets in a cluster can be significantly affected by stellar encounters by performing large scale N-body simulations. \citet{Shara2016} have shown that scatterings due to passing stars in a star cluster can change the orbits of two-planet planetary systems, producing both hot Jupiters and very distant Saturns. Stellar encounters can clearly induce external perturbations that significantly change the architecture of multiple planet systems in clusters/associations, and can thus affect the subsequent internal dynamics within the planetary system
(see also recent works by \citealt{Cai2019,Flammini2019,Li2020,Wang2020a,Wang2020b}). 
We naively predict that multi-planet systems which begin close to orbital resonances will be less susceptible to changes to their orbital configurations due to stellar interactions (i.e., flybys).  This is because (small) integer ratios in the orbital periods mark stabilized subsets of the parameter space.  Many mechanisms can act to stabilize orbital resonances (see \citet{murray99} for a more detailed discussion).  One important example is due to fine-tuning in the degree of the imparted impulse at specific points along the orbits.  Strong perturbations are needed to knock multi-planet systems well away from pre-existing resonances.  Hence, if a system is perturbed only weakly away from an orbital resonance, it is very likely to secularly migrate back to this same orbital resonance (or another nearby resonance).  Conversely, if multi-planet systems do not begin with any such strong resonances, we naively expect to see stronger variations in the final distributions of orbital parameters, relative to cases where one or more initial resonances are present.  In this paper, we perform scattering experiments both close to and far away from strong resonances (i.e., with both large and small ratios between orbital periods), in order to study and quantify these competing effects.

How do stellar encounters alter the efficiency of hot Jupiter formation? To answer this question in full, a sophisticated simulation is needed which includes cluster evolution, stellar evolution, planet formation, single star/binary star scatterings and effects related to the internal dynamical evolution of planetary systems, such as tidal migration and/or planet-planet scatterings. In this paper, we take a first step in this direction by performing scatterings between single-star planetary systems hosting two giant planets, Jupiter-like and Saturn-like. After the scattering, tidal migration models are adopted to track the formation channels of the hot Jupiters with full few-body simulations. 

The paper is organized as follows. In section 2 we describe the numerical models we use. Modifications of planetary systems' architectures due to flybys are described in section 3. In section 4 we describe the formation of hot Jupiters from high eccentricity tidal migration, including Lidov-Kozai and planet-planet scattering effects. We discuss our results and summarize our conclusions in section 5.   

\section{Numerical method} \label{method}
The scattering experiments are performed using the high precision few body code SpaceHub (details in \citealt{Wang2018,Wang2019}). We use the scattering facilities of SpaceHub to generate the initial conditions. The following subsections present the initial conditions and the assumptions going into our numerical scattering experiments.

\begin{figure*}
 \includegraphics[width=2\columnwidth]{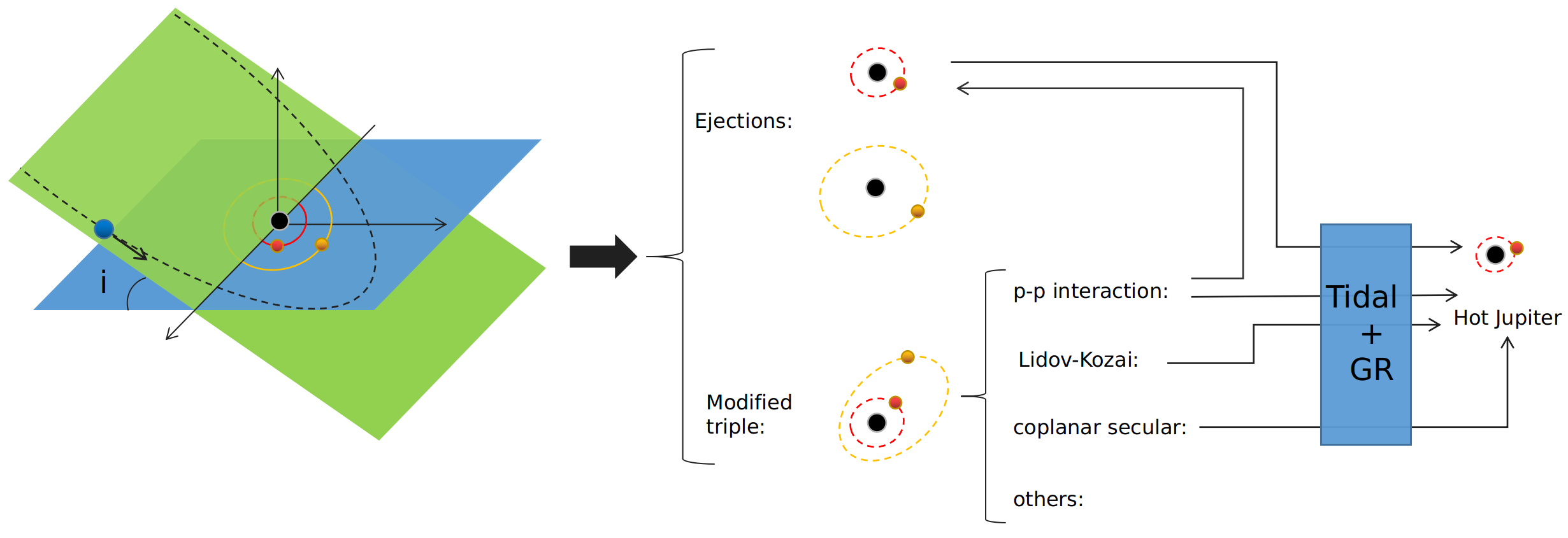}\\
 \caption{Schematic illustration of the flyby and the various outcomes after long timescale interactions. After the flyby, the planet system is modified. If the system remains intact, with both planets still orbiting their host star, the modified triple may form a new configuration in the planet-planet {{red}(p-p)} interaction regime, the Lidov-Kozai (LK) regime and the coplanar secular regime. In those regimes, the eccentricity of the Jupiter can be excited to high values, which makes hot Jupiter formation possible due to tidal circularization with {{red}general relativistic (GR) effects.}}
 \label{fig:schematics}
\end{figure*}

\begin{figure*}
    \includegraphics[width=.45\textwidth]{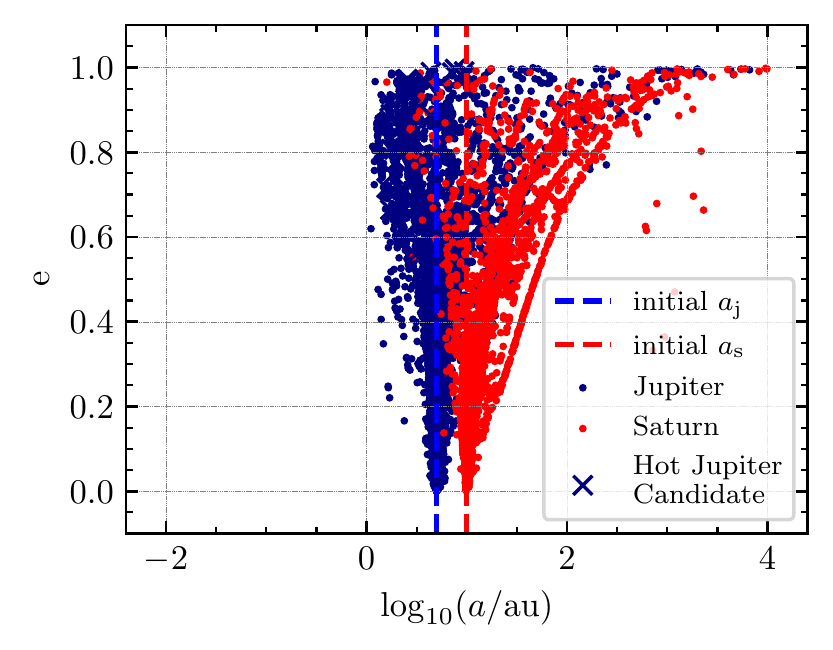}
    \includegraphics[width=.45\textwidth]{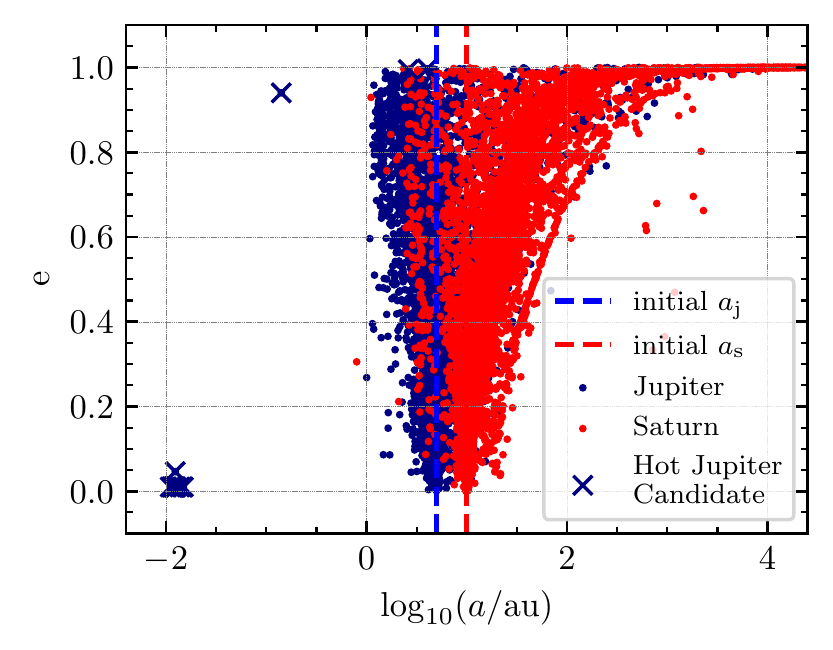}\\
    \includegraphics[width=.45\textwidth]{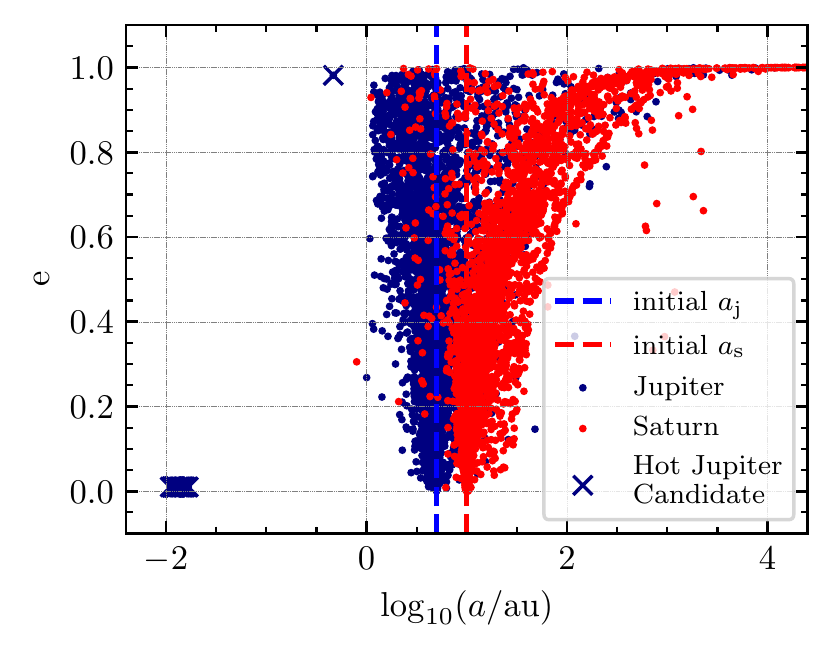}
    \includegraphics[width=.45\textwidth]{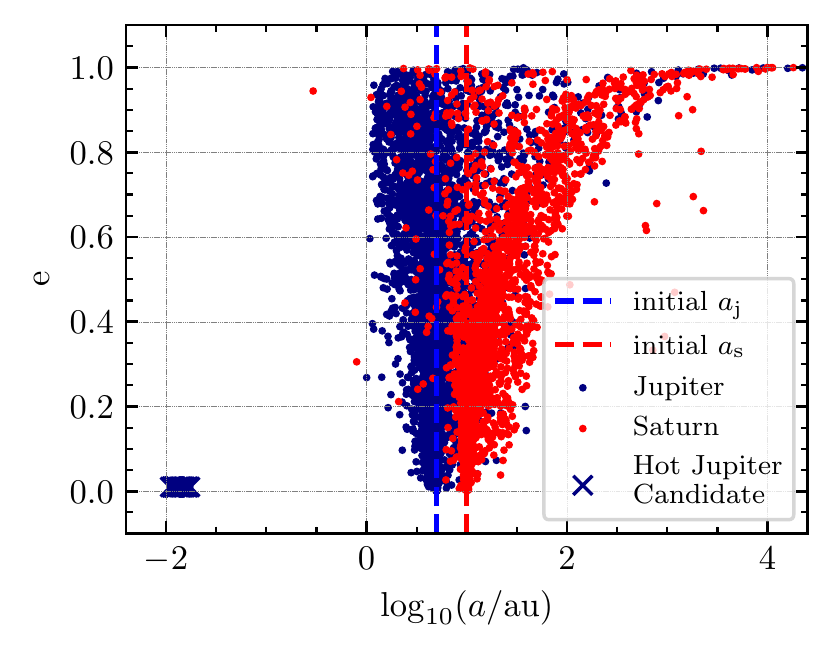}
    \caption{The distribution of orbital parameters of the planets which remain bound at four different times:
    \textit{Upper left}: Immediately after the flyby ; \textit{Upper right}: $10^6$~yr after the flyby; \textit{Bottom left}: $10^7$~yr after the flyby; ; \textit{Bottom right}: $10^8$~yr after the flyby;. The velocity dispersion is $\sigma$=1 km s$^{-1}$ and the initial SMA of the Jupiter and Saturn are 5 AU and 10 AU, respectively.}
    \label{Fig:a-etot-r2}
\end{figure*}

\begin{figure*}
    \includegraphics[width=.45\textwidth]{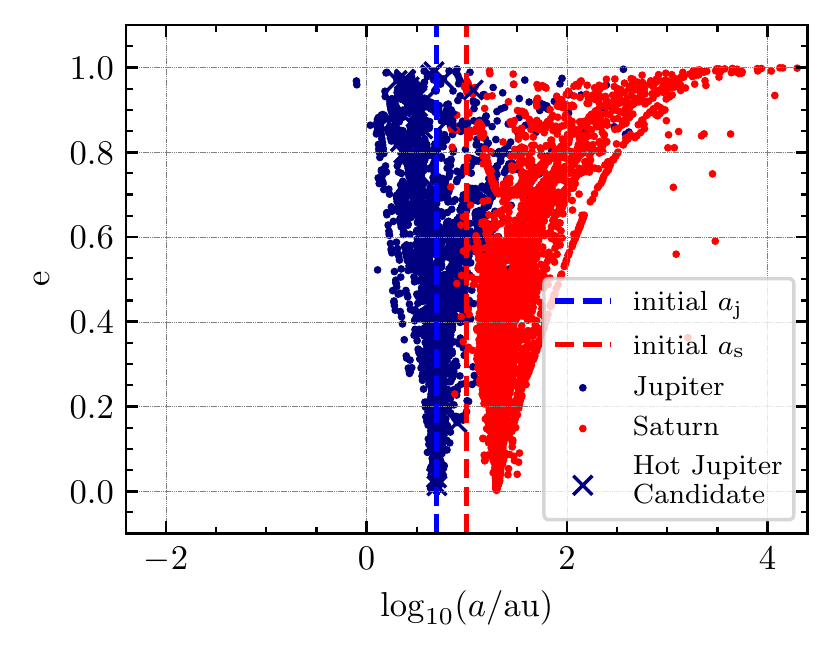}
    \includegraphics[width=.45\textwidth]{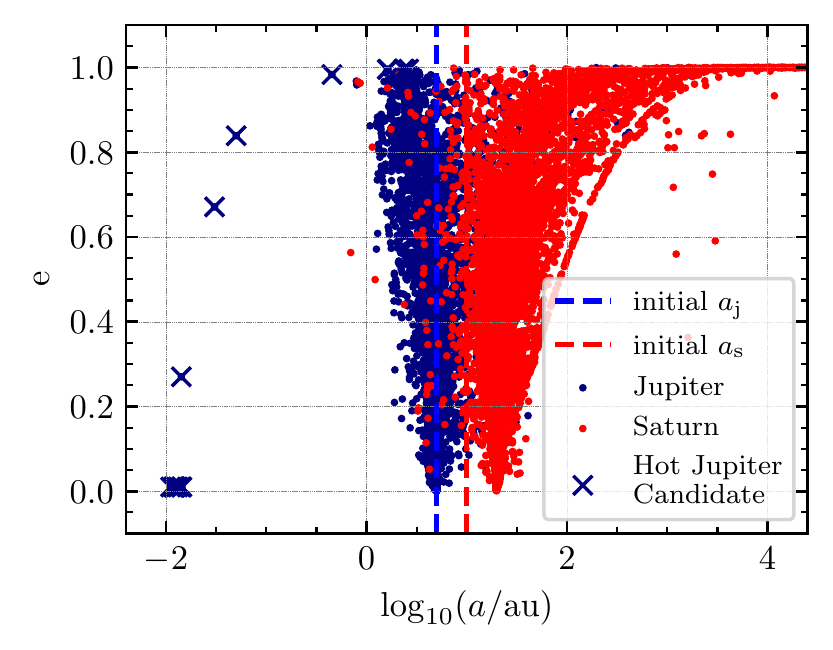}\\
    \includegraphics[width=.45\textwidth]{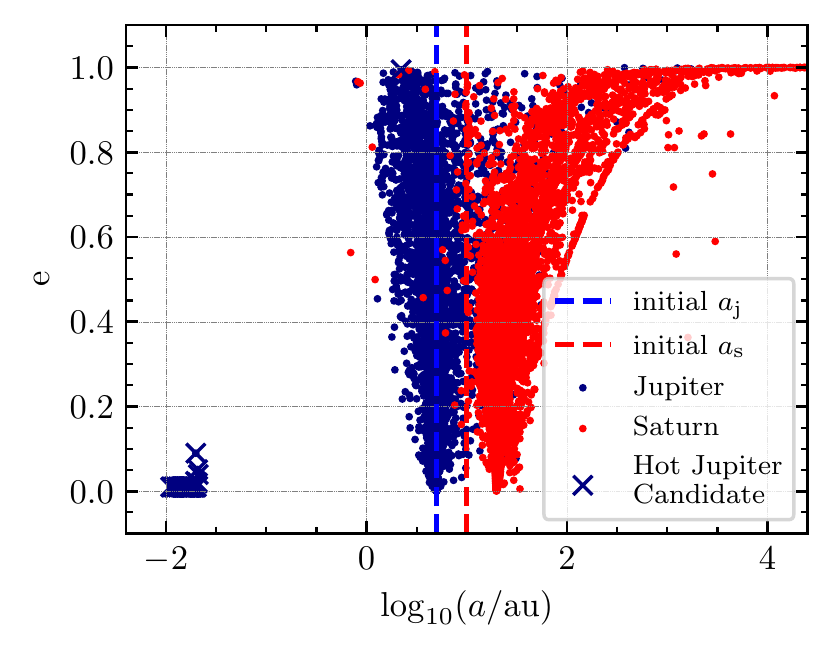}
    \includegraphics[width=.45\textwidth]{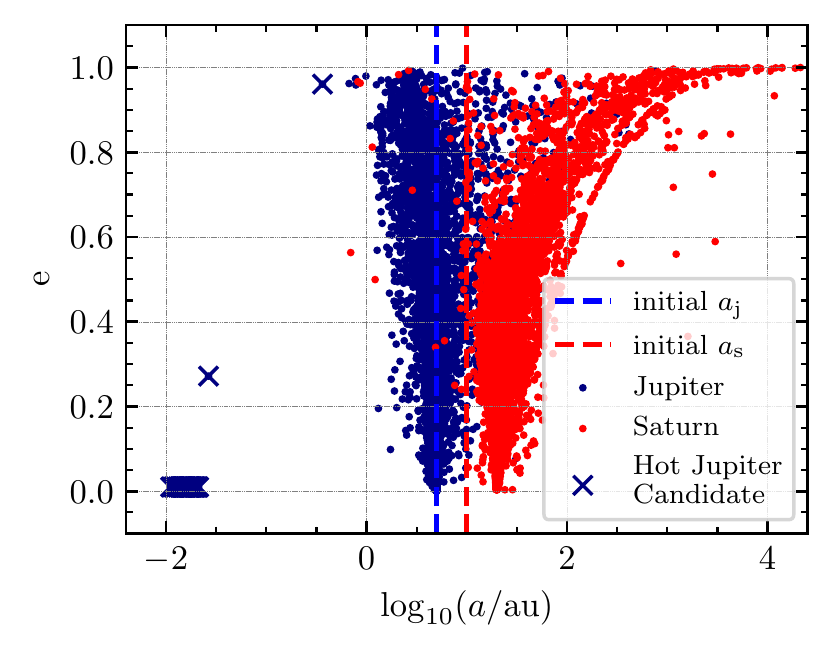}
    \caption{The distribution of orbital parameters of the planets which remain bound at four different times:
    \textit{Upper left}: Immediately after the flyby ; \textit{Upper right}: $10^6$~yr after the flyby; \textit{Bottom left}: $10^7$~yr after the flyby; ; \textit{Bottom right}: $10^8$~yr after the flyby;. The velocity dispersion is $\sigma$=1 km s$^{-1}$ and the initial SMA of the Jupiter and Saturn are 5 AU and 20 AU, respectively.}
    \label{Fig:a-etot-r4}
\end{figure*}

\begin{figure*}
    \includegraphics[width=.45\textwidth]{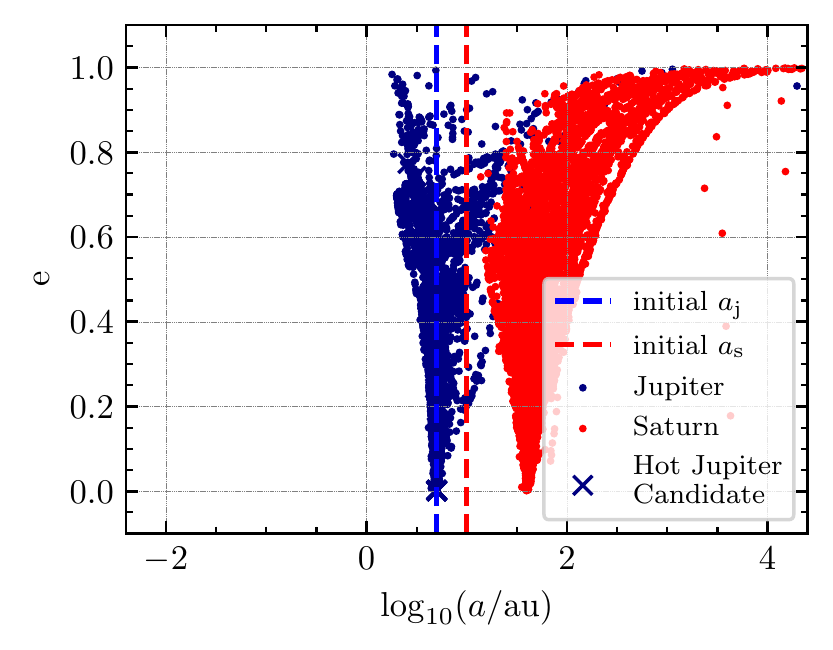}
    \includegraphics[width=.45\textwidth]{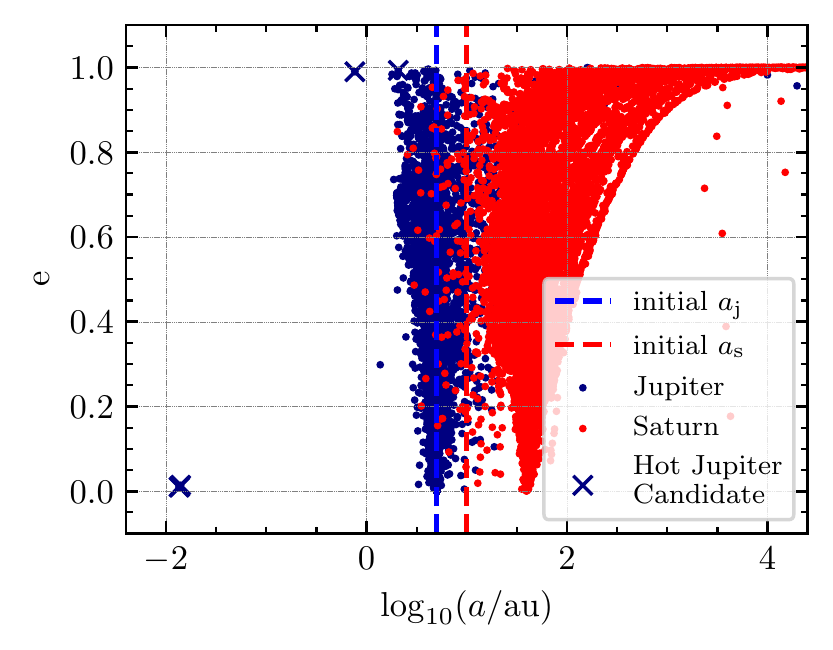}\\
    \includegraphics[width=.45\textwidth]{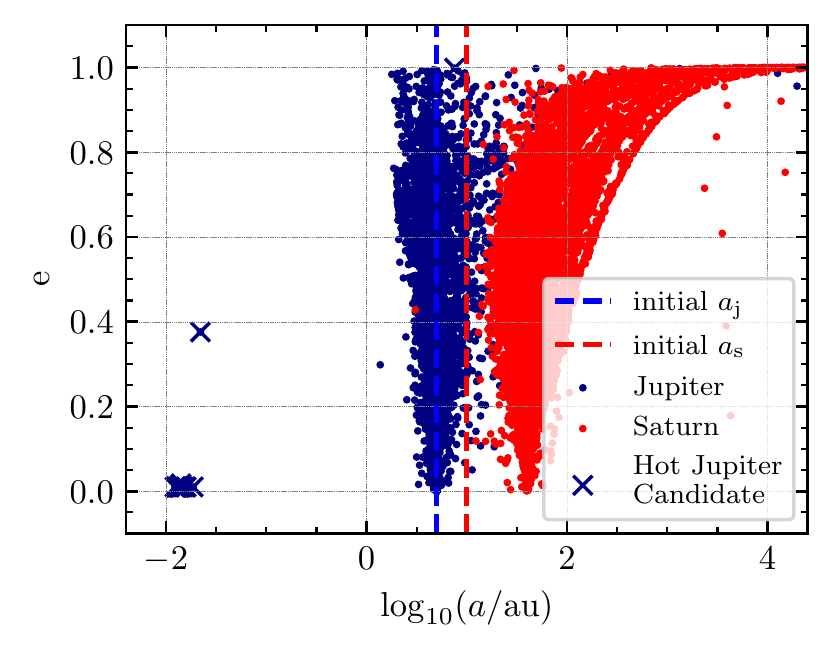}
    \includegraphics[width=.45\textwidth]{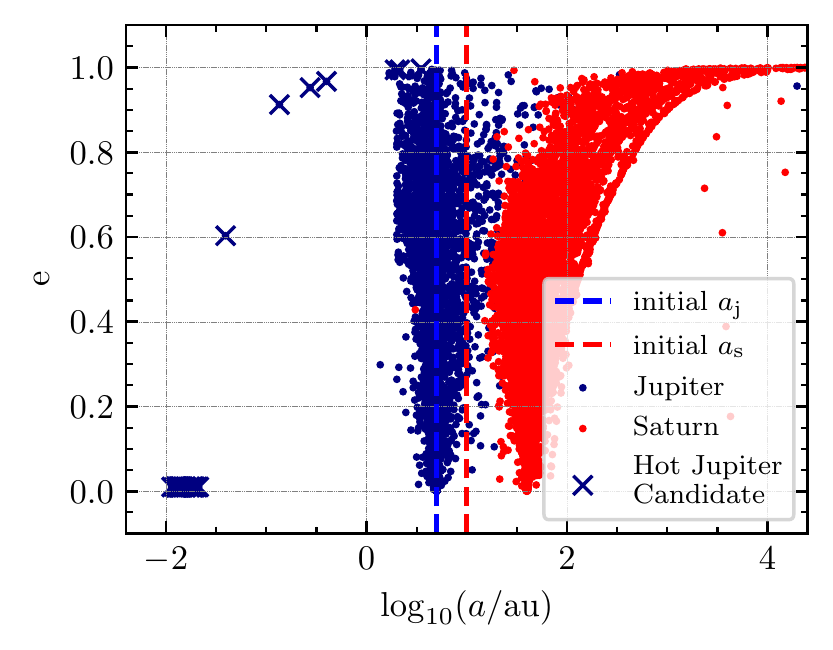}
    \caption{The distribution of orbital parameters of the planets which remain bound at four different times:
    \textit{Upper left}: Immediately after the flyby ; \textit{Upper right}: $10^6$~yr after the flyby; \textit{Bottom left}: $10^7$~yr after the flyby; ; \textit{Bottom right}: $10^8$~yr after the flyby;. The velocity dispersion is $\sigma$=1 km s$^{-1}$ and the initial SMA of the Jupiter and Saturn are 5 AU and 40 AU, respectively.}
    \label{Fig:a-etot-r8}
\end{figure*}

\begin{figure*}
    \includegraphics[width=.45\textwidth]{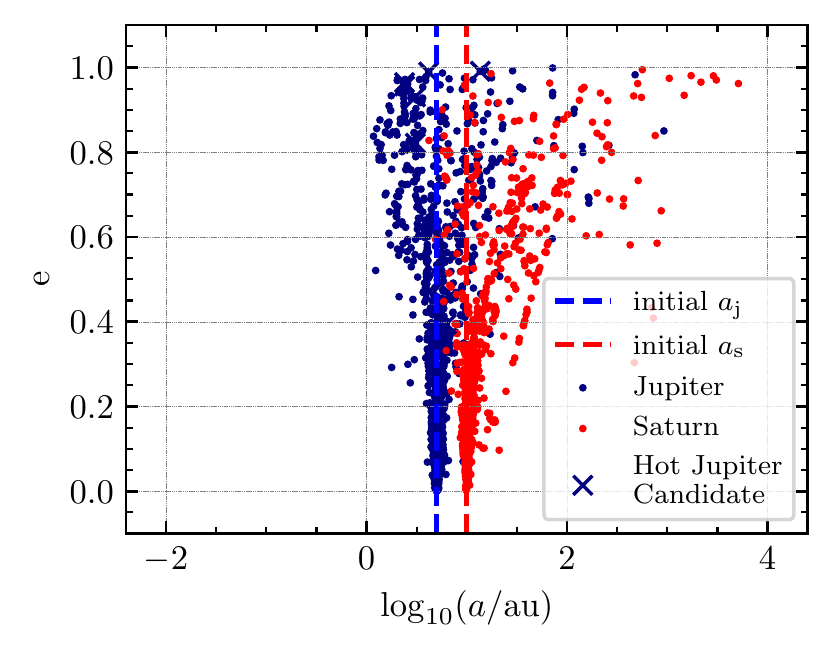}
    \includegraphics[width=.45\textwidth]{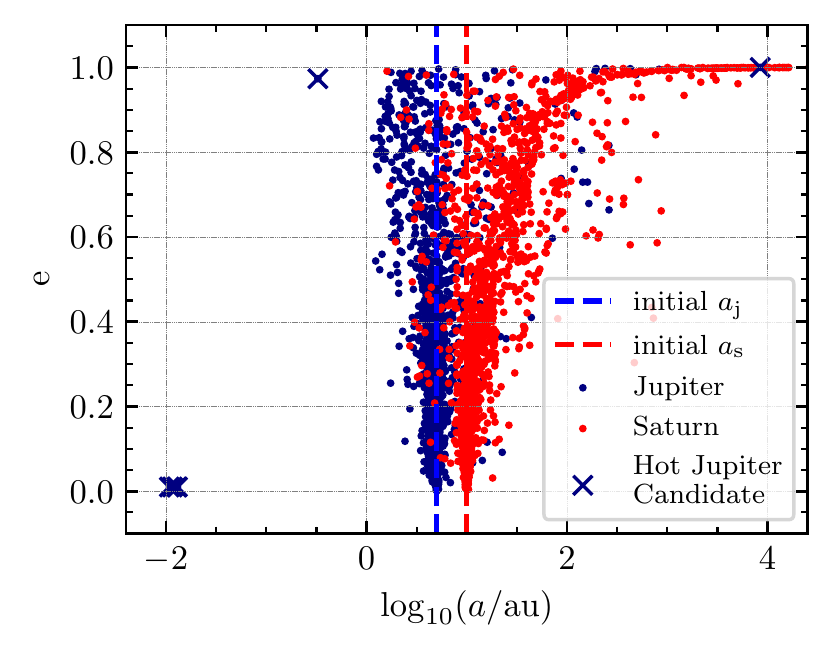}\\
    \includegraphics[width=.45\textwidth]{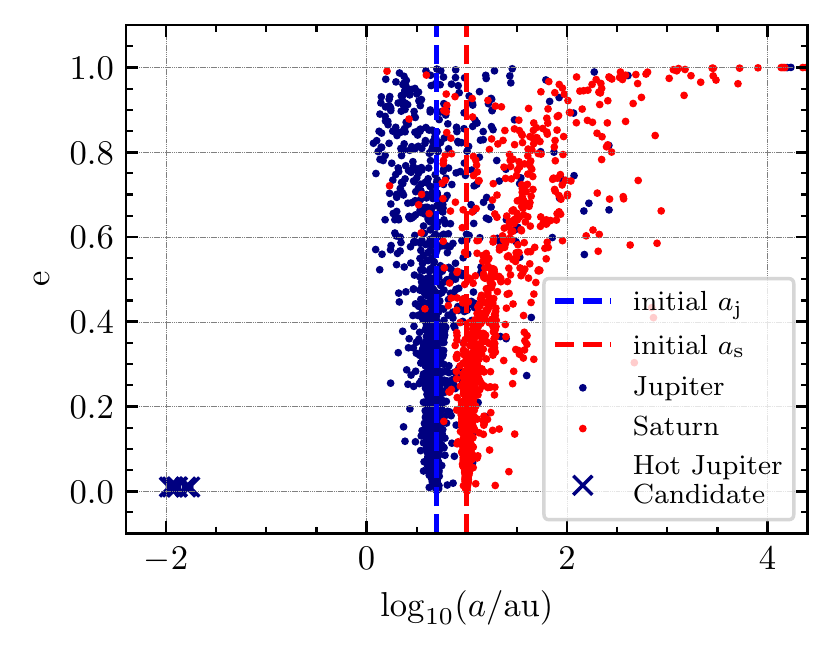}
    \includegraphics[width=.45\textwidth]{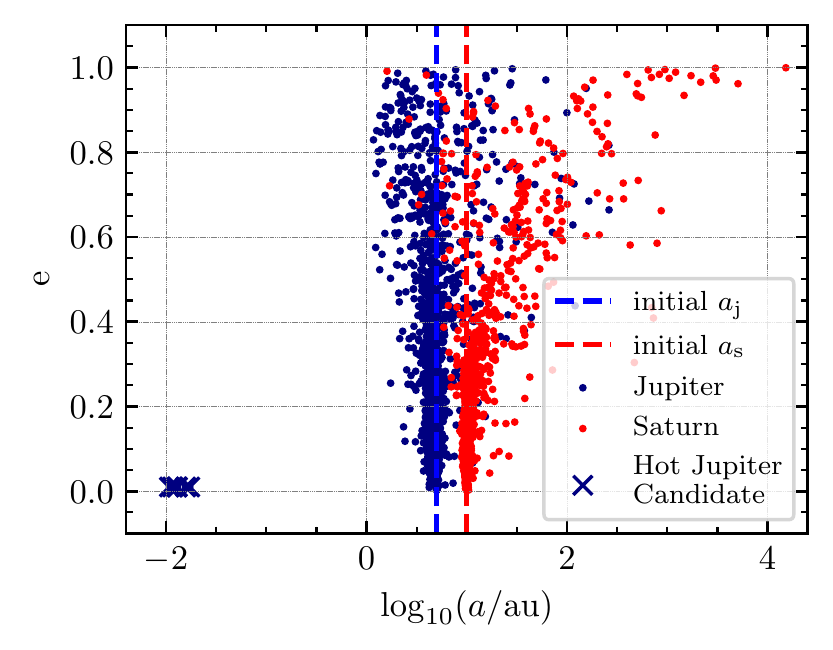}
    \caption{The distribution of orbital parameters of the planets which remain bound at four different times:
    \textit{Upper left}: Immediately after the flyby ; \textit{Upper right}: $10^6$~yr after the flyby; \textit{Bottom left}: $10^7$~yr after the flyby; ; \textit{Bottom right}: $10^8$~yr after the flyby;. The velocity dispersion is $\sigma$=0.2 km s$^{-1}$ and the initial SMA of the Jupiter and Saturn are 5 AU and 10 AU, respectively.}
    \label{Fig:a-etot-s02}
\end{figure*}

\begin{figure*}
    \includegraphics[width=\textwidth]{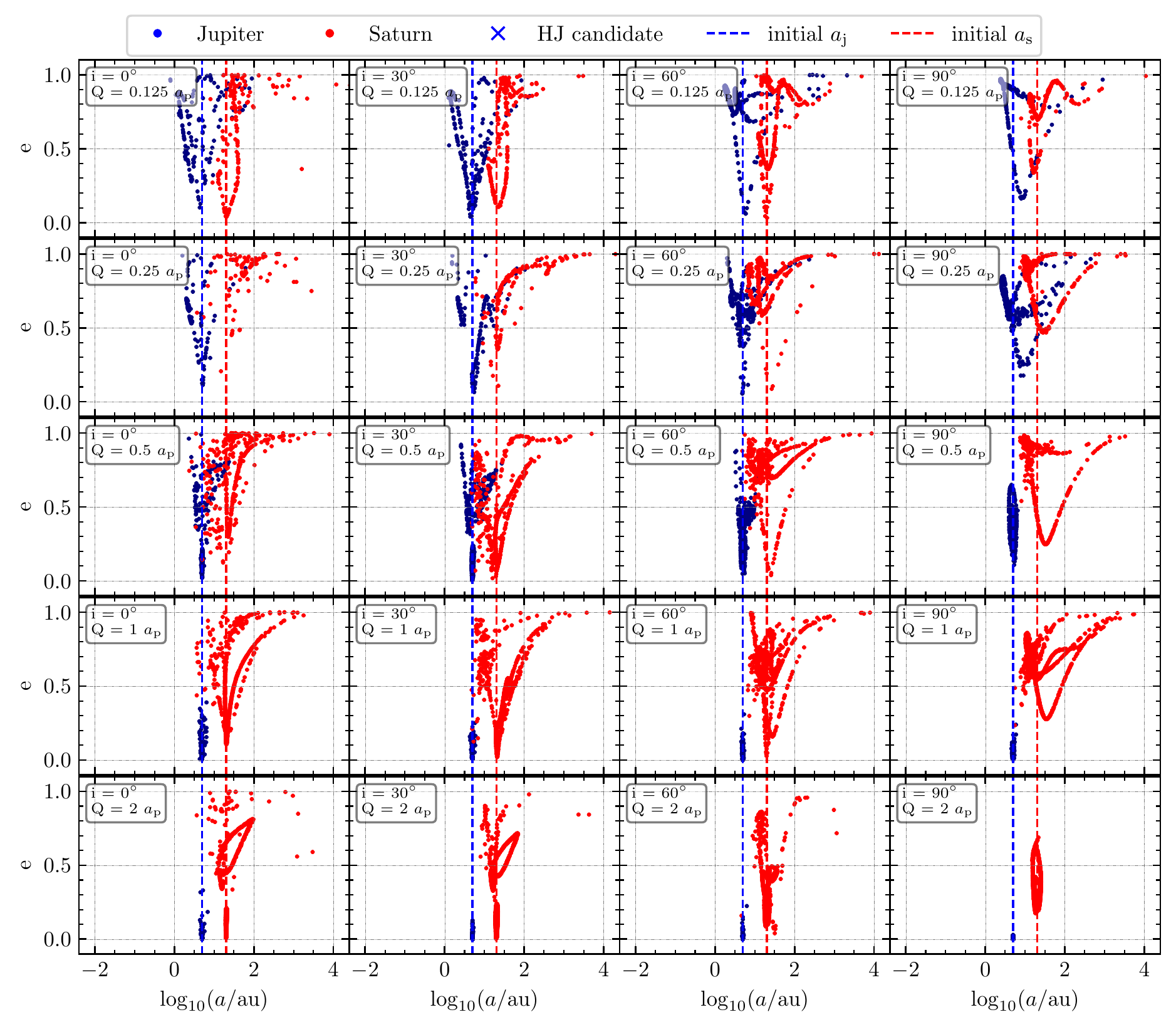}
    \caption{Distribution of orbital parameters right after the flyby. The velocity dispersion is $\sigma$=1 km s$^{-1}$ and the initial SMA of the Jupiter and Saturn are 5 AU and 20 AU, respectively.}
    \label{fig:a-et0}
\end{figure*}

\begin{figure*}
    \includegraphics[width=\textwidth]{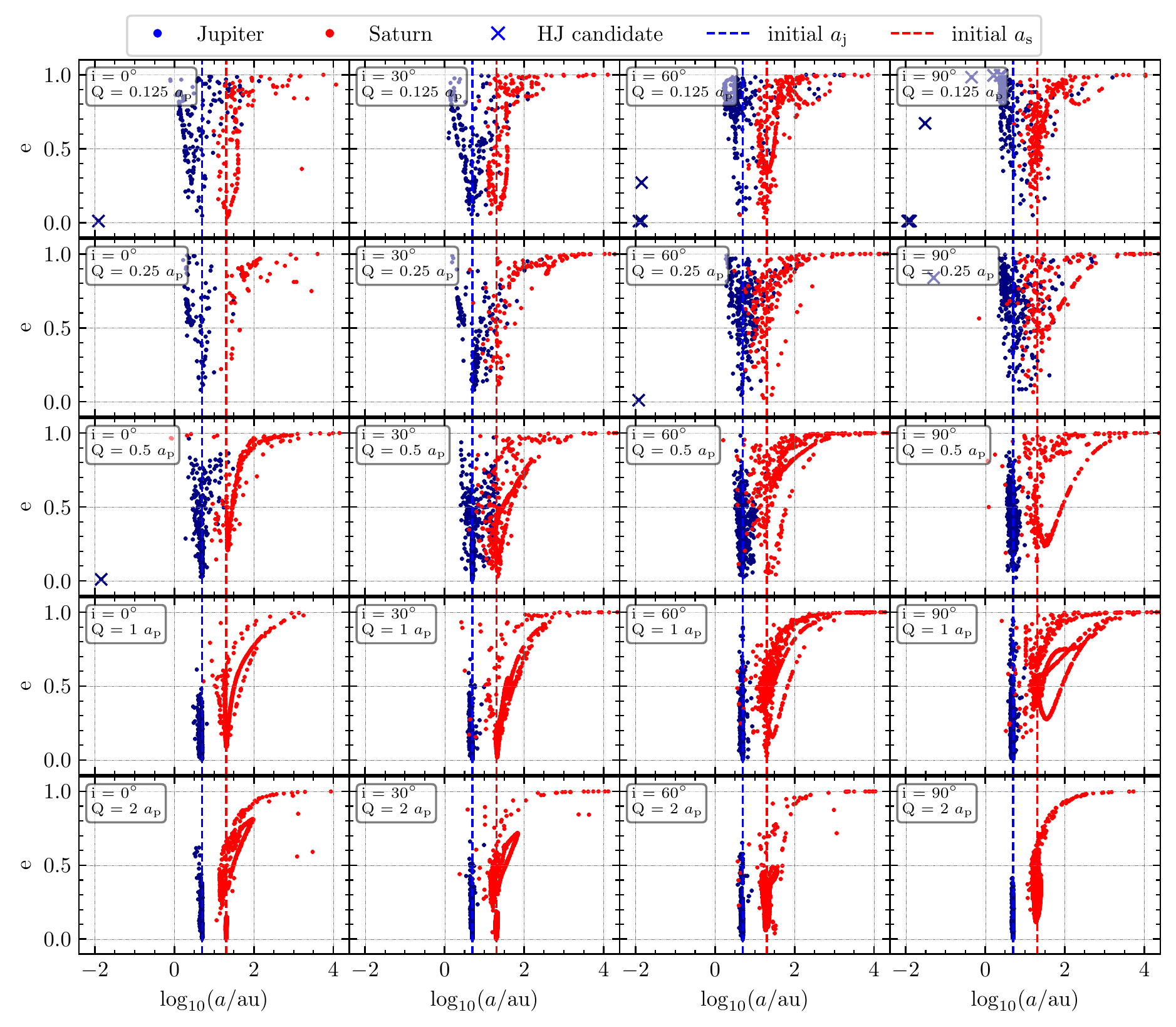}
    \caption{Distribution of orbital parameters at $10^6$~yr after the flyby. The velocity dispersion is $\sigma$=1 km s$^{-1}$ and the initial SMA of the Jupiter and Saturn are 5 AU and 20 AU, respectively.}
    \label{fig:a-et6}
\end{figure*}
\begin{figure*}
    \includegraphics[width=\textwidth]{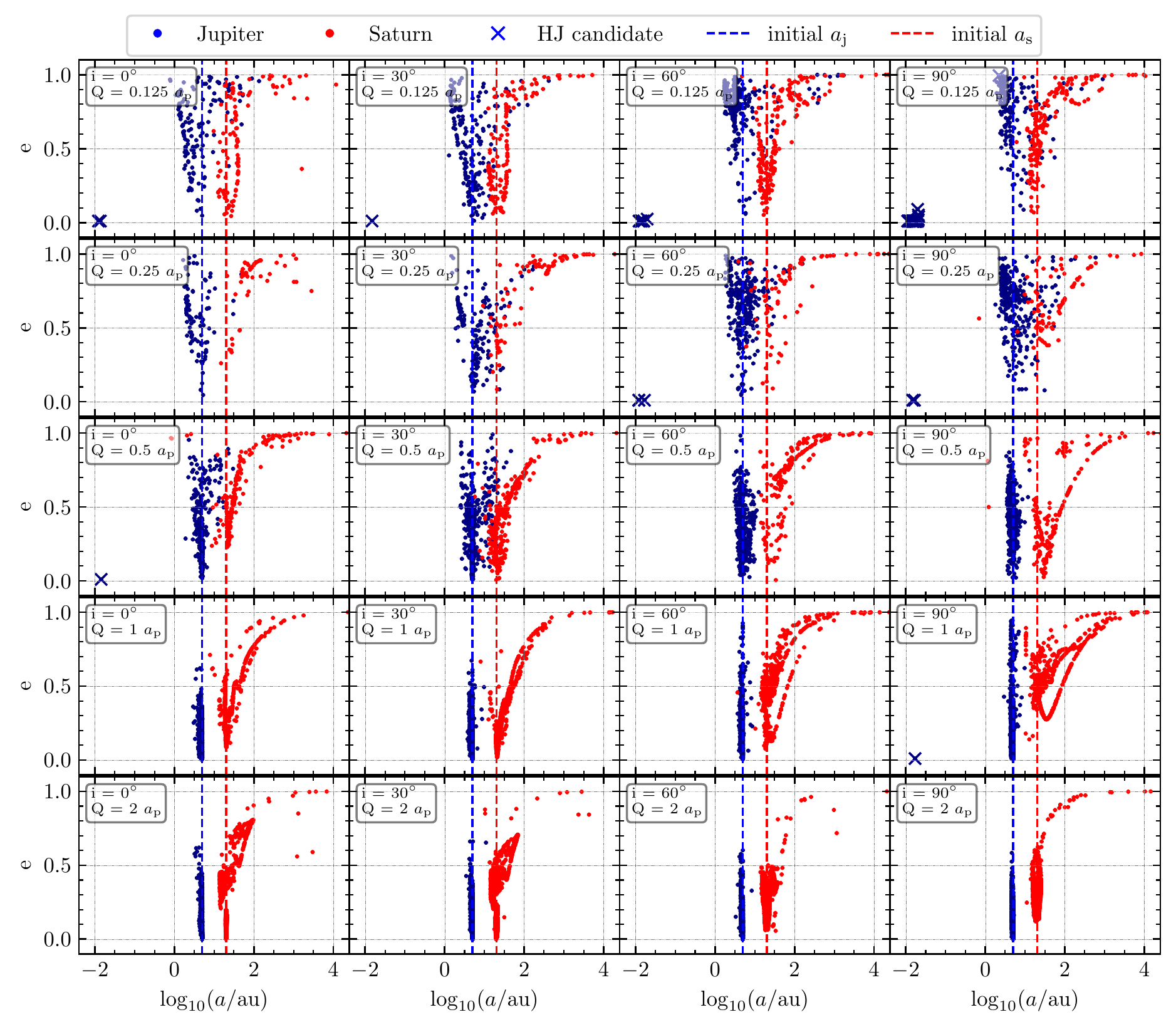}
    \caption{Distribution of orbital parameters at $10^7$~yr after the flyby. The velocity dispersion is $\sigma$=1 km s$^{-1}$ and the initial SMA of the Jupiter and Saturn are 5 AU and 20 AU, respectively.}
    \label{fig:a-et7}
\end{figure*}
\begin{figure*}
    \includegraphics[width=\textwidth]{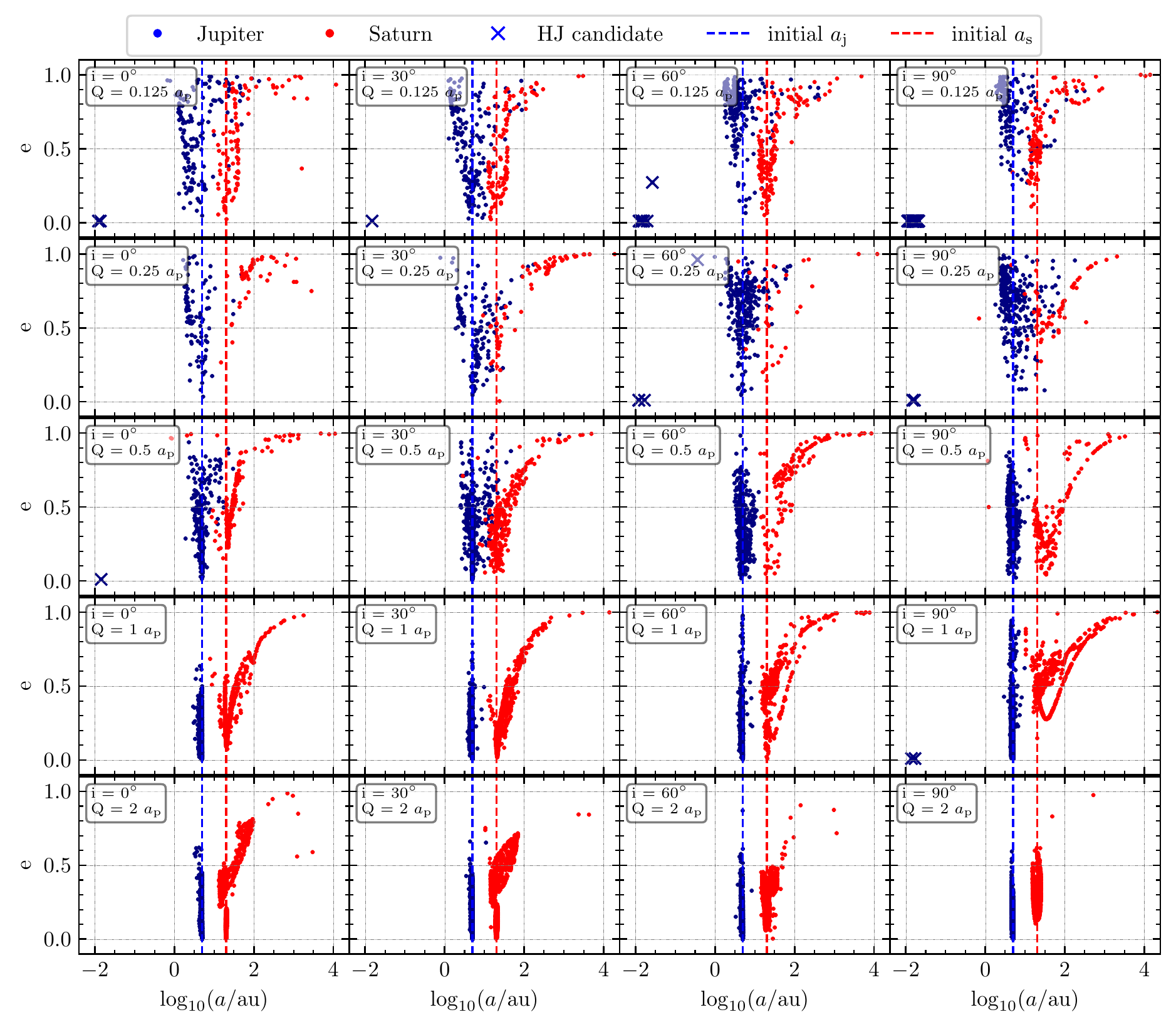}
    \caption{ Distribution of orbital parameters at $10^8$~yr after the flyby. The velocity dispersion is $\sigma$=1 km s$^{-1}$ and the initial SMA of the Jupiter and Saturn are 5 AU and 20 AU, respectively.}
    \label{fig:a-et8}
\end{figure*}

\subsection{Flyby Process}

\subsubsection{Planetary system setups}
The planetary system consists of a one solar mass main sequence star and two giant planets. The inner planet has a mass equal to Jupiter's, and its orbit is circular with a 5 AU radius. The outer planet has the mass of Saturn, in a circular orbit in the same plane as the inner planet. We explore 3 different semi-major axis (SMA) values for the outer planet: 10 AU, 20 AU and 40 AU. The mean anomalies (i.e. the fraction of an orbit's period that has elapsed since the orbiting body passed periapsis) of the two planets are generated uniformly between $0$ and $2\pi$, independently. The uniform distribution of mean anomalies ensures that the phase of the planet is uniformly distributed in time.

\subsubsection{Critical velocity for multi-hierarchy scattering}
The critical velocity $v_{\rm c}$ is an important variable in multi-body scattering. It is defined such that, for a relative velocity at infinity between two objects $v_\infty > v_{\rm c}$ (objects can be isolated bodies or they can have complex internal orbital architectures as is the case for binaries, triples, etc.), the total energy of the system in the centre of mass reference frame is positive \citep[e.g.][]{hut83,fregeau04,leigh12,leigh16a,leigh16b}. Thus, in the regime where the total interaction energy is positive, full ionization is a possible outcome scenario for objects containing internal orbital architectures.  In this regime, the incident object usually does a flyby relative to the scattered object, and interacts with it only once. For $v_\infty < v_{\rm c}$, the total energy of the scattering system is negative. The corresponding scatterings are likely to be resonant in this regime, such that a bound chaotic interaction ensues, with the incident object undergoing repeated flybys relative to the scattered object.  These interactions typically progress until disruption, forming at least one tightly bound system (other orbits need not survive to satisfy the criteria for energy and angular momentum conservation).

We label the masses of the k components of the scattered system to be $m_{00}$, $m_{01}$, ... $m_{0k}$ and the masses of the l components of the incident system to be $m_{10}$, $m_{11}$, ... $m_{1l}$. If the relative velocity between the centres of mass of the two objects is $v_\infty$, the total energy of the scattering system in the centre of mass reference frame is

\begin{eqnarray}
E_{\rm tot}=\frac{1}{2}\frac{M_0M_1}{M_0+M_1} v_\infty^2 + \sum_i^n \frac{m_{0i}(\mathbf{v}_{0i}-\mathbf{V}_0)^2}{2} \nonumber \\ 
- \sum_{i<j\le k}\frac{Gm_{0i}m_{0j}}{|\mathbf{r}_{0i} - \mathbf{r}_{0j}|} 
+ \sum_i^m \frac{m_{1i}(\mathbf{v}_{1i}-\mathbf{V}_1)^2}{2} \nonumber \\
- \sum_{i<j\le l}\frac{Gm_{1i}m_{1j}}{|\mathbf{r}_{1i} - \mathbf{r}_{1j}|}\,,
\end{eqnarray}
where $M_0 = \sum_i m_{0i}$ is the total mass of the scattered object and $M_1 = \sum_i m_{1i}$ is the total mass of the incident object. $V_0$ and $V_1$ are the initial centre of mass velocities of the scattered and incident objects, respectively. The critical velocity between two objects corresponding to a total energy of zero for the system is calculated using the scattering facilities of SpaceHub, specifically via the following equation:
\begin{eqnarray}\label{eq:vc}
v_{\rm c}^2= \frac{M_0+M_1}{M_0M_1}\bigg[ \sum_{i<j\le k}\frac{2Gm_{0i}m_{0j}}{|\mathbf{r}_{0i} - \mathbf{r}_{0j}|} +  \sum_{i<j\le l}\frac{2Gm_{1i}m_{1j}}{|\mathbf{r}_{1i} - \mathbf{r}_{1j}|}\nonumber \\
- \sum_i^k {m_{0i}(\mathbf{v}_{0i}-\mathbf{V}_0)^2}
  - \sum_i^l {m_{1i}(\mathbf{v}_{1i}-\mathbf{V}_1)^2}\bigg]\,.
\end{eqnarray}

\subsubsection{Impact parameter for high mass ratio scattering}\label{sec:impact parameter}
Due to gravitational focusing, incident objects at spatial infinity with impact parameter $b$ will pass the scattered object at closest approach with a corresponding critical pericenter distance for the hyperbolic/parabolic orbit:

\begin{eqnarray}\label{eq:p}
Q_{\rm c} &=& a_{0,1}(1-e_{0,1})\nonumber \\
&=&a_{0,1}\bigg(1-\sqrt{1+\frac{b^2}{a_{0,1}^2}}\bigg)\nonumber\\
&=&-\frac{G(M_0+M_1)}{v_\infty^2}\bigg(1-\sqrt{1+\frac{b^2}{(-\frac{G(M_0+M_1)}{v_\infty^2})^2}}\bigg)\,.
\end{eqnarray}
We only consider interactions satisfying $Q<Q_{\rm c}$, where $p_{\rm c}$ is the critical pericenter distance. Therefore, to cover all parameter space in which the desired outcome is causally and physically permitted, the scattering experiments must be sampled with an impact parameter distribution of $b^2$, uniformly distributed in the range $[0, b_{\rm max}^2]$, where $b_{\rm max}$ is the corresponding impact parameter for $p_{\rm c}$. We can re-write \eqn{eq:p} in the more convenient form:
\begin{eqnarray}\label{eq:bmax}
b_{\rm max} &=& Q_{\rm c}\sqrt{1 - 2\frac{a_{0,1}}{Q_{\rm c}}}\nonumber\\
&=&Q_{\rm c}\sqrt{1 + 2\frac{G(M_0+M_1)}{v_\infty^2Q_{\rm c}}}\,.
\end{eqnarray}
For higher values of $v_\infty$, $b_{\rm max}\rightarrow Q_{\rm c}$, while the lower limit gives $b_{\rm max}\sim \frac{1}{v_\infty}$. In \citet[][]{hut83},
the two limits are combined to construct the following equation:
\begin{equation}\label{eq:bmax_hut}
b_{\rm max}(v_\infty) = \bigg(\frac{C}{v_\infty/v_{\rm c}} + D\bigg)R\,, 
\end{equation}
where $R$ is the contact distance between the two scattering objects (e.g. single-binary scattering is the semi-major axis of the binary; binary-binary scattering is the sum of the semi-major axes of the binaries) and $C$ and $D$ are dimensionless constants that depend on the desired outcome ($DR\sim Q_{\rm c}$ in the high $v_\infty$ limit and $CR\sim Q_{\rm c}$ in the low $v_\infty$ limit). This equation is widely adopted in nearly equal mass single-binary and binary-binary scatterings. However, in this work, the planet mass is much smaller than the star mass, and hence special consideration should be given to this interesting limit. The velocity term $v_\infty$ in the original equation is made dimensionless by the critical velocity $v_{\rm c}$. If we take the limit of high $v_\infty$ in both \eqn{eq:bmax} and \eqn{eq:bmax_hut}, we have:
\begin{equation}
\frac{CR}{v_\infty/v_{\rm c}}   = b_{\rm max} =Q_{\rm c}\sqrt{\frac{2G(M_0+M_1)}{Q_{\rm c}}}/v_\infty\,,
\end{equation}
since $CR\sim Q_{\rm c}$.  The equation above holds only if 
\begin{equation}
v_{\rm c} \sim \sqrt{\frac{2G(M_0+M_1)}{Q_{\rm c}}}=\sqrt{\frac{M_0+M_1}{M_0M_1}\frac{2GM_0M_1}{CR}}.
\end{equation}
\eqn{eq:vc} then requires:
\begin{eqnarray}
\frac{2GM_0M_1}{CR} \sim \sum_{i<j\le k}\frac{2Gm_{0i}m_{0j}}{|\mathbf{r}_{0i} - \mathbf{r}_{0j}|} +  \sum_{i<j\le l}\frac{2Gm_{1i}m_{1j}}{|\mathbf{r}_{1i} - \mathbf{r}_{1j}|} \nonumber\\
- \sum_i^k {m_{0i}(\mathbf{v}_{0i}-\mathbf{V}_0)^2}
  - \sum_i^l {m_{1i}(\mathbf{v}_{1i}-\mathbf{V}_1)^2}\,.
\end{eqnarray}
For single-binary scatterings with component masses ($m_{00},m_{01}$) - $m_1$, this becomes
\begin{equation}
\frac{G(m_{00}+m_{01})m_1}{Ca} \sim  \frac{Gm_{00}m_{01}}{2a}.
\end{equation}
If $m_{00} \sim m_{01} \sim m_1$, this yields $C\sim 4$, which is the frequently adopted value. However, if $m_{01}$ is a Jupiter mass planet, or $m_{00} \sim m_1 \sim 1000\,m_{01}$, we obtain $C \sim 2000$. Indeed,

\begin{equation}
C \propto \frac{M_0M_1}{\sum_{i<j\le k}m_{0i}m_{0j} + \sum_{i<j\le l}m_{1i}m_{1j}}\,.
\end{equation}
Therefore, for high mass ratio systems, we cannot use \eqn{eq:bmax_hut} by normalizing $v_\infty$ by $v_{\rm c}$ and letting $C \sim [4-10]$\citep{hut83,fregeau04} . The scattering facilities of SpaceHub provide both choices given in \eqn{eq:bmax} and \eqn{eq:bmax_hut}.

\subsubsection{Tidal factor between two objects}\label{sec:tidal-factor}
The tidal factor is widely used in scattering experiments to estimate the tidal perturbation between two objects. It is  defined as
\begin{equation}
\delta_{0\rightarrow 1} =  F_{\rm tid,0\rightarrow 1}/F_{\rm rel,1}\,,
\end{equation}
where $F_{\rm tid,0\rightarrow 1}$ is the tidal force that object 0 exerts on object 1 and $F_{\rm rel,1}$ is the internal gravitational force of object 1. If $\delta \rightarrow 0$, then object 0 and object 1 are isolated.  This tidal tolerance factor is often used to save on computational runtime, by switching to integrating Keplerian orbits in the limit of very low tidal tolerance factors (i.e., this option is turned on whenever the tidal tolerance parameter drops below a critical user-defined value).  In this paper, we only use this parameter to generate the initial position of the interloper but we do not invoke this approximation afterwards, and explicitly integrate the N-body problem without any such simplifying approximations.

In binary-binary and binary-single scatterings, as in \citet{fregeau04}, we have
\begin{eqnarray}
F_{\rm tid,0\rightarrow 1} &=& \frac{2GM_0M_1}{|\mathbf{r}_0-\mathbf{r}_1|^3}a_1(1+e_1)\\
F_{\rm rel,1} &=& \frac{Gm_{10}m_{11}}{|a_1(1+e_1)|^2}\,. 
\end{eqnarray}
Indeed, a more general equation for scatterings between any two generic objects is provided by the tidal radius: 
\begin{equation}
r_{\rm tid,0\rightarrow 1}(\delta_{0\rightarrow 1})=|\mathbf{r}_0-\mathbf{r}_1| = \bigg(\frac{M_0}{\delta_{0\rightarrow 1}M_1}\bigg)^{1/3}D_1\,,
\end{equation}
where $D_1$ is the geometric size of object 1 (i.e., the stellar radius for a single star, the orbital separation for a binary star, etc.). For binary-single scatterings we have $D_1 = a_1(1+e_1)$. The typical value $\delta \sim 0.02$ yields the classical tidal disruption equation
\begin{equation}
|\mathbf{r}_0-\mathbf{r}_1| = 3.7\bigg(\frac{M_0}{M_1}\bigg)^{1/3}D_1\,.
\end{equation}
The scattering facilities of SpaceHub take the maximum value of either $r_{\rm tid,0\rightarrow 1}(\delta_{0\rightarrow 1})$ or $r_{\rm tid,1\rightarrow 0}(\delta_{1\rightarrow 0})$ as the final tidal radius $r_{\rm tid}(\delta)$.

For scattering experiments, strong interactions occur if $p<p_{\rm c}\sim r_{\rm tid}(\delta=0.02)$ while perturbative interactions occur if $ r_{\rm tid}(\delta=0.02) < p< r_{\rm tid}(\delta=10^{-4})$.
The tidal radius is also used to calculate the initial relative distance between the scattered and incident objects. In this project, we take $ r_{\rm tid}(\delta=10^{-5})$ as the initial distance separating the centres of mass for the two scattering objects. The trajectories from infinity to this initial distance are calculated analytically by regarding the scattering objects as isolated up until this critical point.

\subsubsection{Initial conditions for orbital angles }
The phases of the scattered and incident objects are isothermally distributed. For each orbit, $\cos(i)$ is uniformly distributed in the range $[-1,1]$, $\Omega$ is uniformly distributed in the range  $[-\pi,\pi]$, $\omega$ is uniformly distributed in the range $[-\pi,\pi]$, and $M$ is uniformly distributed in the range $[-\pi, \pi]$, where $i$ is the orbital inclination, $\Omega$ is the longitude of the ascending node, $\omega$ is the argument of periapsis and $M$ is the mean anomaly.

\subsubsection{Termination time of flyby}
The interloper is dropped at the point where the distance between the centre of mass of the interloper and the centre of mass of the planetary system is $r_{\rm tid}(\delta=10^{-5})$. In the host cluster environment, we adopt the 1-D velocity dispersions $\sigma\sim 0.2$~km s$^{-1}$ and 1 ~km s$^{-1}$, following \citet{Shara2016}, which are typical of open star clusters \citep[e.g.][]{geller15,leigh16b}.  The corresponding relative velocities at infinity are, respectively, $v_\infty\sim 0.73$~km s$^{-1}$ and $\sim 3.6$~km s$^{-1}$. The velocity is not in the resonant interaction regime, i.e $v_\infty > v_{\rm c}$. Therefore, we terminate the flyby process at $t=2t_{\rm drop}$, where the drop-in time $t_{\rm drop}$ is the time that the interloper takes to travel from the starting point to the point of closest approach. In the hyperbolic asymptotic limit,
\begin{equation}
t_{\rm drop}=M\sqrt{\frac{a^3}{GM_{\rm tot}}}\,,
\end{equation}
where $M$ is the relative mean anomaly between the interloper dropping point and the distance of closest approach, $a$ is the corresponding semi-major axis of the incident hyperbolic orbit and $M_{\rm tot}$ is the total mass of the system.

\subsection{Tidal migration process}\label{sec:tidal}
After the flyby, the planetary system is perturbed, and high eccentricity excitations can occur due to several channels including planet-planet scatterings \citep{Rasio1996,Weidenschilling1996, Ford2006,Chatterjee2008}, Lidov-Kozai oscillations \citep{Wu2003, Fabrycky2007, Katz2011,Naoz2012,Petrovich2015,Naoz2011,Teyssandier2013} and coplanar secular interactions\citep{Li2014}. The (excited) high-eccentricity effectively increases the rate of tidal dissipation due to the tidal force at pericenter, which makes planet migration possible. We integrate the after-flyby planetary systems up to 1 Gyr, and discuss the results of the flybys at 0.01~Gyr, 0.1~Gyr and 1~Gyr.

\subsubsection{Tidal dissipation}
We adopt the static tidal mode by implementing the tidal perturbation acceleration between stellar particles and planet particles in the simulation. The tidal acceleration that the star exerts on the planet is
\begin{equation}
\mathbf{a}_{\rm tid} = -3k_{\rm AM,p}\frac{GM_*^2}{M_{\rm p}r^2}\bigg(\frac{R_{\rm p }}{r} \bigg)^5\bigg[ \bigg(1+\tau_{\rm p} \frac{\dot{r}}{r}\bigg)\hat{\mathbf{r}}-(\Omega_{\rm p} - \dot{\theta})\tau_{\rm p}\hat{\mathbf{\theta}} \bigg]\,,
\end{equation}
where $r$, $R_{\rm p}$, $k_{\rm AM,p}$ and $\tau_{\rm p}$ are the particle distance, the planetary radius, the apsidal motion constant and the tidal time lag, respectively. $\Omega_{\rm p}$ is the planetary spin frequency, where we use the orbit-averaged assumption \citep{Hut1981,Hamers2017},
\begin{equation}
\Omega_{\rm p} = n\frac{1+\frac{15}{2}e^2 +\frac{45}{8}e^4+\frac{5}{16}e^6}{(1+3e^2+\frac{3}{8}e^4)(1-e^2)^{3/2}}\,,
\end{equation}
where $n=\sqrt{GM_*/a^3}$ is the mean motion of the planetary orbit and $a$ and $e$ are the semi-major axis and the eccentricity, respectively. In this work, we use $k_{\rm AM,p}=0.25$ and $\tau_{\rm p} = 0.66$~s \citep{Hamers2017}.
Typically, the time scale for stellar tides to operate is much longer than for planetary tides. Therefore, we do not take it into consideration.

{
In the N-body code, the variables $a$ and $e$ are calculated from the instant relative positions and velocities between particles, that is
\begin{eqnarray}
a&=& \frac{\mu}{\mu/|d\mathbf{r}|-|d\mathbf{v}|^2}\\
\mathbf{e}&=& (\frac{|d\mathbf{v}|^2}{\mu}-\frac{1}{|d\mathbf{r}|})d\mathbf{r} - \frac{d\mathbf{r}\cdot d\mathbf{v}}{\mu}d\mathbf{dv}\,,
\end{eqnarray}
where $\mu=GM_{\rm tot}$ is the gravitational parameter of the planet-star pair and $d\mathbf{r}$ and $d\mathbf{v}$ are the relative position and velocity between the planet-star pair, respectively. Since the accelerations are calculated between particle pairs in the code, the variables $d\mathbf{r}$ and $d\mathbf{v}$ are always calculated between particles. However, for systems with more than two particles, once a hierarchical triple system is formed, the outer object will orbit around the centre of mass of the inner binary. Therefore, to calculate $a$ and $e$ of the outer orbit, the $d\mathbf{r}$ and $d\mathbf{v}$ must be the relative position and velocity between the outer object and the centre of mass of the inner binary. In our case, the inner binary is a star-planet system, and hence the centre of mass of the inner binary is close to the star. Therefore, it is a reasonably good approximation to simply use position and velocity of the star to calculate $a$ and $e$ of the outer orbit.
}
\subsubsection{General Relativistic (GR) effects}
In secular eccentricity excitation processes (including the standard Lidov-Kozai and coplanar secular processes), the periapsis of the inner orbit precesses around its centre of mass. However, the general relativity effect precesses the orbit in the opposite direction of the LK oscillations, and can thus significantly suppress the eccentricity excitation \citep{Liu2015}. In this work, we implement GR precession via the first order post-Newtonian (PN) method. The corresponding pair-wise acceleration is
\begin{equation}
\mathbf{a}_{\rm 1PN}=\frac{GM_*}{c^2r^3}\bigg[ 4(\mathbf{r}\cdot \dot{\mathbf{r}})\dot{\mathbf{r}}  + 4GM_*\hat{\mathbf{r}} - (\dot{\mathbf{r}}\cdot \dot{\mathbf{r}})\mathbf{r} \bigg].
\end{equation}

The second order PN term contributes very little to the GR precession, which can be regarded as a correction to the first order PN term with the efficiency being lower by an order of magnitude in $v/c$. Thus, it can be safely ignored in our calculations. The 2.5 PN gravitational wave radiation term in the planetary system can also be ignored due to its very long timescale. The timescale for merger due to gravitational radiation in a circular isolated binary with initial SMA $a_0$ is \citep{Peters1964}

\begin{eqnarray}
     &&\tau_{\rm GW}(e=0)=\frac{5c^5a_0^2}{256G^3m_{12}m_1m_2}\\
    &&=3.4\times10^{19}\bigg( \frac{M_\odot}{m_{12}}\bigg)\bigg( \frac{M_{J}}{m_1m_2/m_{12}}\bigg) \bigg(\frac{a_{0}}{\rm au} \bigg)^4\, {\rm yr}\,,\nonumber
\end{eqnarray}
where $m_{12}$ is the total mass of the isolated binary and $M_{J}$ is Jupiter mass. For an eccentric isolated binary, the corresponding time scale is
\begin{equation}
    \tau_{\rm GW}(e\sim 1)=\frac{768}{425}(1-e^2)^{7/2}\tau_{\rm GW}(e=0)\,.
\end{equation}
In order for the timescale $\tau_{\rm GW}(e\sim 1)$ to be smaller than the age of the Universe, $e$ needs to be larger than $\sim 0.999$, which happens in only a tiny fraction of our simulations even when eccentricity excitation mechanisms are important.

\section{Flyby modifications of planetary systems}\label{sec:flyby}
Flybys can significantly change the architecture of the target planetary system such that the modified system may undergo tidal migration due to several different mechanisms. While the single interloper flyby problem and the mechanisms that lead to high eccentricity have previously been studied \citep[e.g.][]{hamers19a,hamers19b}, it is instructive to see the fractions of hot Jupiters that emerge via different formation channels in interacting stellar environments.  These effects are quantified below in the following sections.

We mainly consider planet-planet LK oscillations, planet-planet scattering and coplanar secular interactions. In order to operate, all of these mechanisms require that the planetary system retains two planets. However, the flyby star may cause ejections of one or both planets, especially when the impact parameter is small. Clearly we must first answer the question: what is the fraction of systems that retain two planets immediately after the flyby? How does that fraction vary with impact parameter, incident angles and planet SMA ratio? 

To answer these questions, we perform groups of flyby scatterings with $a_{\rm s}/a_{\rm j}=[2,4,8]$, $i=[0^\circ, 30^\circ, 60^\circ, 90^\circ]$, $Q=[1/8, 1/4, 1/2, 1, 2, 4, 8]$ $a_{\rm s}$ and $\sigma=[1,0.2]$ km s$^{-1}$, where $a_{\rm s}$ is the SMA of the outer (Saturn) planet, $a_{\rm j}$ is the SMA of the inner (Jupiter) planet, $i$ is the incident inclination of the interloper, $Q$ is the closest approach distance between the interloper and the planetary system and $\sigma$ is the velocity dispersion of the stellar cluster. In each setup, prograde and retrograde interlopers are uniformly generated in equal numbers.

The outcomes immediately after the flyby are classified into seven sets:

\textbf{Jupiter ejection:} The inner planet is ejected due to the stellar flyby, but the outer planet remains bound to its original host star.

\textbf{Saturn ejection:} The outer planet is ejected due to the stellar flyby, but the inner planet remains bound to its original host star.

\textbf{Both ejected:} Both  planets are ejected due to the stellar flyby.

\textbf{Both remain:} Both planets remain bound to the original host star.

\textbf{Star-star collisions:} Collisions between two stars when the radii of the objects overlap.

\textbf{Star-planet collisions:} Collisions between a star and a planet when the radii of the objects overlap.

\textbf{Planet-planet collisions:} Collisions between two planets when the radii of the objects overlap.

\subsection{Orbital and System architectures after the flyby for different velocity dispersions}

\begin{figure*}
    \includegraphics[width=\textwidth]{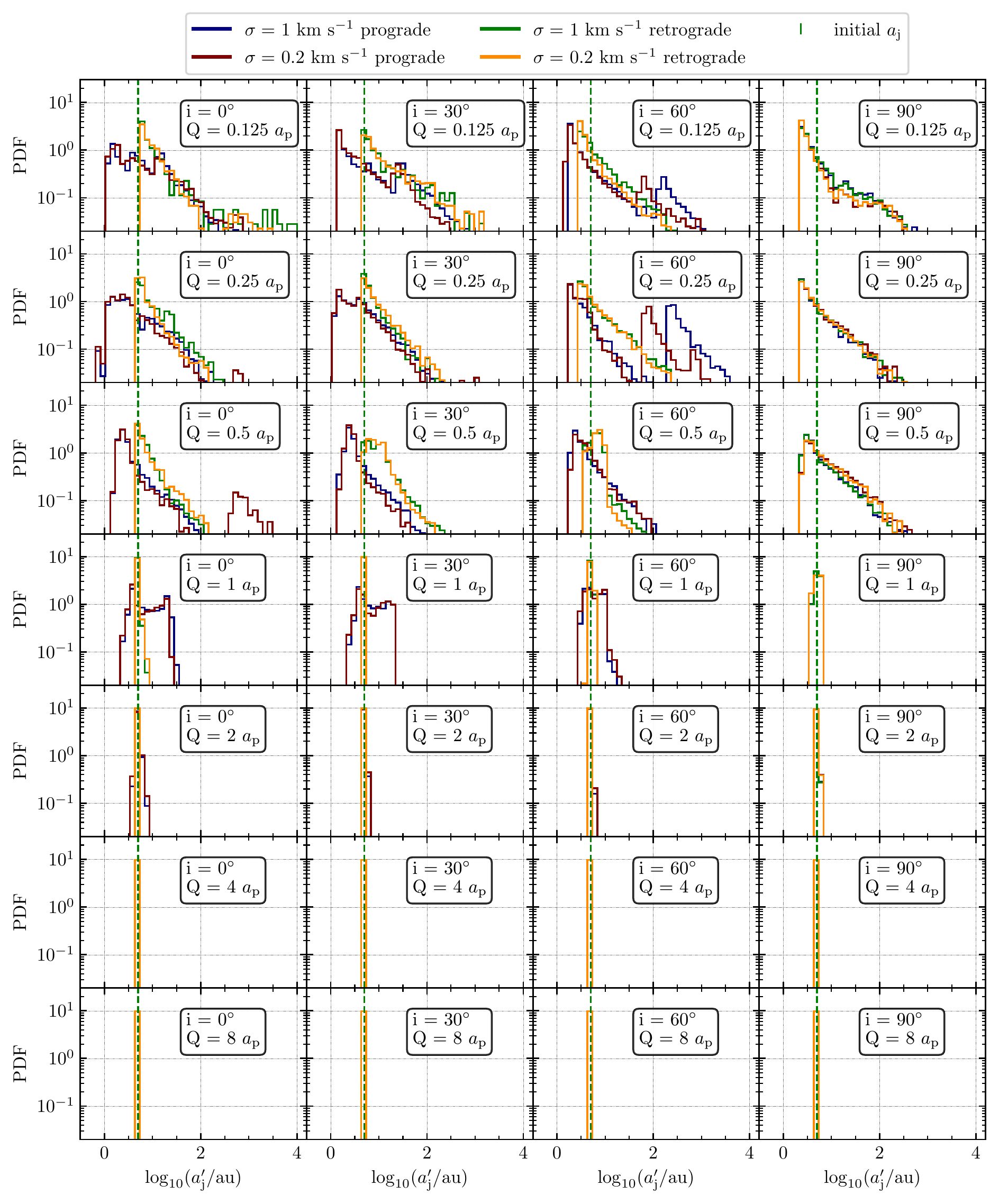}
    \caption{The distribution of the SMA of the Jupiter right after the flyby. The green vertical line shows the initial distance. The initial SMA ratio $a_{\rm s}/a_{\rm j}=2$ is fixed with $a_{\rm j}$ = 5AU. Each subplot shows the result for fixed incident inclination angle $i$ and distance of closest approach $Q$.}
    \label{fig:aj-flyby-sigma}
\end{figure*}

 We have performed more than one million scattering experiments with fixed incident angles, closest approach distances, semimajor axis ratios, and cluster velocity dispersions as described above. The orbits of all planets that remain bound to their host stars, described by their eccentricity - semimajor axis distributions, are shown in \fgr{Fig:a-etot-r2} to \fgr{Fig:a-etot-s02}.  
 
 Several features of \fgr{Fig:a-etot-r2} to \fgr{Fig:a-etot-s02} encapsulate some of the key findings of this paper. They illustrate the time evolution of the orbits of Jupiters and Saturns which survive an encounter, for the cases $a_{\rm s}/a_{\rm j}$ = 2, 4 and 8. The numbers of Jupiters reaching Saturn's orbit, and vice versa, is largest for $a_{\rm s}/a_{\rm j}$ = 2. Perhaps most striking is that hot Jupiters are {\it never} immediately formed as an aftermath of a close flyby. In contrast, both warm Jupiters (with $a_{\rm j}$ $\sim$ 1 AU) and "ultra-cold" Saturns (with $a_{\rm s}$ $\sim$ 100 - 10,000 AU {\it are} immediately form, and they persist for longer than 100 Myr. By $10^{6}$ yr hot Jupiters have been formed, and these increase in number at $10^{7}$ and $10^{8}$ yr due to tidal circularization. {{red} Ultra-cold} Saturns decrease in number as these weakly bound objects occasionally acquire enough energy to escape their host stars. 
 
 The different orbital outcomes for a range of incident angles and closest approaches by the intruder star, immediately after the flyby, and at $10^{6}$, $10^{7}$ and $10^{8}$ yr, respectively, are shown in \fgr{fig:a-et0} to \fgr{fig:a-et8}. Both Saturn and Jupiter's initial orbits are negligibly perturbed when the intruder's closest approach is further than twice Saturn's semimajor axis, and these are omitted from the figures. 
 
 As Figure 6 shows, a flyby star approaching as close as Saturn strongly perturbs that planet on the flyby timescale. It sends some Saturns inwards as close to their host stars as Jupiter, and flings others outwards as far as 10,000 AU, beginning their {{red} ultra-cold} Saturn existence. Strongly preferred and avoided regions in a-e phase space are obvious in this figure. On timescales of $10^{6} - 10^{8}$ years both planets migrate enough, under their mutual gravitational influence and tides, to smear out the sharp a-e features of \fgr{fig:a-et0}, as shown in \fgr{fig:a-et6} to \fgr{fig:a-et8}. 
 
 \fgr{fig:a-et6} shows that hot Jupiters are produced within $10^{6}$ years of the encounter. \fgr{fig:a-et6} suggests, and \fgr{fig:a-et7} and \fgr{fig:a-et8} confirm, that high inclination flybys with closest approaches inside Jupiter's orbit are by far the most effective way to make hot Jupiters. 
 
 In each subplot of \fgr{fig:a-et0} to \fgr{fig:a-et8}, with fixed $i$ and $Q$, the initial phases of the two planets are uniformly generated in time with eccentricity equal to zero. We see that for perturbations with $Q>2a_{\rm p}$, where $a_{\rm p}$ is the semi-major axis of the outer planet (Saturn), the flyby star hardly changes the semi-major axis of the Jupiter, regardless of the incident inclinations. With $Q<2a_{\rm p}$ and even $Q<a_{\rm p}$, where the interloper penetrates the planetary system, there are no directly formed hot Jupiters with $a_{\rm j}^\prime < 0.09$ AU. This is because the flyby star must extract most of the energy from the Jupiter orbit to make it shrink below $\sim$ 0.09 AU, which is almost impossible within a single flyby. From this plot we also see that, for a given $i$ and $Q$, the cluster velocity dispersion has little effect on the distribution of the SMA of the Jupiter right after the flyby. This indicates that,  in the strong interaction regime $Q<a_{\rm p}$, the energy change of the inner planet is insensitive to the velocity dispersion of the environment. This is because the velocity of the interloper at the closest approach is
\begin{equation}\label{eq:vq}
v_{Q} = \sqrt{v_\infty^2 + \frac{2GM_{\rm tot}}{Q}}\,,
\end{equation}
where $M_{\rm tot}$ is the total mass of the scattering system. For a velocity dispersion $\sigma$\,=\,0.2 or 1~km s$^{-1}$, $v_\infty$ is negligible compared to ${2GM_{\rm tot}}/{Q}$ with small $Q$. Therefore, the term $v_{Q}$ is not sensitive to the environmental velocity dispersion in the strong interaction regime.

In each subplot, we also see clear differences between the prograde and retrograde orbits. For scatterings with fixed $Q$ but different $i$, we see that in the prograde scatterings, the Jupiter shrinks its SMA more than in the case of retrograde scatterings. Since in the prograde scattering the interloper has more time to interact with the Jupiter at the point of closest approach due to the low relative velocity between the interloper and the planets, the interloper is able to extract more energy from Jupiter's orbit. This trend begins to disappear as the inclination $i$ increases to  $90^\circ$. In the $90^\circ$ case, with the interloper traveling perpendicular to the planet orbital plane, the interaction time between the interloper and the planet is the same in the prograde and retrograde scatterings.

As shown in each row of  \fgr{fig:aj-flyby-sigma}
that, as  $i$ increases, the SMA changes become smaller, due to the shorter interaction time between the interloper and planet at the point of closest approach. As  $i$ increases, the maximum energy that can be extracted from the planetary system becomes smaller.

We also see in many of the subplots in \fgr{fig:aj-flyby-sigma} that there are two peaks in both prograde and retrograde scatterings. In total, we have four peaks from different relative positions at closest approach: {\em (i)} Prograde scattering with interloper and planet at the same side of the star (interloper and planet move in the same direction with small relative distance, thus contributing to the dominant peak). {\em (ii)} Retrograde scattering with the interloper and planet at the opposite side of the star (interloper and planet move in the same direction with large relative distance), which contributes to the secondary dominant peak. {\em (iii)} Prograde scattering with the interloper and planet at opposite sides of the star (interloper and planet move in opposite directions with large relative distance), which contribute to the third prominent peak. {\em (iv)} Retrograde scattering with interloper and planet at the same side of the star (interloper and planet move in the opposite direction with small relative distance), which contributes to the weakest peak that is almost invisible.

The distance between the peaks becomes smaller as $Q$ increases. The reason is that, for large $Q$, the distance between the interloper and planet, $Q\pm a_{\rm j}$, is dominated by $Q$ instead of by $\pm a_{\rm j}$. Thus, no matter which side of the star the planet is located on, i.e.  $+a_{\rm j}$ or $-a_{\rm j}$, the distance between the interloper and the planet in either case becomes similar.

\begin{figure*}
    \includegraphics[width=\textwidth]{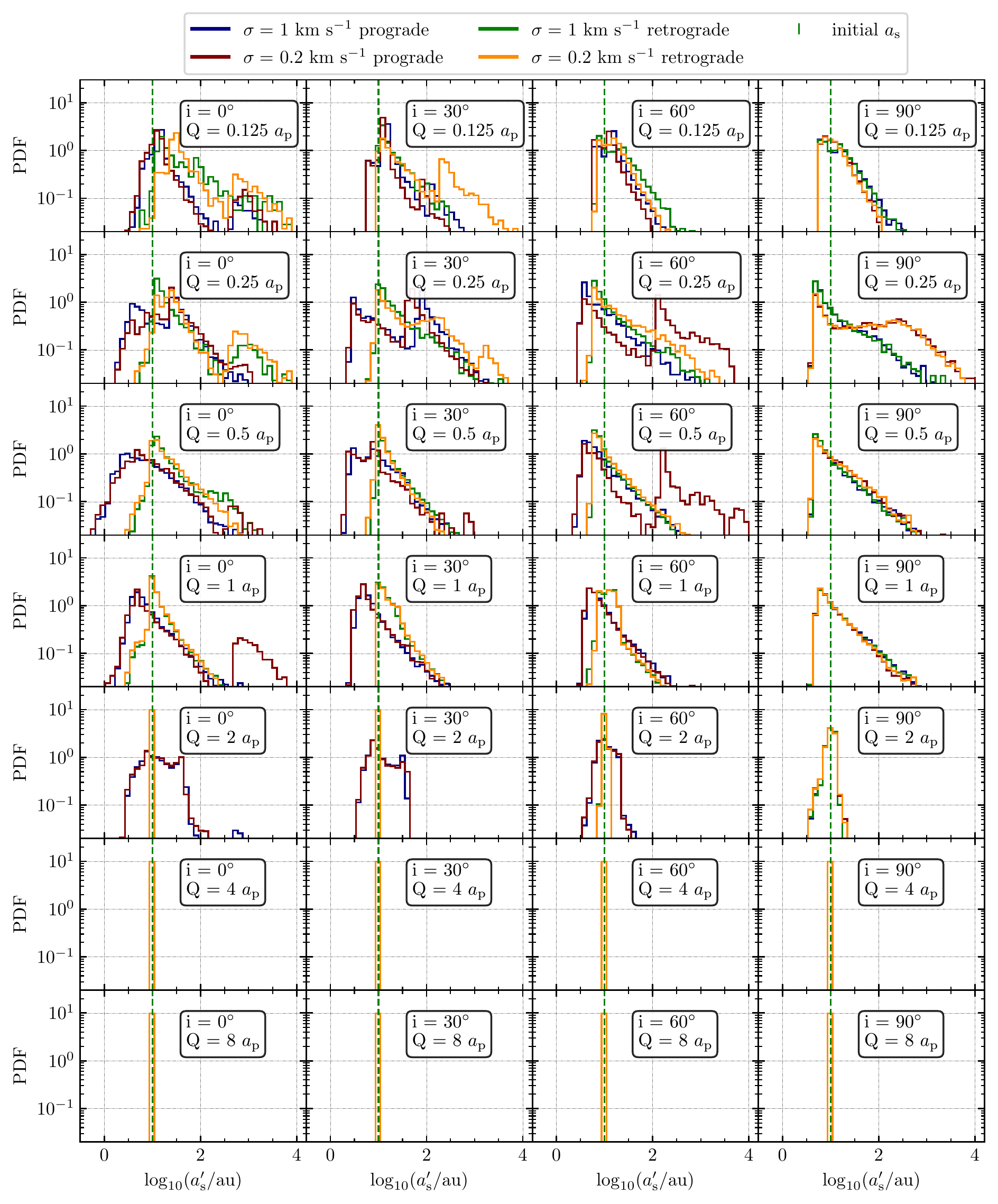}
    \caption{The distribution of the SMA of the Saturn right after the flyby. The initial conditions are the same as in \fgr{fig:aj-flyby-sigma}.}
    \label{fig:at-flyby-sigma}
\end{figure*}

\fgr{fig:at-flyby-sigma} shows the distribution of the SMA of the Saturn right after the flyby. Because the outer planet orbit is more weakly bound, it is easier to modify the orbit of the Saturn relative to that of the Jupiter. In general, the orbit of the Saturn right after the flyby shows similar properties to that of the Jupiter. However, we see more differences arising when adopting different velocity dispersions, especially in prograde scatterings. This is because, with fixed $Q$, the relative velocity between the interloper and the planets determines the interaction time between the interloper and the planet, which affects the properties of the planet right after the flyby. The relative velocity between  interloper and planet at the distance of closest approach is determined by the planet's orbital velocity and the velocity of the interloper at closest approach $v_{\rm Q}$. In the case of Jupiter, due to the tight orbit, the orbital velocity is dominant over the velocity of the interloper, thus the velocity dispersion does not affect much the relative velocity between the interloper and the planet at the point of closest approach. However, for the Saturn with smaller orbital velocity, the orbital velocity becomes less dominant over the velocity of the interloper. Thus, the velocity dispersion has a larger impact on the relative velocity between the interloper and the Saturn at the point of closest approach.

The SMA change of the planets can be estimated by 
\begin{equation}
    dE = E_{\rm p } - E_{\rm p ,0} = -\frac{GM_*M_{\rm p}}{2a_{\rm p}^\prime} + \frac{GM_*M_{\rm p}}{2a_{\rm p}}\,,
\end{equation}
where $a_{\rm p}$ is the initial SMA and $a_{\rm p}^\prime$ is the SMA immediately after the flyby. It is straightforward to note that, if $(E_{\rm p} - E_{\rm p,0})/|E_{\rm p,0}| < 0$, the corresponding orbit shrinks, while if $0 < (E_{\rm p} - E_{\rm p,0})/|E_{\rm p,0}| < 1$, the orbit expands. For $(E_{\rm p} - E_{\rm p,0})/|E_{\rm p,0}| > 1$, the planet is ejected from its host star. As the incident angle increases, the maximum relative binding energy shift becomes smaller as the interaction time at the point of closest approach decreases. Therefore, the binding energy shift becomes weaker as $Q$ increases.

The prograde scatterings are responsible for the majority of the planet ejections where $(E_{\rm p} - E_{\rm p,0})/|E_{\rm p,0}| > 1$. The prograde scatterings with interloper and planets at the same side of the host star have the longest interaction times (and hence impart the largest impulse) needed to remove the binding energy from the planet. Since the retrograde encounter is not as efficient as the prograde case in removing the binding energy of the planet due to its shorter interaction time, the retrograde scattering can hardly change the SMA of the Jupiter at $Q\sim 4a_{\rm j}$ and the SMA of the Saturn at $Q\sim 4a_{\rm s}$, whereas prograde scatterings do change the SMA of the planet significantly (by up to about one order of magnitude).

\begin{figure*}
    \includegraphics[width=\textwidth]{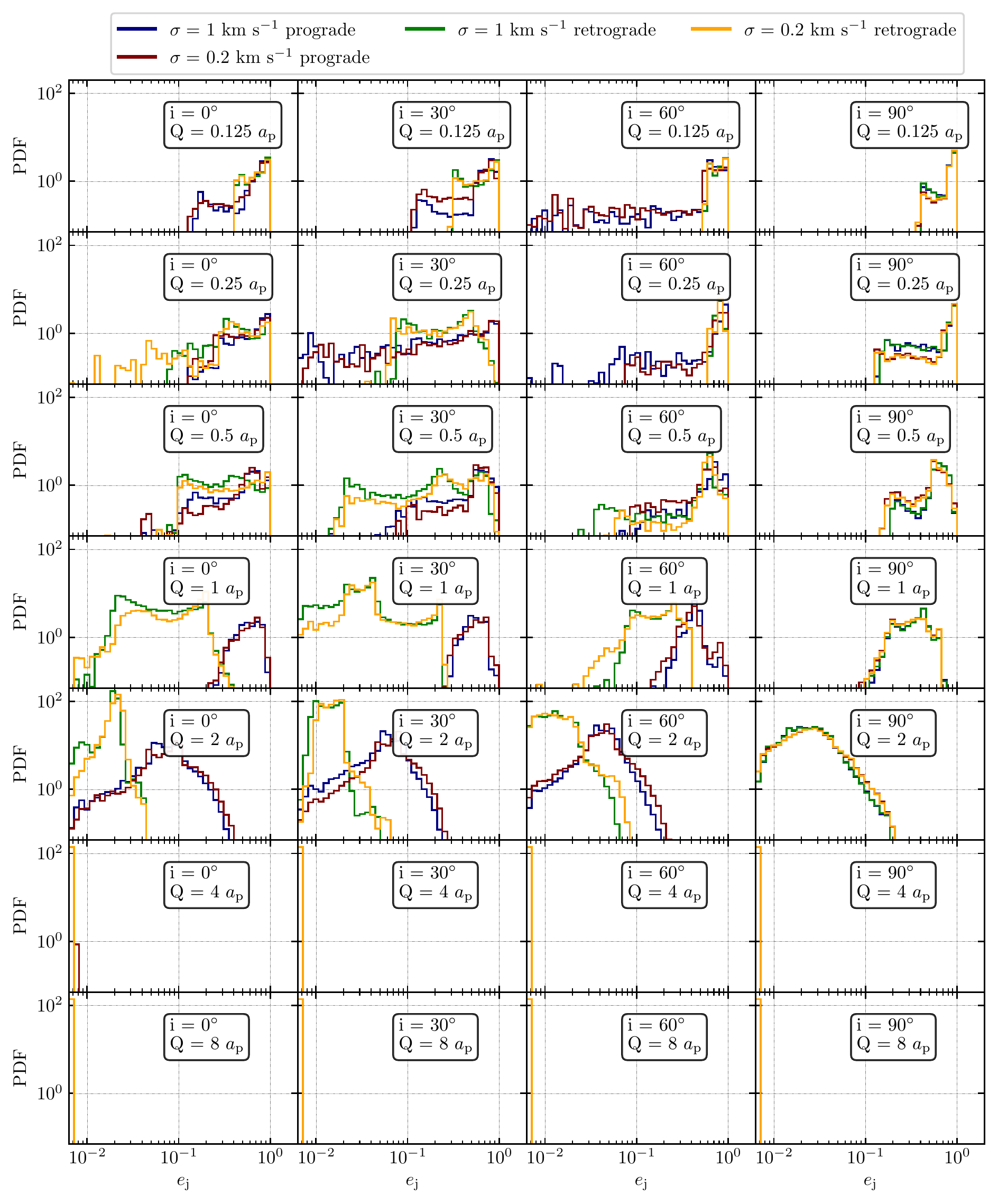}
    \caption{The distribution of the eccentricity of the Jupiter immediately after the flyby.  The initial conditions are the same as in \fgr{fig:aj-flyby-sigma}.} 
    
    \label{fig:ej-flyby-sigma}
\end{figure*}

\begin{figure*}
    \includegraphics[width=\textwidth]{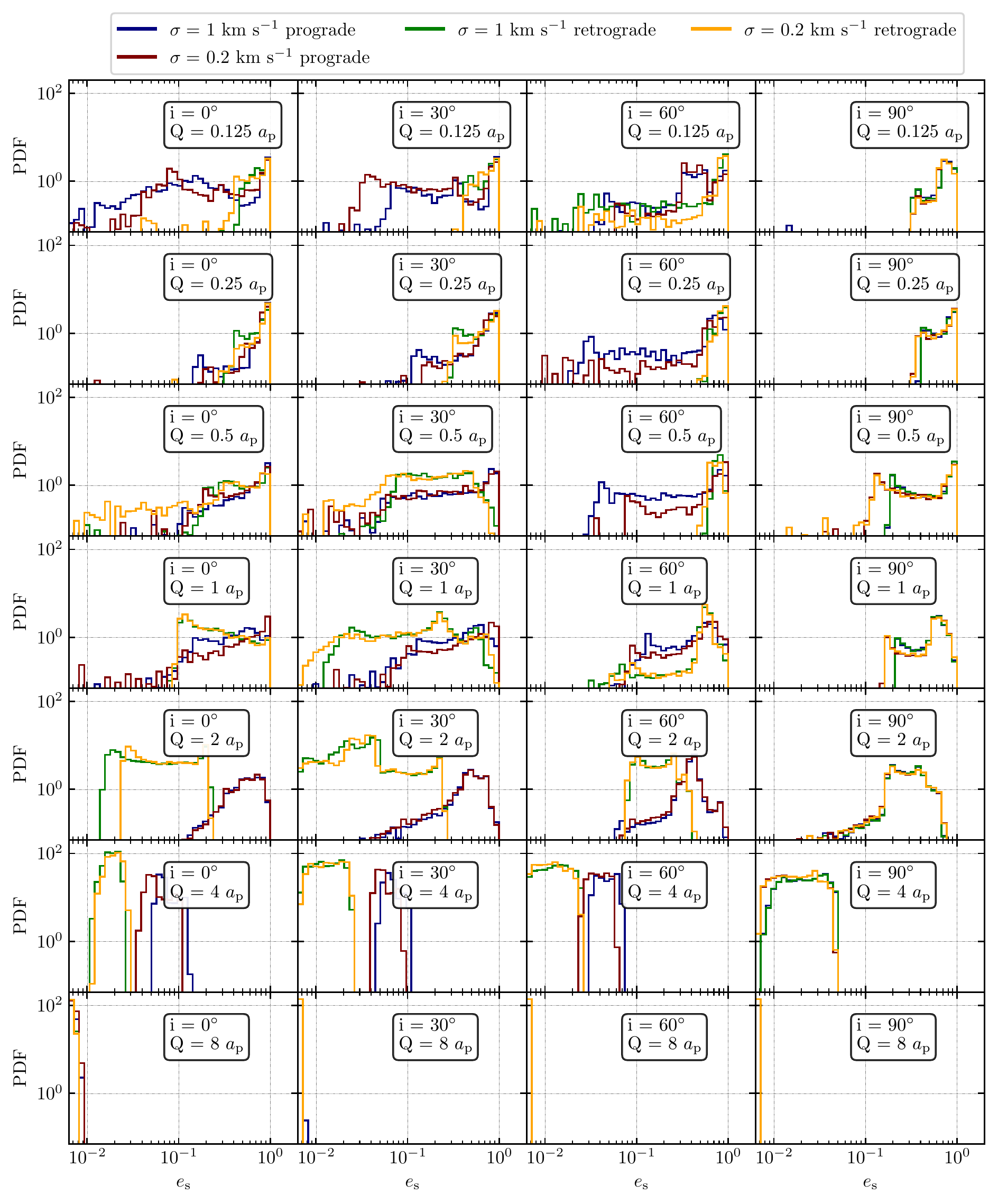}
    \caption{The distribution of the eccentricity of the Saturn right after the flyby.  The initial conditions are the same as in \fgr{fig:aj-flyby-sigma}.} 
    
    \label{fig:es-flyby-sigma}
\end{figure*}

\fgr{fig:ej-flyby-sigma} and \fgr{fig:es-flyby-sigma} show the eccentricity of the Jupiter and Saturn right after the flyby. Since the initial value of the eccentricity of the planets is zero, this also indicates the eccentricity change right after the first flyby. For $Q<4 a_{\rm j}$ and $Q < 4a_{\rm s}$ (i.e., the strong interaction regime) a flyby can significantly kick the eccentricity of a planet to a large  value. For $Q < a_{\rm j}$ or $Q<a_{\rm s}$, where the interloper penetrates within the orbit of the planetary system, the majority of the planets are perturbed into the ``near ejection" state where $e\sim 1$.  As for $Q>4 a_{\rm j}$ and $Q > 4a_{\rm s}$, where scattering happens in the perturbative regime, the eccentricity shift due to the flyby rapidly drops below 0.1. These results indicate that just one flyby does not efficiently increase the eccentricity of a planet, which is essential to enable the subsequent high eccentricity tidal migration process.

From \fgr{fig:ej-flyby-sigma} and \fgr{fig:es-flyby-sigma}, we can also see that as the inclination $i$ increases, retrograde scatterings become similar to the prograde scatterings in removing the angular momentum of the planets. 

\begin{figure*}
    \includegraphics[width=\textwidth]{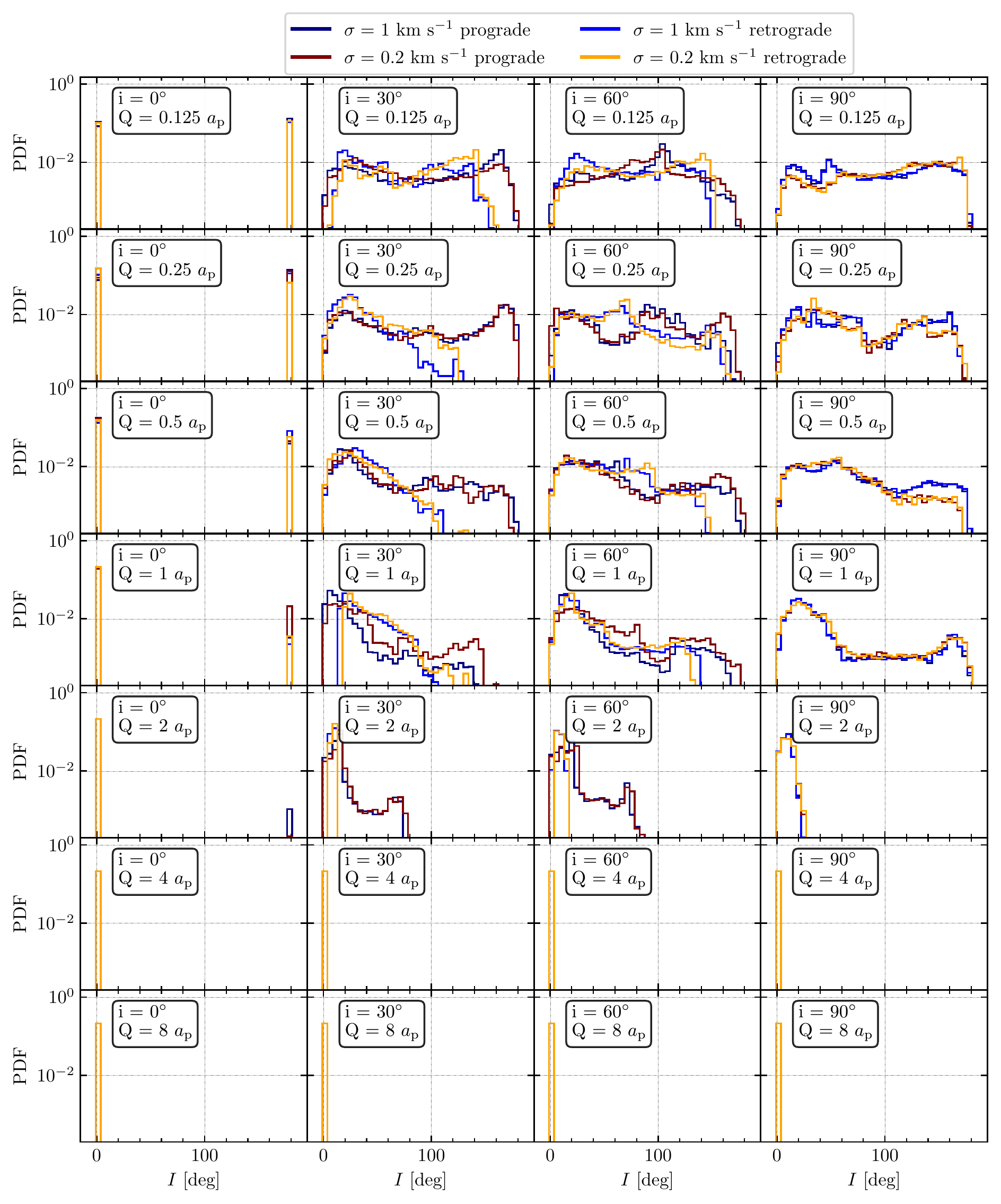}
    \caption{The distribution of the inclination between the two planet orbits right after the flyby.  The initial conditions are the same as in \fgr{fig:aj-flyby-sigma}.}  
    \label{fig:inc-flyby-sigma}
\end{figure*}

\fgr{fig:inc-flyby-sigma} shows the inclination between the Jupiter and Saturn orbits right after the flyby if both the Jupiter and Saturn still remain bound to their original host star. When the closest approach Q is smaller than $a_{\rm p}$, we see that it is possible for the interloper to flip the orbital orientation of the Saturn to the opposite direction by transferring sufficient angular momentum to its orbit. Also, as the incident inclination $i$ increases, we see that the probability of creating an inclined Saturn orbit with respect to the Jupiter orbit increases as well, especially in the prograde flyby case.
It is difficult to get a highly inclined orbit when $Q>a_{\rm p}$, such that Lidov-Kozai oscillations are unlikely to operate on the system after the flyby to help make hot Jupiters.

From \fgr{fig:aj-flyby-sigma} to \fgr{fig:inc-flyby-sigma}, we explore the properties of the planetary system right after the flyby with two different values of the environmental velocity dispersion, $\sigma$=0.2 or 1~km s$^{-1}$. We find that close flybys create an ideal situation for tidal migration to produce hot Jupiters. Close flybys can excite the eccentricity of the planets by extracting significant angular momentum from planetary systems and kick the original coplanar planetary system into an inclined planetary system. These two processes increase the efficiency of the tidal migration process.

We find that the orbital properties are insensitive to the low velocity dispersion. The velocity of the interloper at infinity contributes very little to the velocity of the interloper at the distance of closest approach for small $Q$ due to the strong gravitational focusing, as indicated in \eqn{eq:vq}.

\subsection{System architecture after the flyby with different initial SMA ratios}\label{sec:sma-ratio}
Besides exploring how the cluster velocity dispersion changes the orbital properties of the planetary system after the flyby, we also explore the flyby scattering with different initial SMA ratios. We fix the velocity dispersion to be 1 km~s$^{-1}$ and perform scatterings with initial SMA ratios equal to 2, 4 and 8.

Before proceeding, we briefly comment on the possible impact of orbital resonances on our results.  The naive expectation here is that, for smaller initial SMA ratios, such resonances should be more important as the initial SMA ratio is closer to a strong resonance (e.g., the 2:1 resonance for an initial SMA ratio of 2).  As the initial ratio of SMAs increases, we expect resonances to become less important, as these resonances would tend to operate at very high ratios of the orbital periods, where we expect resonances to be very weak and have little to no dynamical effects. In our simulations, we see only mild evidence for systems becoming locked into orbital resonances for initial SMA ratios of 2, with at most of order a few percent of our total simulations ending up stabilized in a resonant state.  For higher initial SMA ratios, these effects disappear.  This is expected since the perturbed planetary systems will be scattered closest to mainly higher-order resonances, which tend to get weaker as the ratio of periods increases.  In the end, we find that orbital resonances have a very negligible effect on our overall results, statistics and conclusions, at least for the initial conditions considered here.  Thus, these effects can safely be ignored in this section when trying to understand how the product distributions from our simulations depend on the initial SMA ratio.

\begin{figure*}
    \includegraphics[width=\textwidth]{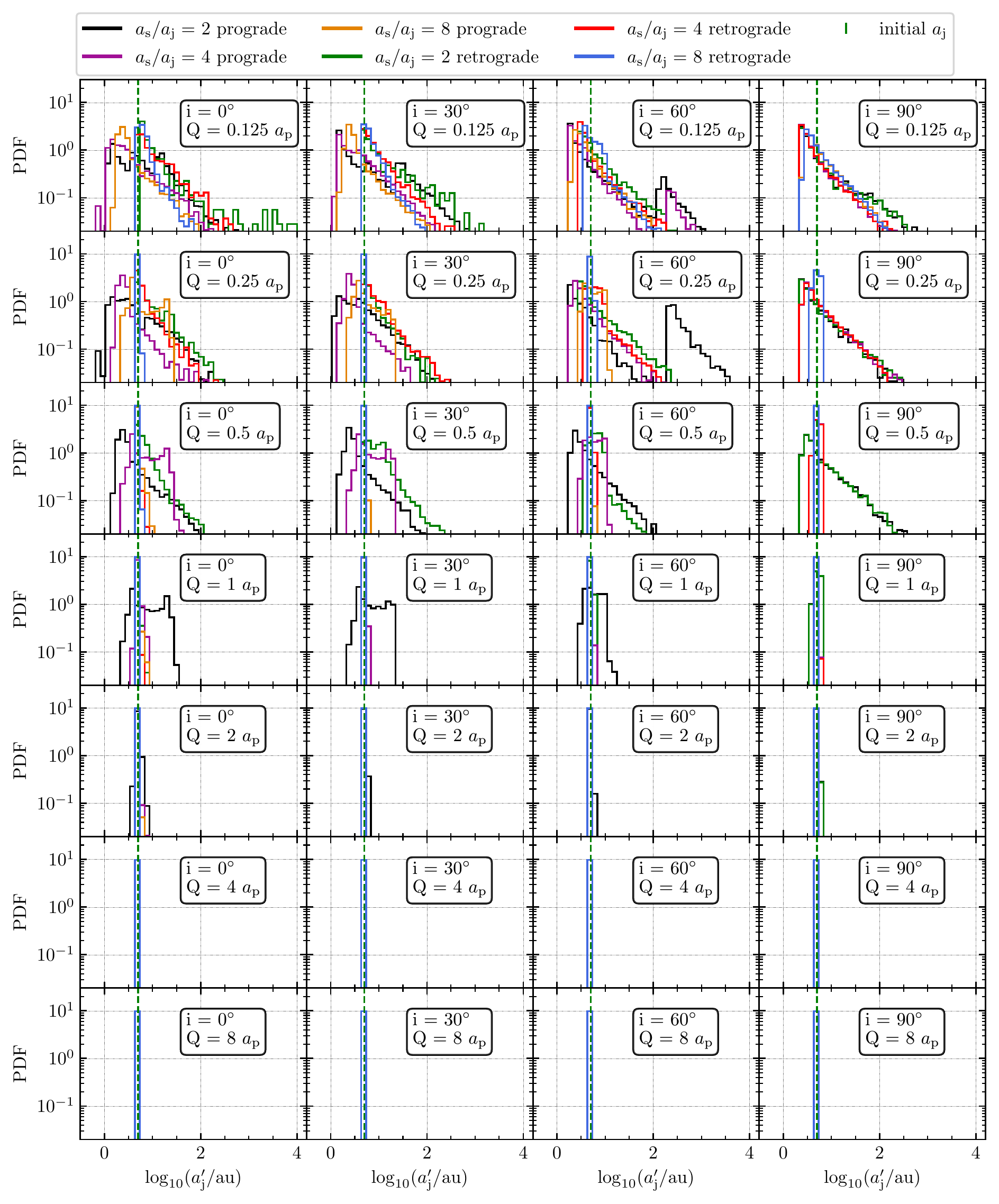}
    \caption{The distribution of the SMA of the Jupiter right after the flyby, for different initial SMA of the Saturn. The velocity dispersion is $\sigma=$1 km s$^{-1}$ and the initial SMA of Jupiter is $a_{\rm j}$ = 5 AU. Each subplot shows the result for fixed incident inclination angle $i$ and distance of closest approach $Q$.}
    \label{fig:aj-flyby-ratio}
\end{figure*}

\begin{figure*}
    \includegraphics[width=\textwidth]{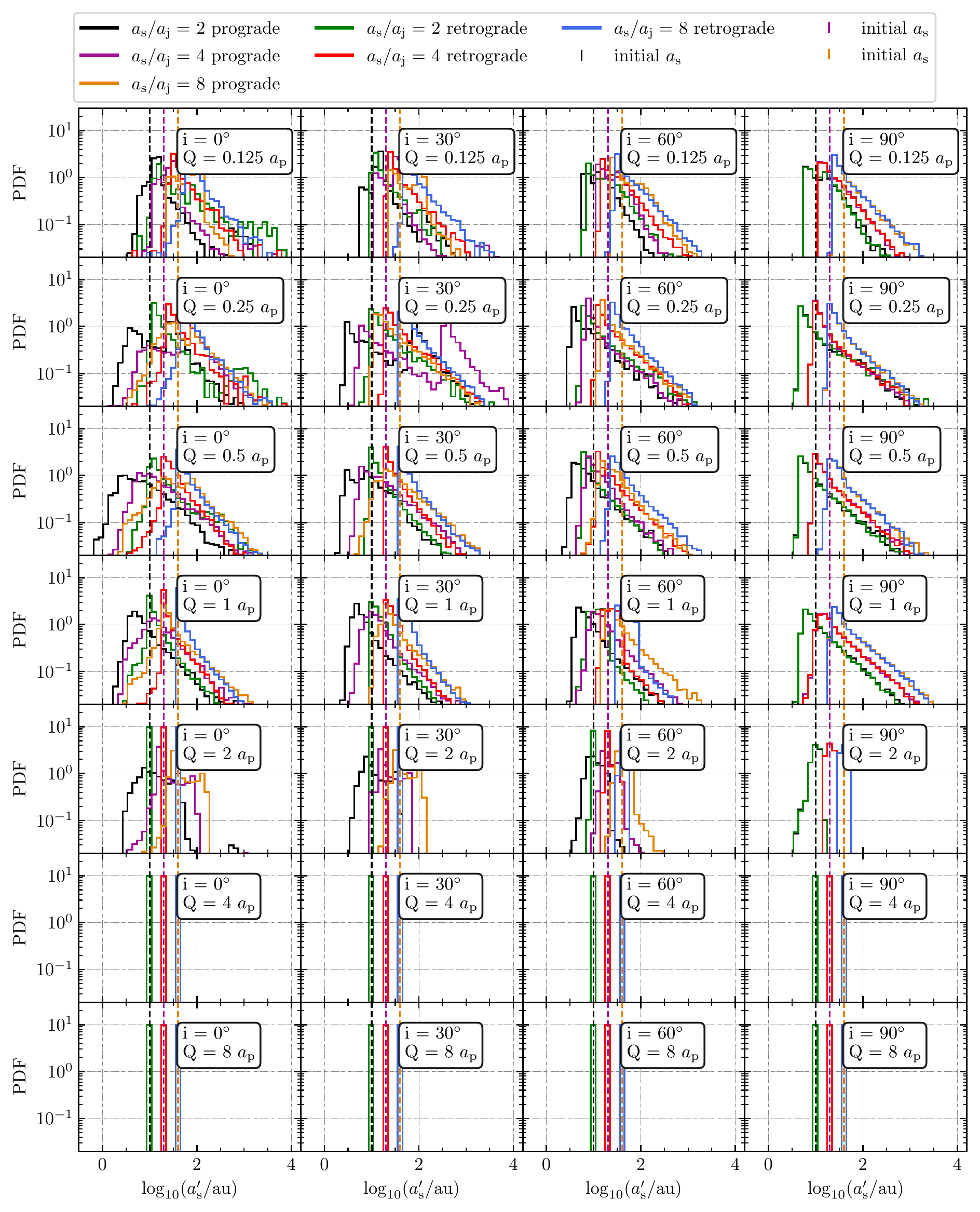}
    \caption{The distribution of the SMA of the Saturn right after the flyby.  The initial conditions for $\sigma$, $a_j$, and the three values of $a_s$ are the same as in \fgr{fig:aj-flyby-ratio}.} 
    \label{fig:as-flyby-ratio}
\end{figure*}

\begin{figure*}
    \includegraphics[width=\textwidth]{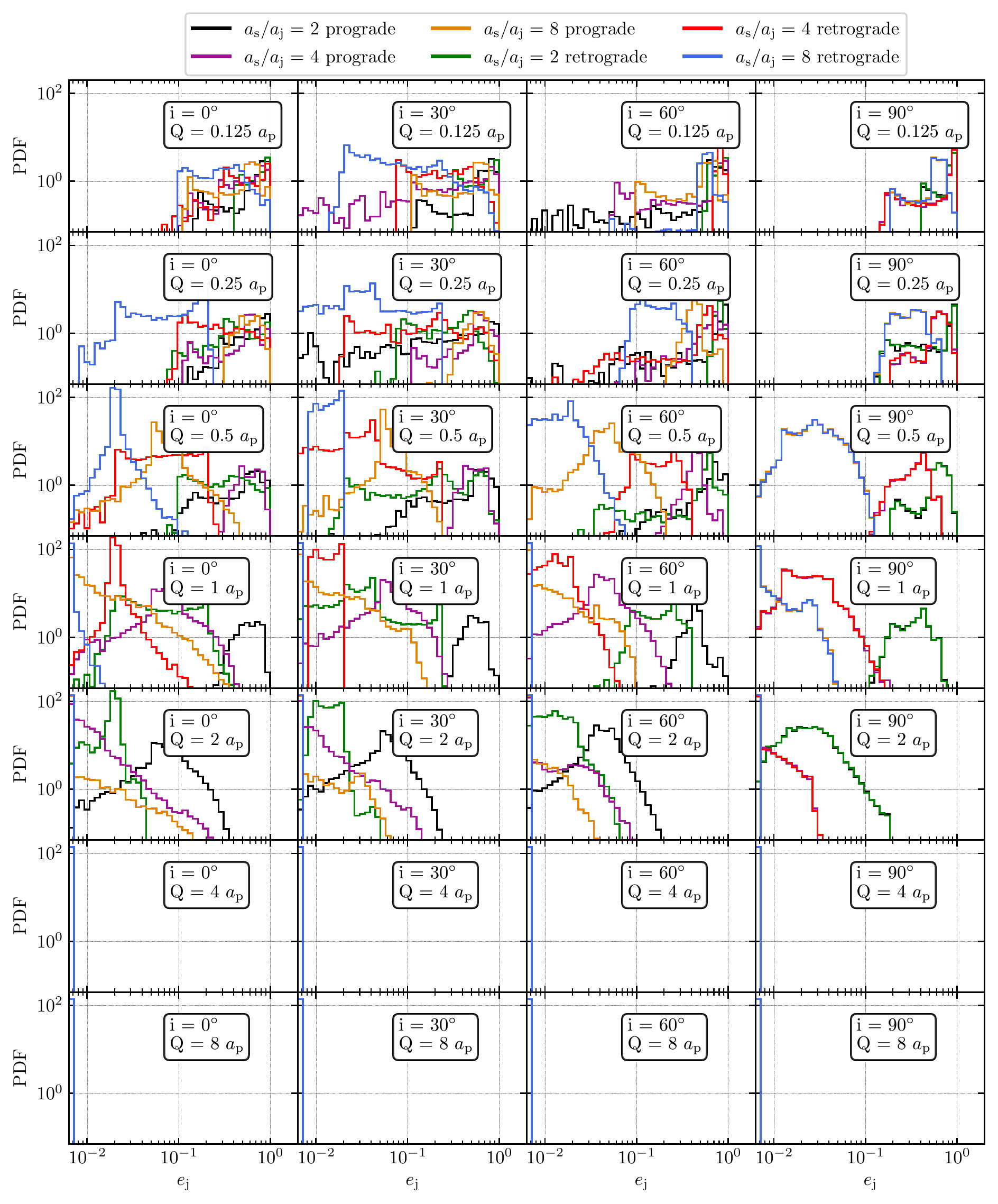}
    \caption{The distribution of the eccentricity of the Jupiter immediately after the flyby. The initial conditions are  the same as in \fgr{fig:aj-flyby-ratio}.} 
    \label{fig:ej-flyby-ratio}
\end{figure*}

\begin{figure*}
    \includegraphics[width=\textwidth]{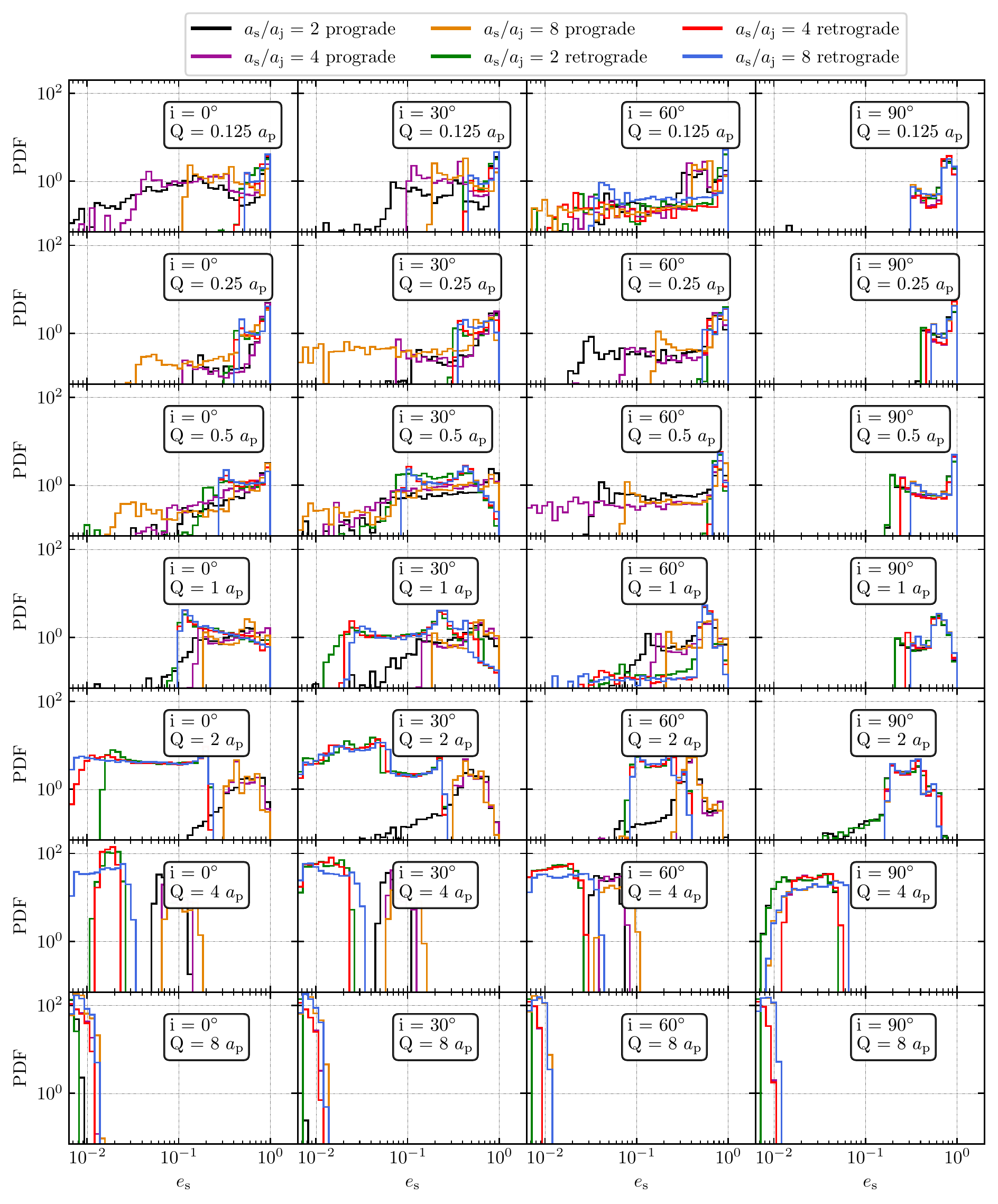}
    \caption{The distribution of the eccentricity of the Saturn right after the flyby. The initial conditions are the same as in \fgr{fig:aj-flyby-ratio}.} 
    \label{fig:es-flyby-ratio}
\end{figure*}

As in \fgr{fig:aj-flyby-sigma}, \fgr{fig:aj-flyby-ratio} shows the semi-major axis of the Jupiter right after the flyby for different initial planet semi-major axis ratios. It is indicated in the figure that, as we double the semi-major axis ratio, the distribution of $a_{\rm j}$ is  almost identical to the distribution of $a_{\rm j}$ with the original ratio but with $Q$ doubled, e.g. the $a_{\rm s}/a_{\rm j}=8$ with $Q=0.125a_{\rm p}$ is almost identical to $a_{\rm s}/a_{\rm j}=4$ with $Q=0.25a_{\rm p}$. It is not surprising that the absolute closest approach determines the post-flyby semi-major axis and eccentricity (as shown in \fgr{fig:ej-flyby-ratio}) of the Jupiter. The reason is that the inter-planet interaction in the close flyby is negligible due  to the short interaction flyby time. Thus, we can treat the perturbation exerted on each planet almost independently.

As already noted and discussed, orbital inclinations are created by highly inclined flybys. The changing z-direction to the interloper changes the direction and magnitude of the angular momentum of the planets. For fixed $Q$, as the SMA ratio decreases, the SMA of the Jupiter decreases while keeping the SMA of the Saturn constant. Since the angular momentum of the planet is 
\begin{equation}
    L_{\rm p} \propto \sqrt{a(1-e^2)},
\end{equation}
then, for a smaller SMA, it is easier to change the direction of the inner planet by importing/extracting the same angular momentum to/from the planet's orbit. Hence we can see in \fgr{fig:inc-flyby-ratio} that, with a smaller SMA ratio, the planetary system tends to be more inclined after the flyby. This argument is also valid in coplanar flybys: smaller SMA ratios also create more retrograde orbits in all those interactions corresponding to the closest approaches.

We can further analyze the SMA ratio after the flyby with the same method by treating the perturbation exerted on each planet independently. The binding energy of the planet is
\begin{equation}
    E_{\rm p}\propto 1/a.
\end{equation}
Therefore, it is harder to change the SMA of the planet with smaller initial SMA by importing/extracting the same energy to/from the planet's orbit. With fixed $Q$, a smaller initial SMA ratio gives a smaller absolute SMA for the Jupiter. It becomes more difficult to change the SMA of the Jupiter. Thus, the final SMA ratio right after the flyby depends more on the change in SMA of the Saturn. 

Since the absolute SMA of the planets determines if the interaction between the planet and the interloper is in the fast/slow regime, it is difficult to make a simple argument on how the final SMA ratio depends only on the initial SMA ratio. For instance, if the initial SMA ratio is 4, where both Saturn and Jupiter reside in the slow regime where the flyby tends to increase the SMA, then the final SMA tends to become larger. The reason is that, while the SMA of both planets tends to become larger, the change in Saturn's SMA is more significant. If the initial SMA ratio is 2, Jupiter's orbit may enter into the faster regime, where the flyby tends to tighten  Jupiter's orbit while keeping the Saturn in the slow regime. Thus, although smaller initial SMA ratios make the final SMA ratio more weakly dependent on Jupiter's orbit, the final SMA ratio tends to become larger than in the case where the initial SMA ratio is equal to 4 because the two planets migrate in opposite directions.

To make the planets interact more efficiently, the SMA ratio of the planets right after the flyby should not be too large nor too small. Two close orbits make planet-planet scattering possible during the subsequent long term evolution after the flyby and shorten the timescale for secular processes to operate (under the condition that the system is stable).

\begin{figure*}
    \includegraphics[width=\textwidth]{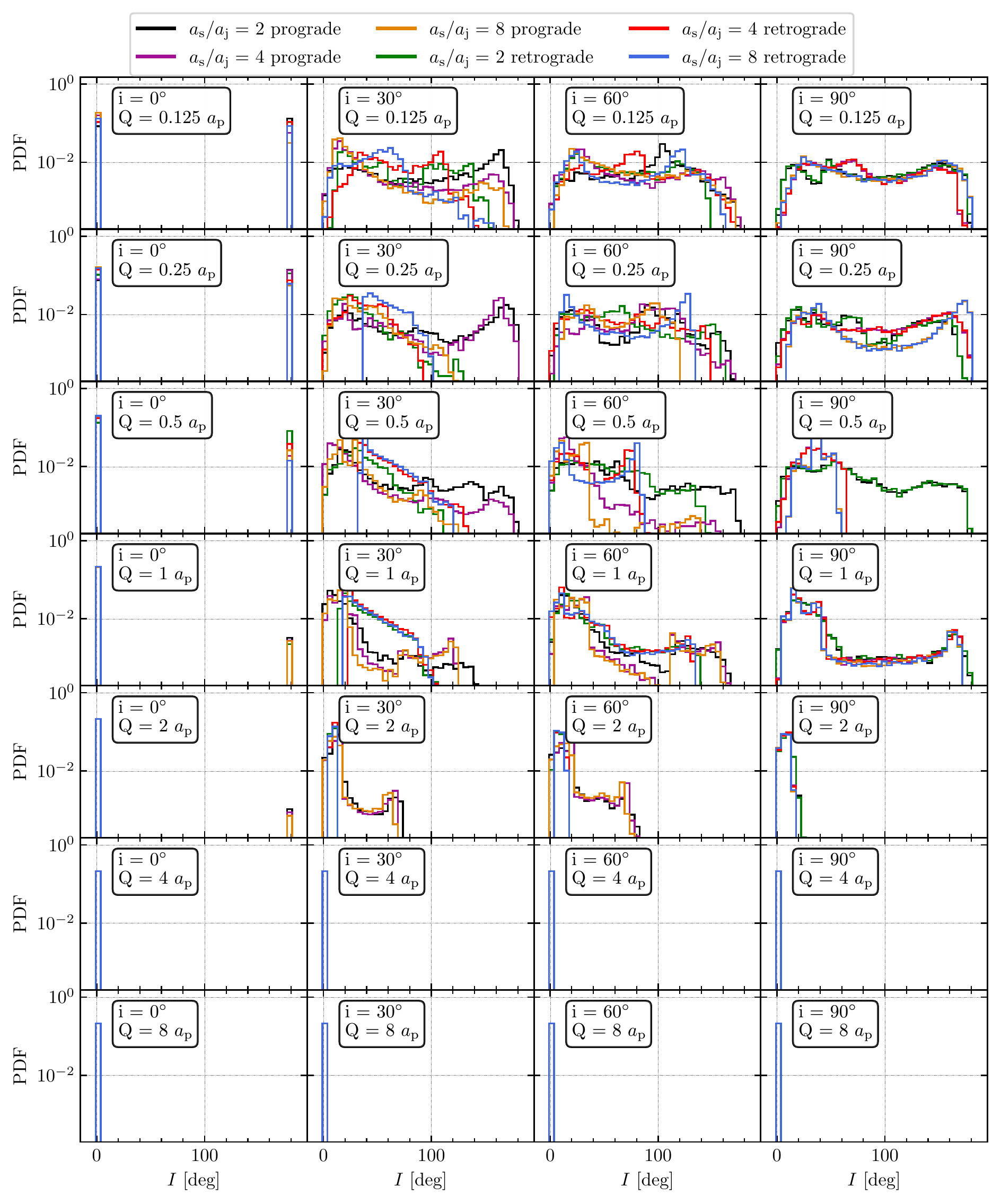}
    \caption{The distribution of the inclination between the two planet orbits right after the flyby. The initial conditions are the same as in \fgr{fig:aj-flyby-ratio}.}
    \label{fig:inc-flyby-ratio}
\end{figure*}

Because of the short flyby time, the interaction between the planets during the flyby at the point of closest approach can be safely ignored. Hence, we can treat the perturbation exerted by the interloper on the two planets independently. Therefore, the scattering with different initial SMA ratios affects the properties of the planets independently, as shown in \fgr{fig:aj-flyby-ratio} to \fgr{fig:es-flyby-ratio}.

However, the initial SMA ratio affects the inclination and final SMA of the planetary system right after the flyby, thus influencing the probability and timescale for subsequent tidal migration. We will discuss the corresponding `sweet spot' for the initial SMA ratio in the next section.

\subsection{Event fraction right after the flyby}
As listed in \sectn{sec:flyby}, the outcomes after the flyby are classified into `planet-planet collision', `star-planet collision', `star-star collision', `Jupiter ejection', `Saturn ejection', `Both ejection' and `Both remain'. In this subsection we will discuss the fraction of each outcome for different environmental velocity dispersions and initial SMA ratios. We pay special attention to the `Both-remain' systems since these are obviously required for post-interaction tidal migration.

\begin{figure*}
    \includegraphics[width=\textwidth]{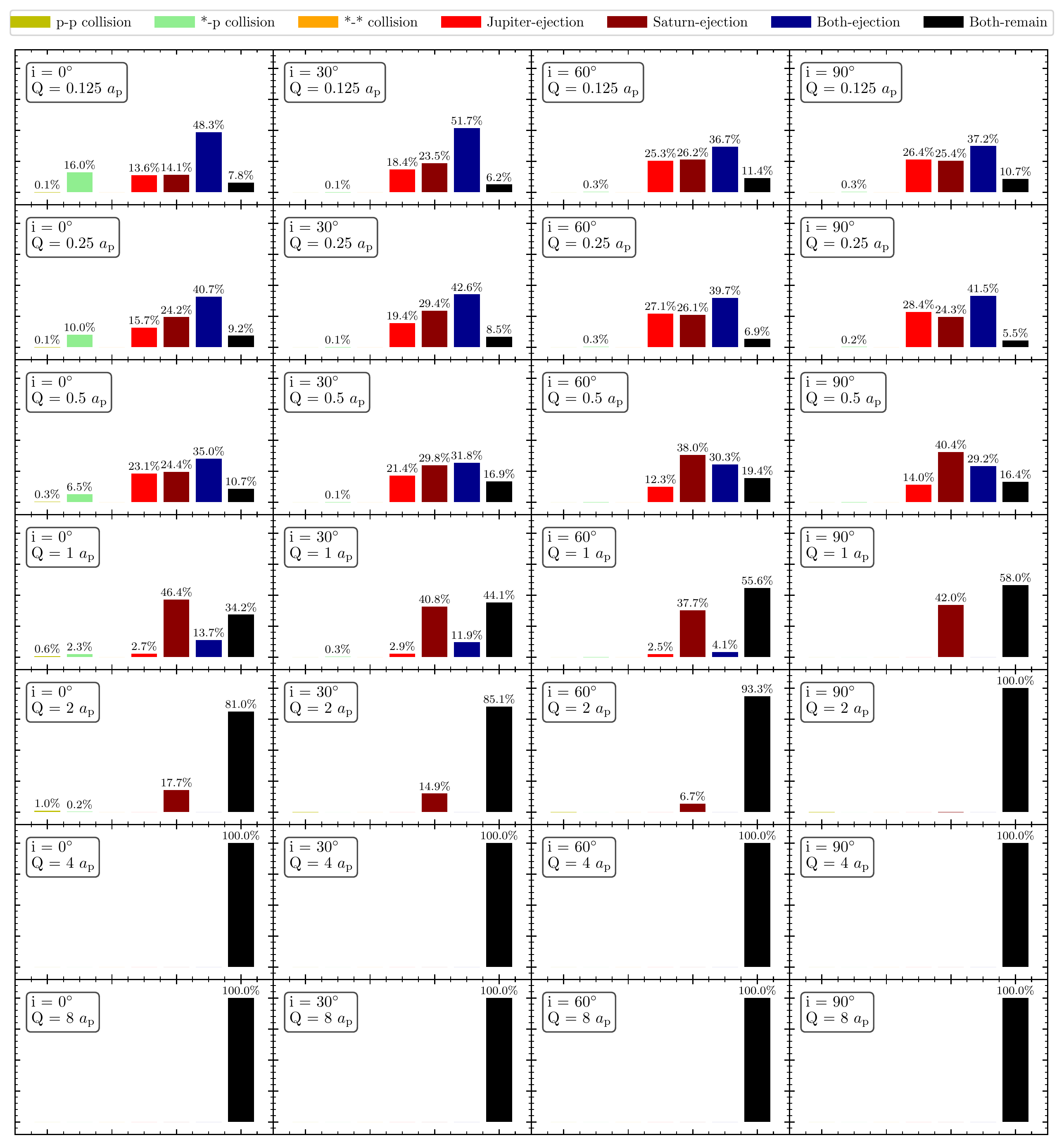}
    \caption{The event fraction right after the flyby. Each subplot shows the result for fixed incident inclination $i$ and closest approach $Q$. The initial SMA ratio is $a_{\rm s}/a_{\rm j}=2$ with $a_{\rm j}$ = 5 AU and the velocity dispersion $\sigma$ is 1~km~s$^{-1}$.}
    \label{fig:flyby-frac-s1-r2}
\end{figure*}
\fgr{fig:flyby-frac-s1-r2} shows the standard case where $a_{\rm s}/a_{\rm j}=2$ and $\sigma=1$~km~s$^{-1}$. We can see that for $Q>4a_{\rm p}$, the flyby will never change the architecture of the planetary system, and therefore all the outcomes are `Both-remain'. As the closest approach comes close to the planetary system with $Q\sim 2a_{\rm p}$, the flyby begins to eject the outer planet Saturn and then eject the inner planet Jupiter. As shown in the row with $Q=2a_{\rm p}$, high inclination incident angles where the interaction time is shorter decrease the ejection fraction for Saturn alone. 

The fraction of the `Both-remain' in the close flyby regime where $Q<a_{\rm p}$ is roughly 10\%. We also find that most of the collisions are planet-star collisions that only appear in coplanar scatterings.  The reason for this is related to the volume-filling factor of the objects, which is much smaller for coplanar configurations relative to the isotropic case \citep{leigh17,leigh18}.  In other words, there is more free space in the isotropic case, increasing the timescale for two objects to coincide in both time and space.

As an aside we note that planet-star collisions can give rise to a rich set of possible outcomes \citep{schuler2011,demarco2011,
kratter2012,ginsburg2012,veras2013,deal2015}. In particular, planets that are engulfed may shape the ejecta of their host stars, while swallowed and disrupted planets may enrich the outer layers of stars in detectable lithium, calcium, magnesium, and iron.  

\begin{figure*}
    \includegraphics[width=\textwidth]{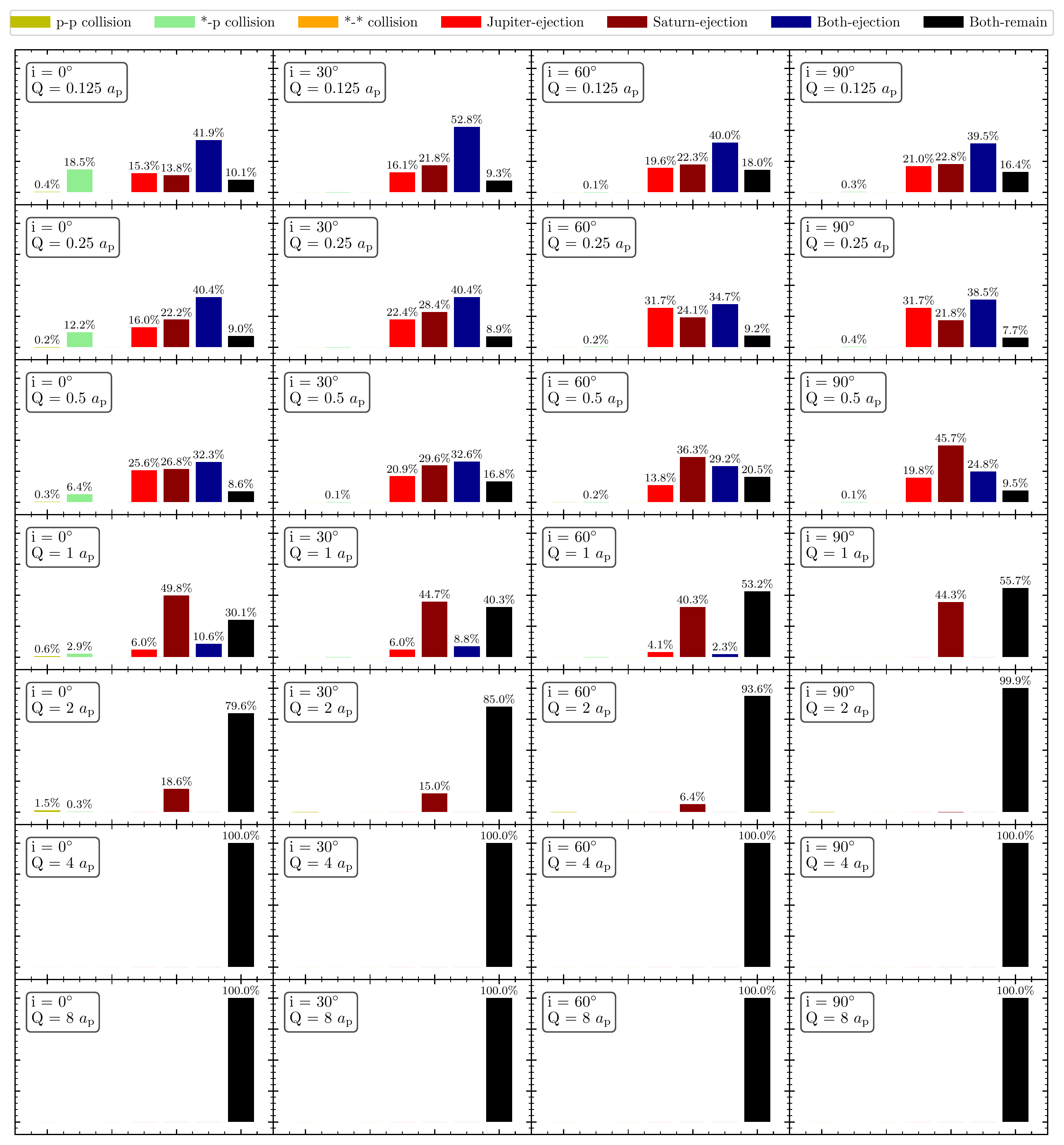}
    \caption{Same as in \fgr{fig:flyby-frac-s1-r2} but with a velocity dispersion $\sigma$ of 0.2~km~s$^{-1}$.}
    \label{fig:flyby-frac-s02-r2}
\end{figure*}
\fgr{fig:flyby-frac-s02-r2} shows results for the same initial conditions as in \fgr{fig:flyby-frac-s1-r2}, but with $\sigma$ equal to 0.2 km s$^{-1}$. As discussed before, $v_\infty$  contributes only minimally to the velocity of the interloper at the distance of closet approach $v_{\rm Q}$. Therefore, the fractions are very similar to the standard case with $\sigma$ equal to 1 km s$^{-1}$. Hence we may conclude that in stellar clusters with low velocity dispersions where gravitational focusing is significant, the fraction of planetary systems that may form hot Jupiters after a close flyby is not particularly sensitive to the velocity dispersion.

\begin{figure*}
    \includegraphics[width=\textwidth]{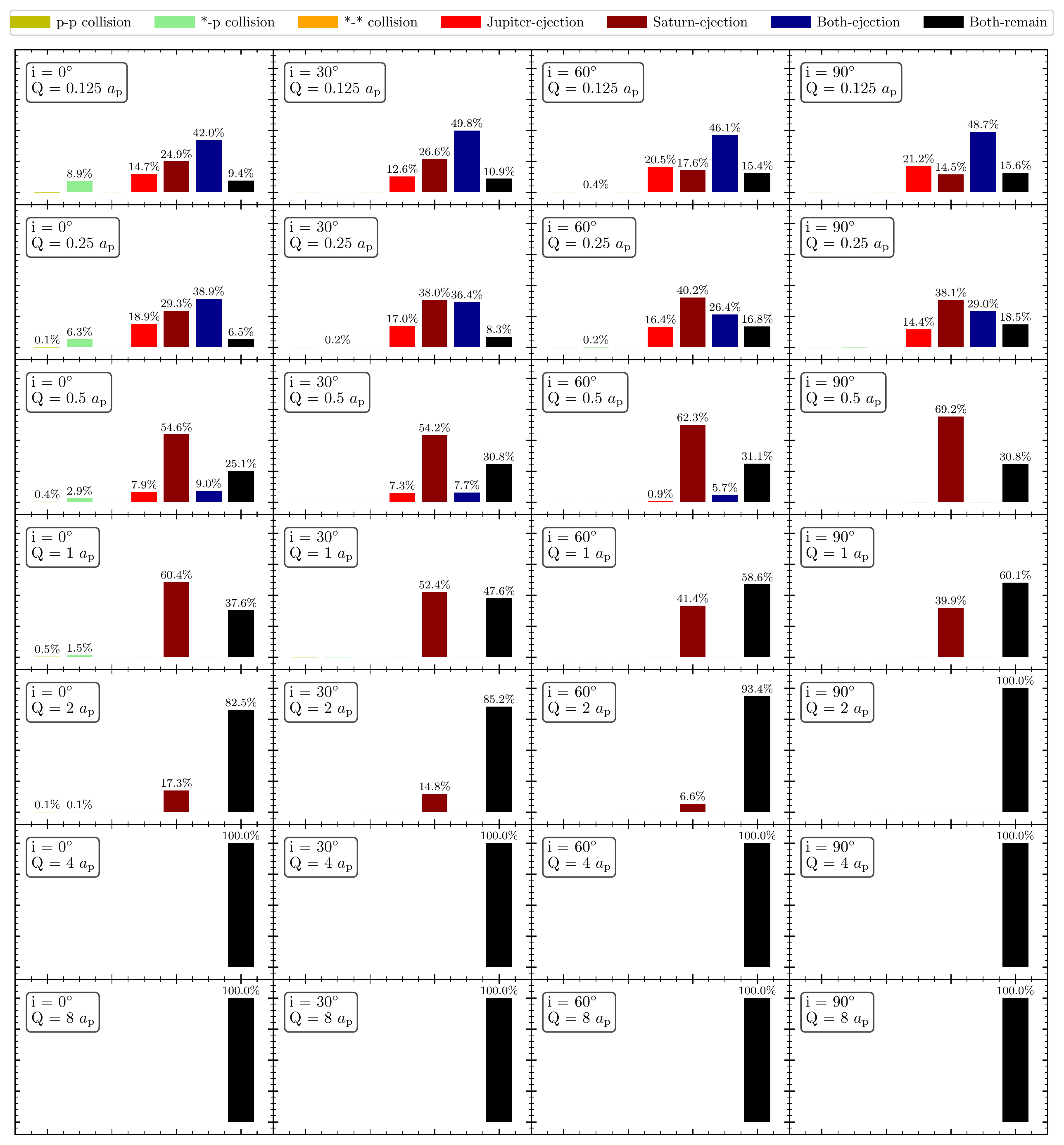}
    \caption{Same as in \fgr{fig:flyby-frac-s1-r2} but with an initial SMA ratio $a_{\rm s}/a_{\rm j}=4$.}
    \label{fig:flyby-frac-s1-r4}
\end{figure*}

\begin{figure*}
    \includegraphics[width=\textwidth]{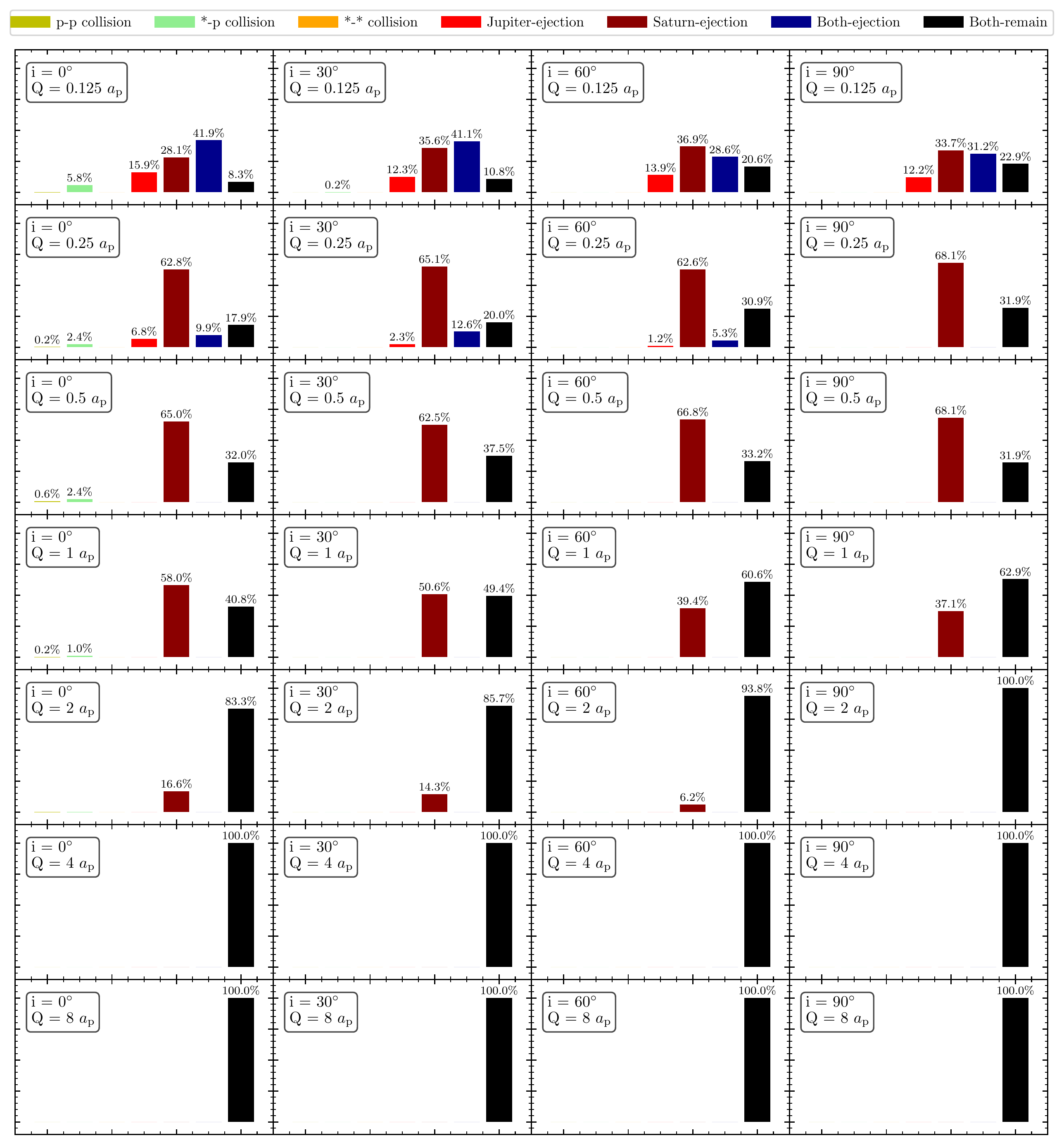}
    \caption{Same as in \fgr{fig:flyby-frac-s1-r2} but with an initial SMA ratio $a_{\rm s}/a_{\rm j}=8$.}
    \label{fig:flyby-frac-s1-r8}
\end{figure*}

\fgr{fig:flyby-frac-s1-r4} and \fgr{fig:flyby-frac-s1-r8} show the cases where $a_{\rm s}/a_{\rm j}=4$ and 8, respectively. With a larger SMA ratio, we see a larger fraction in the ratio of Saturn ejections,  and a much lower `Both-ejection' fraction. Since planet ejection requires a close encounter with the interloper, it is more difficult for the interloper to have a close encounter with both planets if the two planets are far apart.

\section{Hot Jupiters from high-eccentricity tidal migration}
Flybys alone cannot directly create hot Jupiters, as this would require the flyby star to extract large amounts of angular momentum without draining much of the orbital energy of the planet. However, flybys can significantly perturb the architecture of the planetary system into regimes where several mechanisms can then excite the eccentricity to very high values which enable efficient tidal dissipation. Hot Jupiters can then be formed from high-eccentricity tidal migration. Isolated planet tidal dissipation can be described by \citep[e.g.][]{murray99,samsing18,hamers18},
\begin{equation}
\frac{da_{\rm p}}{dt}=-21k_{\rm AM,p}n^2\tau_{\rm p }\frac{M_{*}}{M_{\rm p}}\bigg(\frac{R_{\rm p }}{a_{\rm p}}\bigg)^5 a_{\rm p}e_{\rm p}^2\frac{f(e_{\rm p})}{(1-e_{\rm p}^2)^{15/2}}
\end{equation}
\begin{equation}
\frac{de_{\rm p}}{dt}=-\frac{21}{2}k_{\rm AM,p}n^2\tau_{\rm p }\frac{M_{*}}{M_{\rm p}}\bigg(\frac{R_{\rm p }}{a_{\rm p}}\bigg)^5 e_{\rm p}\frac{f(e_{\rm p})}{(1-e_{\rm p}^2)^{13/2}}\,,
\end{equation}
where
\begin{equation}
f(e_{\rm p})=\frac{1+\frac{45}{11}e_{\rm p}^2+8e_{\rm p}^4 +\frac{685}{224}e_{\rm p}^6 +\frac{255}{448}e_{\rm p}^8+\frac{25}{1792}e_{\rm p}^{10}}{1+3e_{\rm p}^2 + \frac{3}{8}e_{\rm p}^4}\,.
\end{equation}
The two equations converge to $a_{\rm p}\rightarrow a_{\rm p,0}(1-e_{\rm p,0}^2)$ and $e_{\rm p}\rightarrow 0$ as $t\rightarrow+\infty$. Due to the $(1-e_{\rm p}^2)$ term in the denominator, the dissipation rate increases rapidly as $e_{\rm p,0}\rightarrow 0$. High eccentricities are clearly critical in hot Jupiter formation via dynamical processes. Several  mechanisms can excite the eccentricity of the planet to high values, as discussed below.

\subsection{Lidov-Kozai}
\begin{figure}
    \includegraphics[width=.5\textwidth]{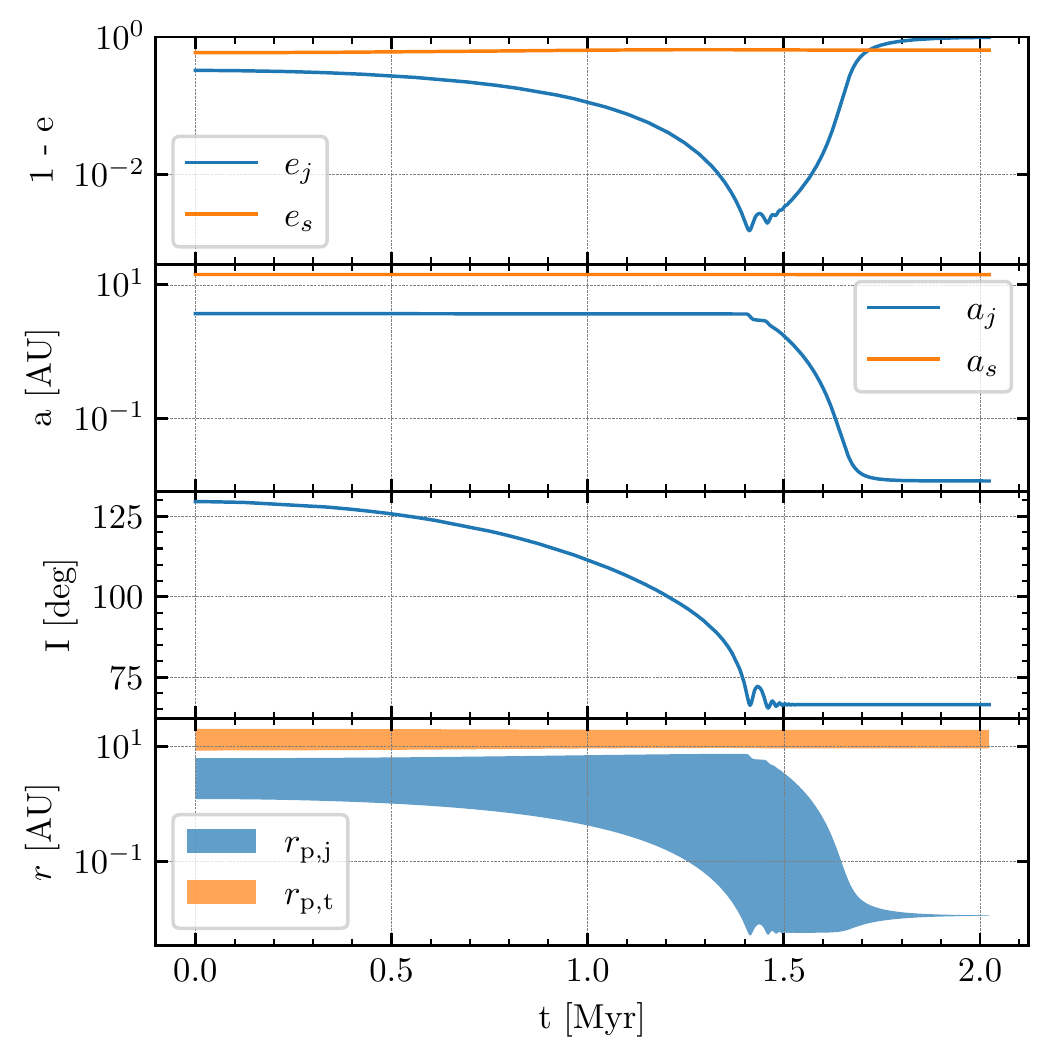}
    \caption{An example of Hot Jupiter formation from the Lidov-Kozai effect. This is clearly visible as operating most effectively between 1 and 1.5 Myr, after which time Jupiter's orbit circularizes.}
    \label{fig:LK}
\end{figure}
In a hierarchical triple system, the inner orbit (here, the star-Jupiter binary) can undergo Lidov-Kozai (LK) oscillations. In LK oscillations, the energy ($\propto -M_{\rm tot,in}/a$) of each orbit is conserved but the inner and outer orbits continuously exchange angular momentum ($\propto \sqrt{a(1-e^2)}$). Therefore, the eccentricity of the inner orbit can be excited to very high values as its angular momentum reaches a minimum, which is balanced to conserve angular momentum by reducing the orbital inclination between the inner and outer orbits. The time scale for the quadrupole LK cycle to operate is
\begin{equation}\label{eq:time-LK}
\tau_{\rm LK}=\frac{1}{n_{\rm in}}\bigg(\frac{M_{\rm tot,in}}{M_{\rm out}} \bigg)\bigg( \frac{a_{\rm out}}{a_{\rm in}}\bigg)^3(1-e_{\rm out}^2)^{3/2}\,,
\end{equation}
where $n_{\rm in}$ is the mean motion of the Jupiter orbit, $M_{\rm tot, in}$ is the total mass of the inner orbit and $M_{\rm out}$ is the mass of the perturber. LK oscillations require the triple system to be hierarchical and stable. These oscillations operate in the range where $-\sqrt{3/5}<\cos(I)<\sqrt{3/5}$, with $I$ being the inclination between the two planets' orbits.

In this work, the flyby perturbation generates the necessary inclination, and the two giant planet orbits ensure that the timescale $\tau_{\rm LK}$ is shorter than the age of the Universe. This combination of factors allows the eccentricity of the Jupiter to be increased to a high enough value that tidal dissipation becomes efficient enough to make a hot Jupiter.

An example of hot Jupiter formation from the Lidov-Kozai effect is shown in Figure~\ref{fig:LK}. It can be seen that the effect starts to operate on a timescale of about 1-1.5~Myr.

\subsection{Planet-planet scattering}

Close planet-planet scatterings can convert Keplerian shear into an angular momentum deficit, triggering high eccentricity migration \citep[e.g.][]{Rasio1996,Weidenschilling1996, Ford2006,Chatterjee2008}. In planet-planet scattering, the planets typically undergo several close encounters to grow their eccentricities to high values. However, if the orbital velocity of a planet is larger than the escape velocity at the planet's surface, the cross section for collision would be larger than the cross section for scattering. This may lead to a planet collision rather than just eccentricity growth. \citet{Goldreich2004,Ida2013} and \citet{Petrovich2014} give the upper limit of the eccentricity from planet-planet scattering as
\begin{equation}
e_{\rm p-p}\sim\frac{\sqrt{GM_{\rm p}/R_{\rm p }}}{2\pi a_{\rm p}/P}\,,
\end{equation}
where $M_{\rm p}$ is the mass of planet, $R_{\rm p}$ is the radius of the planet, $a_{\rm p}$ is the SMA of the planet and $P$ is the orbital period of the planet. A Jupiter mass planet with $a_{\rm p}\sim$ 1 AU has roughly $e_{\rm p-p}\sim$ 1, which makes high eccentricity tidal migration possible.

\begin{figure}
    \includegraphics[width=.5\textwidth]{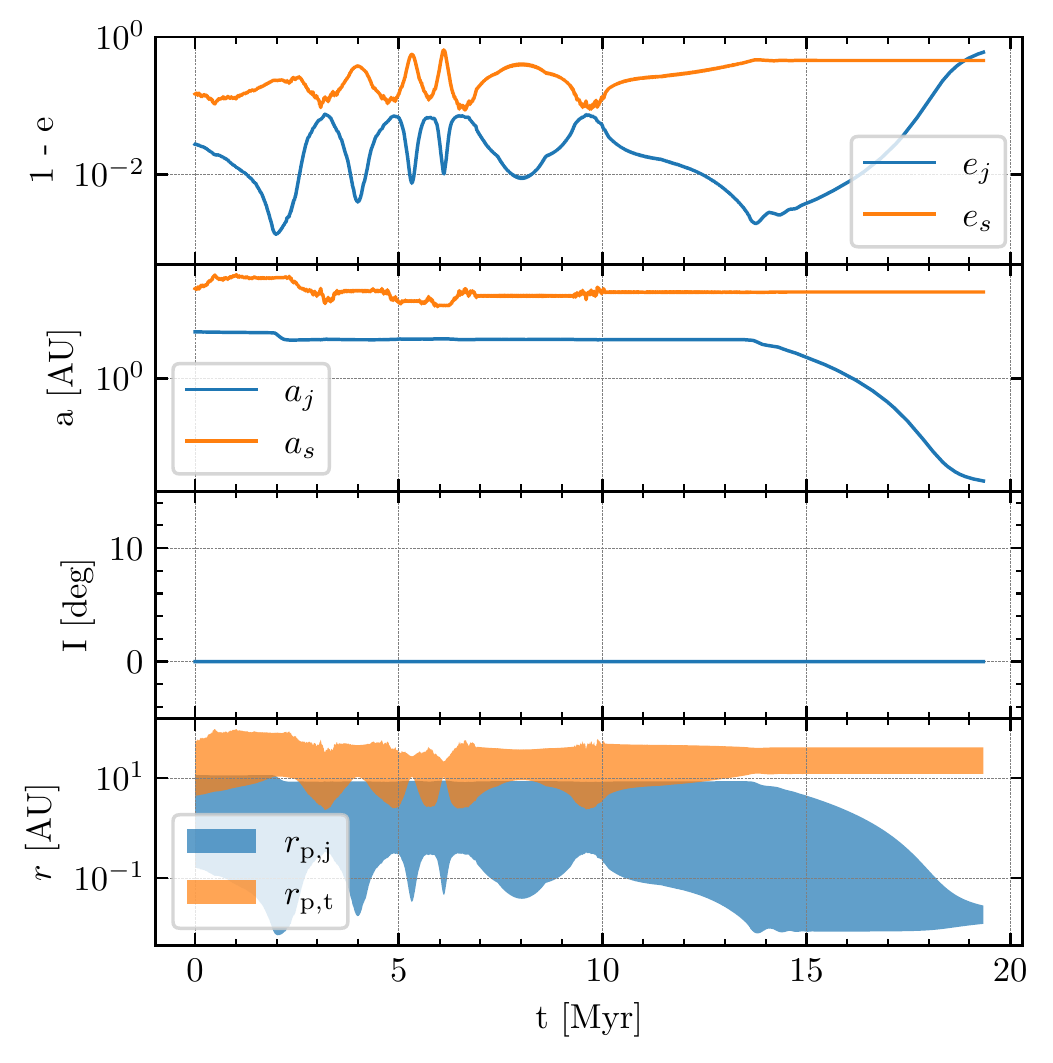}
    \caption{An example of Hot Jupiter formation from planet-planet scattering.  This operates relatively continuously up until about 13 Myr, after which time Jupiter's orbit circularizes, terminating the strong interactions between Jupiter and Saturn.}
    \label{fig:scattering}
\end{figure}

An example of hot Jupiter formation via planet-planet scattering is shown in Figure~\ref{fig:scattering}. 
This occurs rather continuously up to a time of about 13~Myr, when the orbit of Jupiter achieves circularization. After that time, Jupiter and Saturn no longer interact strongly.

\subsection{Coplanar secular effects-octupole LK}
LK oscillations operate only when the orbital inclination is within a certain range. However, \citet{Li2014} found that a perturber with a highly elliptical orbit, where the octupole LK effect is strong, can also excite the eccentricity of an initially coplanar system. Thus tidal migration may also occur in this regime. Indeed, we find that usually the flyby star extracts the angular momentum of the outer planet rather than that of the inner planet. This preferentially forms an elliptical outer orbit for the Saturn, which drives the further evolution of the system. The coplanar octupole effect comes into play when
\begin{equation}\label{eq:oct}
    \epsilon > \frac{8}{5}\frac{1-e_{\rm in}^2}{7-e_{\rm in}(4+3e_{\rm in}^2)\cos(\omega)}\,,
\end{equation}
where $e_{\rm in}$ is the initial eccentricity of the inner binary, $\omega$ is the argument of the periapsis of the outer orbit and $\epsilon$ is the LK octupole parameter
\begin{equation}
\epsilon =\frac{a_{\rm in}}{a_{\rm out}}\frac{e_{\rm out}}{1-e_{\rm out}^2}\,.
\end{equation}

No hot Jupiters were found to be formed via this channel in our simulations.

\subsection{Other mechanisms}
There are other mechanisms that can excite the eccentricity of a planet, including LK oscillations from a stellar companion \citep{Naoz2012,Anderson16} and secular chaos due to multiple LK timescales. Stellar LK oscillations demand a companion star, either from binary star formation as an outcome of the flyby (which is very unlikely) or a flyby binary interloper which is beyond the scope of this work. Secular chaos generally demands more than two planets, also beyond the scope of this paper. Thus we do not consider eccentricity excitation from these mechanisms.

\subsection{Hot Jupiter formation fraction from different channels}
After the flyby, the modified planetary system may fall into the eccentricity excitation regime, where hot Jupiter formation is possible. For all the `Both remain' systems, as discussed in more detail in the preceding section, we divide the outcomes into several different regimes based on the criteria below:

\textbf{LK:} The inclination between the two planets' orbits satisfies $-\sqrt{3/5}<\cos(I)<\sqrt{3/5}$ and the quadrupole timescale is $\tau_{\rm LK} <$1~Gyr.

\textbf{Planet-planet scattering:} The pericenter of the outer planet / the apocenter of the inner planet satisfy $\frac{a_{\rm s}(1-e_{\rm s})}{a_{\rm j}(1+e_{\rm j})}<1$, where the planets could undergo close encounters.

\textbf{Coplanar octupole:} The inclination between the two planet orbits is within $5^\circ$ and \eqn{eq:oct} is satisfied.

\subsection{Long-term Jupiter evolution after the flyby}
After the flyby, we perform long-time integrations of the perturbed planetary systems up to 1~Gyr 

including tidal dissipation and GR effects as described in \sectn{sec:tidal}. Since for $Q> 2a_{\rm p}$ the perturbation becomes too weak to transition the planetary system into the high eccentricity tidal migration regime, we only perform long integrations for systems with $Q<2a_{\rm p}$. 

\begin{figure*}
    \includegraphics[width=\textwidth]{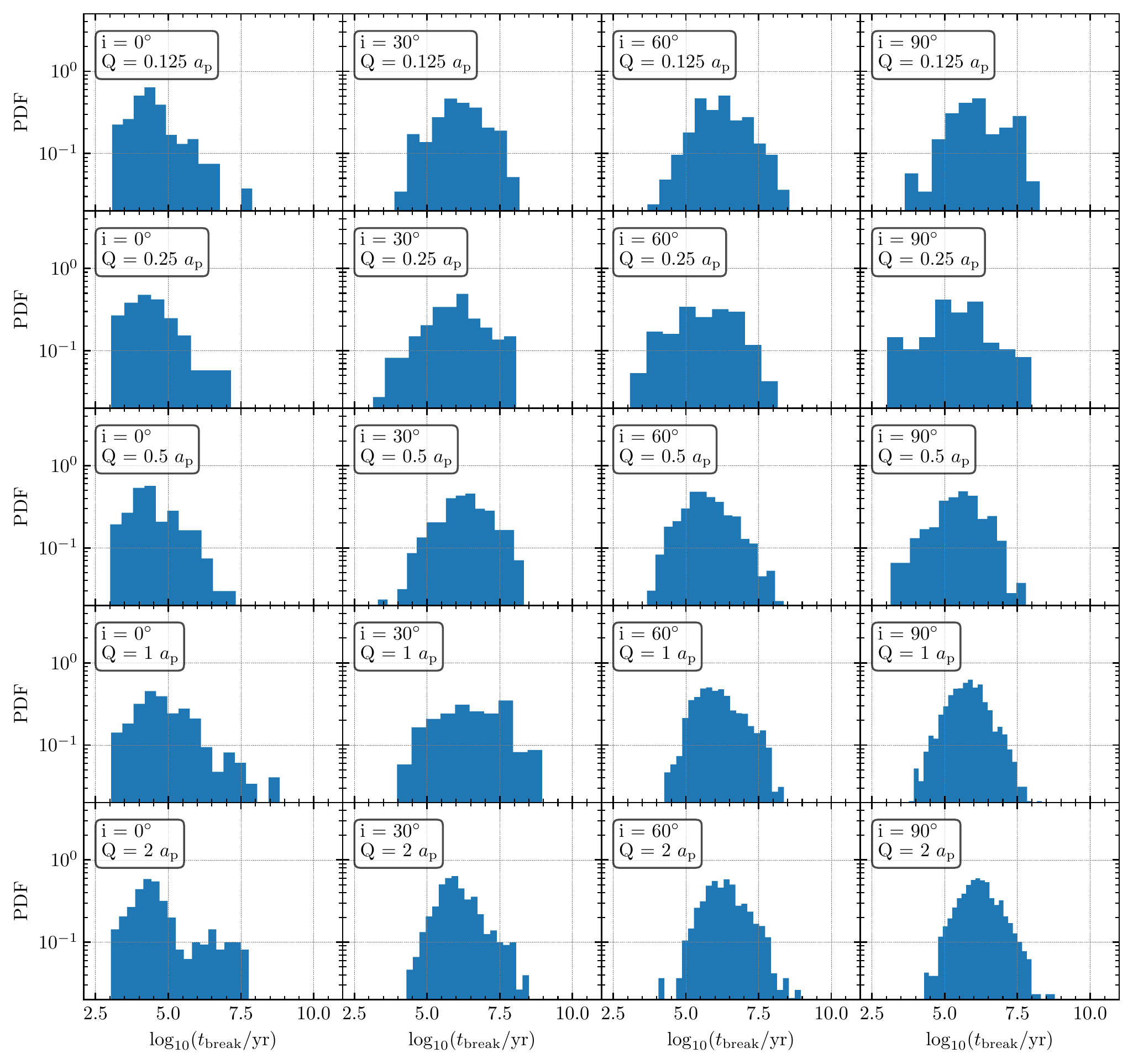}
    \caption{The distribution of the planetary system disruption time in the long interaction time regime. The velocity dispersion of the cluster is 1 km s$^{-1}$, the initial SMA of the Jupiter is 5 AU and the initial SMA of the Saturn is 10 AU.}
    \label{fig:tbreak}
\end{figure*}

During the long time evolution, tidal dissipation and GR effects may break the planetary system by ejecting either the Jupiter or Saturn through a close planet-planet encounter or collision. If the eccentricity of the remaining planet is not high enough to make tidal dissipation efficient, then we will not observe it to form a hot Jupiter within the age of the Universe.

\fgr{fig:tbreak} shows the distribution of the disruption time (i.e. the timescale over which the planetary system is disrupted) in log scale. For the coplanar case, this timescale is much shorter than for the other cases. The planetary system tends to disrupt in $\sim$ 10$^4$ years while the other cases more likely disrupt at around 10$^6$ years. This is because for the coplanar case, there is a much higher probability for the two planets to undergo close encounters. The distribution is truncated by the flyby time on the left side and by the integration upper limit on the right side.

\begin{figure*}
    \includegraphics[width=\textwidth]{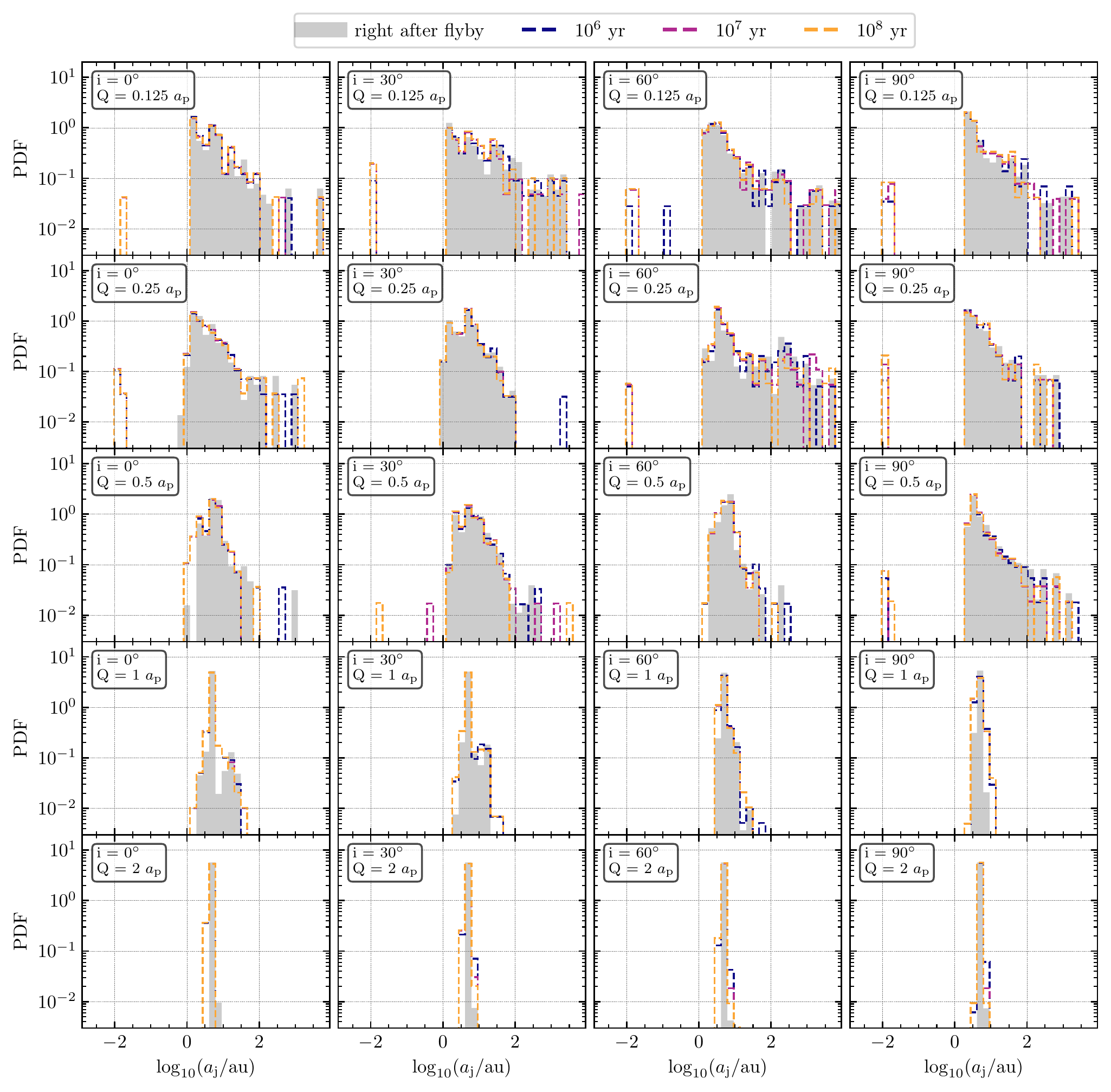}
    \caption{Snapshots of the distribution of the SMA of the Jupiter right after the flyby, after $10^6$~yr  evolution, after $10^7$~yr evolution, and after $10^8$~yr evolution. The velocity dispersion of the cluster is 1 km s$^{-1}$, the initial SMA of the Jupiter is 5 AU and the initial SMA of the Saturn is 10 AU. }
    \label{fig:hj-aj-s1-r2}
\end{figure*}
\fgr{fig:hj-aj-s1-r2} shows the semi-major axis of the Jupiter right after the flyby (grey), 10$^6$ years after the flyby (blue), 10$^7$ years after the flyby (purple) and 10$^8$ years after the flyby (orange). We see that, for $Q<0.5a_{\rm p}$,  a non-negligible fraction of the Jupiters evolve into hot Jupiters with $a_{\rm j}$ between 10$^{-2}$~AU and 10$^{-1}$~AU due to tidal dissipation and the aforementioned high eccentricity excitation mechanisms. For $Q>1~a_{\rm p}$, it is extremely difficult to form a hot Jupiter progenitor. We do not see any hot Jupiters even 10$^{8}$ years after the flyby. 

\begin{figure*}
    \includegraphics[width=\textwidth]{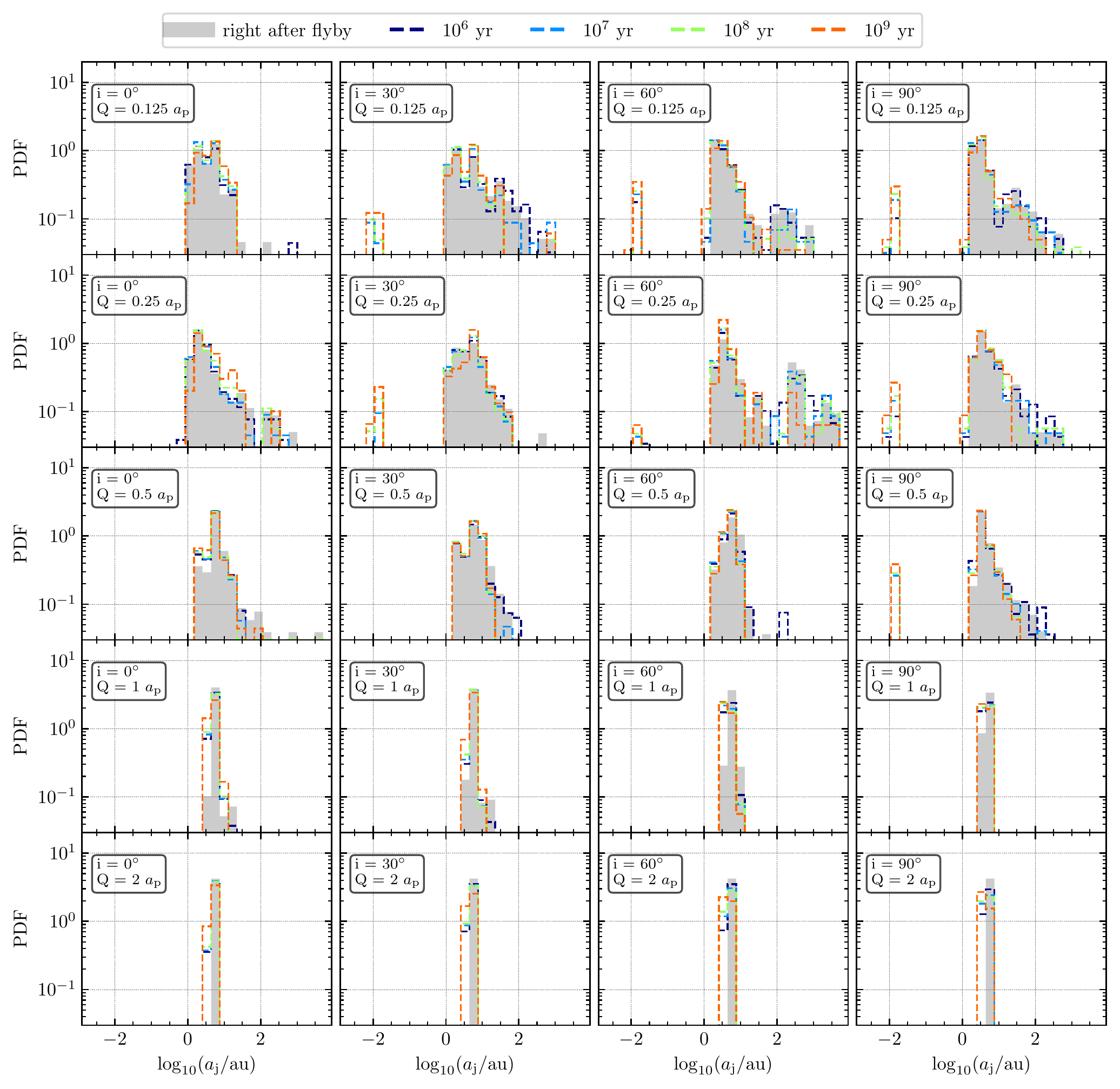}
    \caption{Evolution of the orbital SMA of Jupiter with time, shown at some representative times up to 1 Gyr.}
    \label{fig:convergence}
\end{figure*}

\begin{figure*}
    \includegraphics[width=\textwidth]{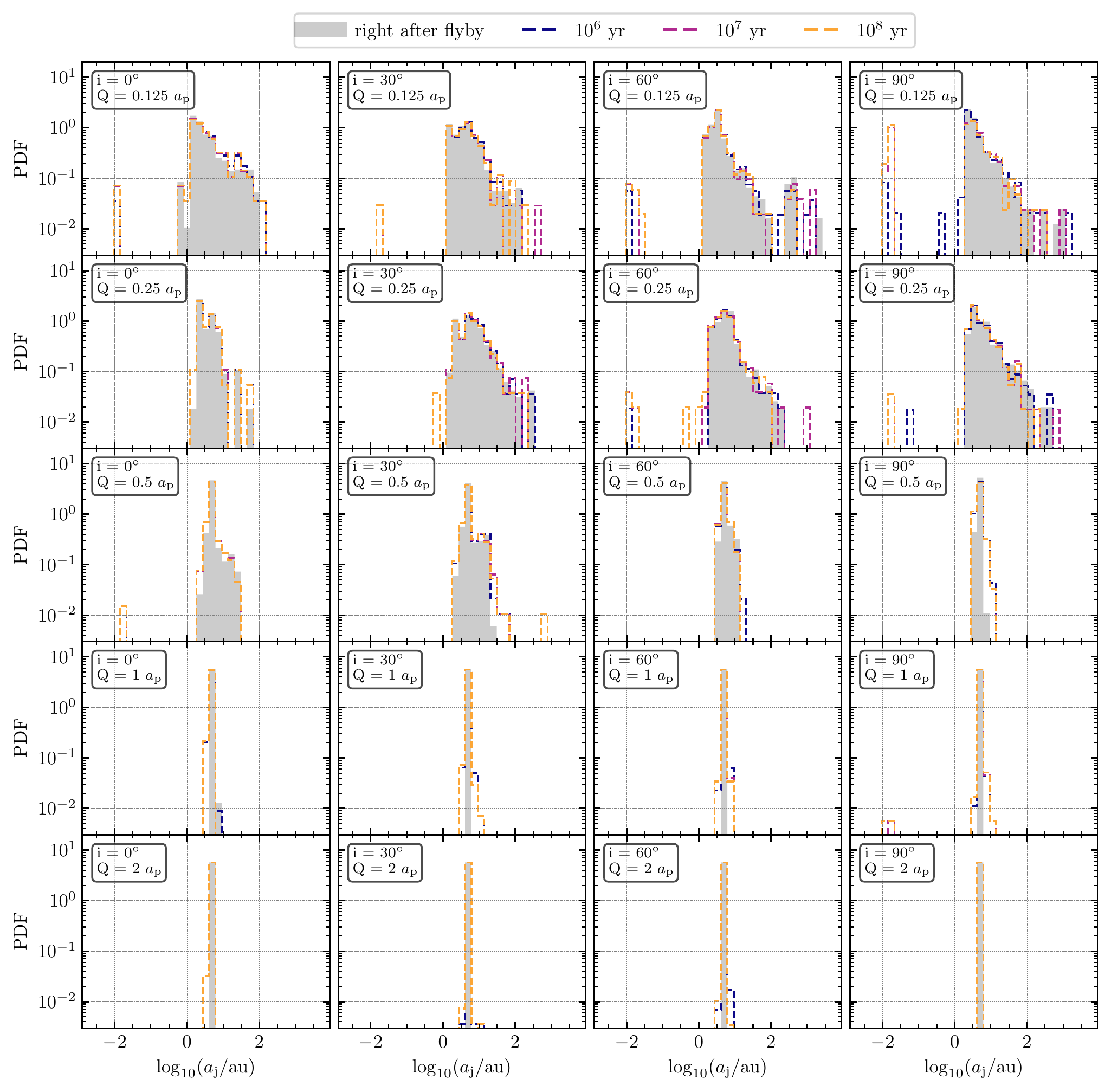}
    \caption{Snapshots of the distribution of the SMA of the Jupiter right after the flyby, after $10^6$, $10^7$ and $10^8$ years of evolution. The velocity dispersion of the cluster is 1 km s$^{-1}$, the initial SMA of the Jupiter is 5 AU and the initial SMA of the Saturn is 20~AU. }
    \label{fig:hj-aj-s1-r4}
\end{figure*}

The evolution of the orbital SMA of Jupiter with time, at several representative times up to 1 Gyr, is shown in Figure~ \ref{fig:convergence}. For most configurations, hot Jupiter formation happens promptly, already present at 1~Myr.

\fgr{fig:hj-aj-s1-r4} shows the same distribution as \fgr{fig:hj-aj-s1-r2} but with an initial SMA ratio equal to 4. A large fraction of the Jupiters evolve into hot Jupiters with $Q=0.125a_{\rm p}$ and $i=90^\circ$. This is the ideal parameter space where the flyby has a large probability to create a highly inclined system with an appropriate SMA ratio that boosts the LK effect. For an initial SMA ratio equal to 2, as indicated  in \fgr{fig:hj-aj-s1-r2}, the two giant planets are too close to maintain the architecture of the planetary system. Thus, there is a higher chance for the planetary system to undergo close planet-planet interactions rather than secular LK excitations. For an initial SMA ratio equal to 8, as shown in \fgr{fig:hj-aj-s1-r8}, we see that the fraction of the hot Jupiters drops again in highly inclined flybys. The timescale of the quadrupole LK effect is given by \eqn{eq:time-LK}. Unlike the LK mechanism in binary star systems, where the perturber is a star, the perturber in the planetary system is only a giant planet. Thus the system requires a much smaller SMA ratio to make the LK effect operate within the age of the cluster. Therefore, with the larger SMA ratio equal to 8, the LK effect becomes less efficient than for the SMA ratio equal to 4. For a Jupiter mass planet-solar mass host star binary with $a_{\rm j}=5$~AU and with a Saturn mass perturber at 10~AU, the typical  timescale given by \eqn{eq:time-LK} that is sensitive to the SMA ratio is $\sim$ 2.5$\times$10$^6$~years. With an initial SMA ratio equal to 8, the timescale is about one order of magnitude longer. This explains why the hot Jupiter formation fraction drops quickly for $a_{\rm j}/a_{\rm s}=8$. However, the SMA ratio cannot be too small in the LK regime, otherwise it breaks the stability of the planetary system. Therefore,  for  $a_{\rm j}/a_{\rm s}=2$, the main source of hot Jupiters is from planet-planet scatterings.

\begin{figure*}
    \includegraphics[width=\textwidth]{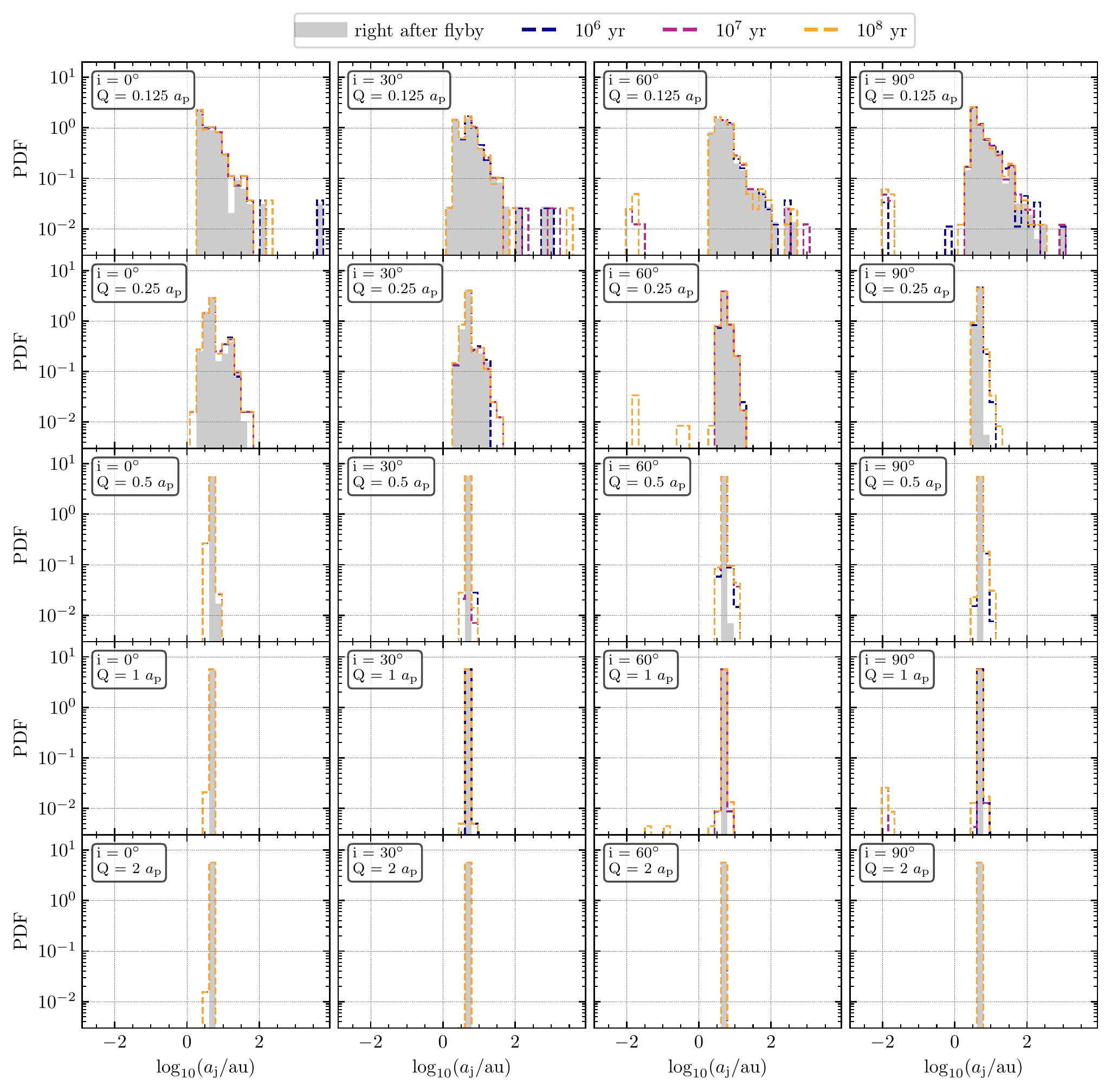}
    \caption{Snapshots of the distribution of the SMA of the Jupiter right after the flyby, after $10^6$, $10^7$ and $10^8$ years of evolution. The velocity dispersion of the cluster is 1 km s$^{-1}$, the initial SMA of the Jupiter is 5 AU and the initial SMA of the Saturn is 40~AU. }
    \label{fig:hj-aj-s1-r8}
\end{figure*}

\begin{figure*}
    \includegraphics[width=\textwidth]{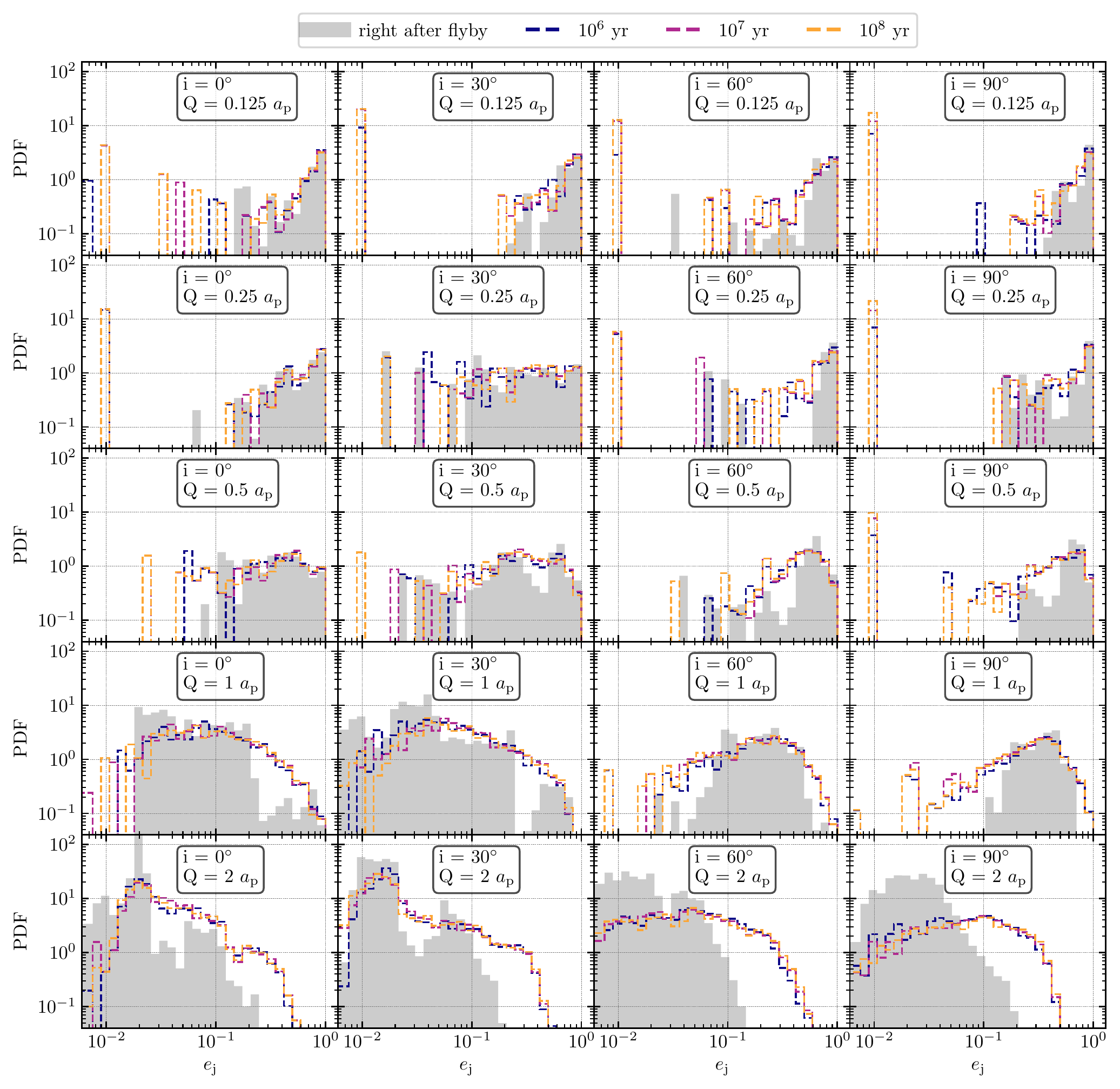}
    \caption{Snapshots of the distribution of the eccentricity of the Jupiter right after the flyby, after $10^6$, $10^7$ and $10^8$ years of evolution. The velocity dispersion of the cluster is 1 km s$^{-1}$, the initial SMA of the Jupiter is 5 AU and the initial SMA of the Saturn is 10 AU. The initial eccentricities are zero.}
    \label{fig:hj-ej-s1-r2}
\end{figure*}

\begin{figure*}
    \includegraphics[width=\textwidth]{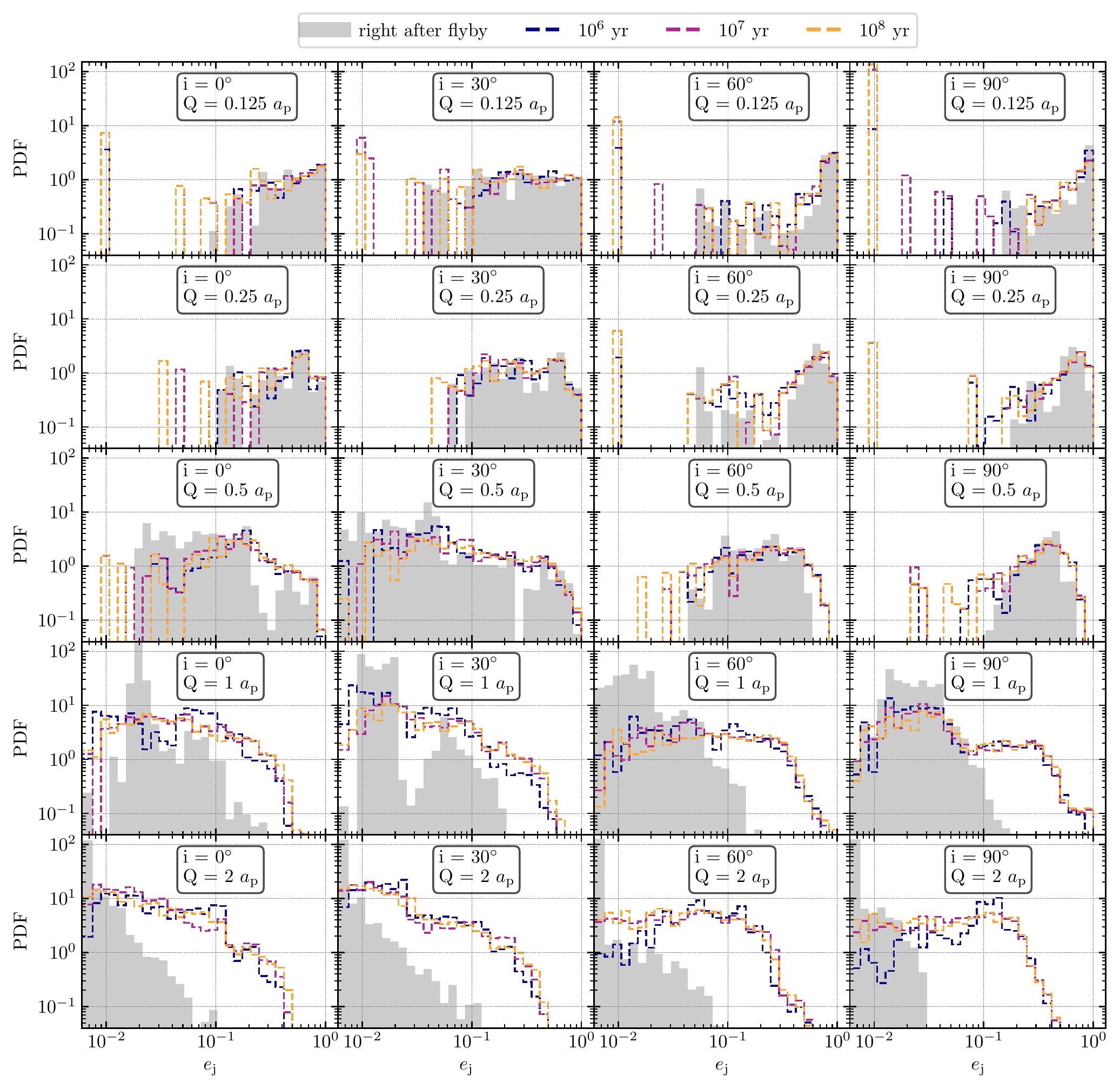}
    \caption{Snapshots of the distribution of the eccentricity of the Jupiter right after the flyby, after $10^6$, $10^7$ and $10^8$ years of evolution. The velocity dispersion of the cluster is 1 km s$^{-1}$, the initial SMA of the Jupiter is 5 AU and the initial SMA of the Saturn is 20 AU. The initial eccentricities are zero.}
    \label{fig:hj-ej-s1-r4}
\end{figure*}

\begin{figure*}
    \includegraphics[width=\textwidth]{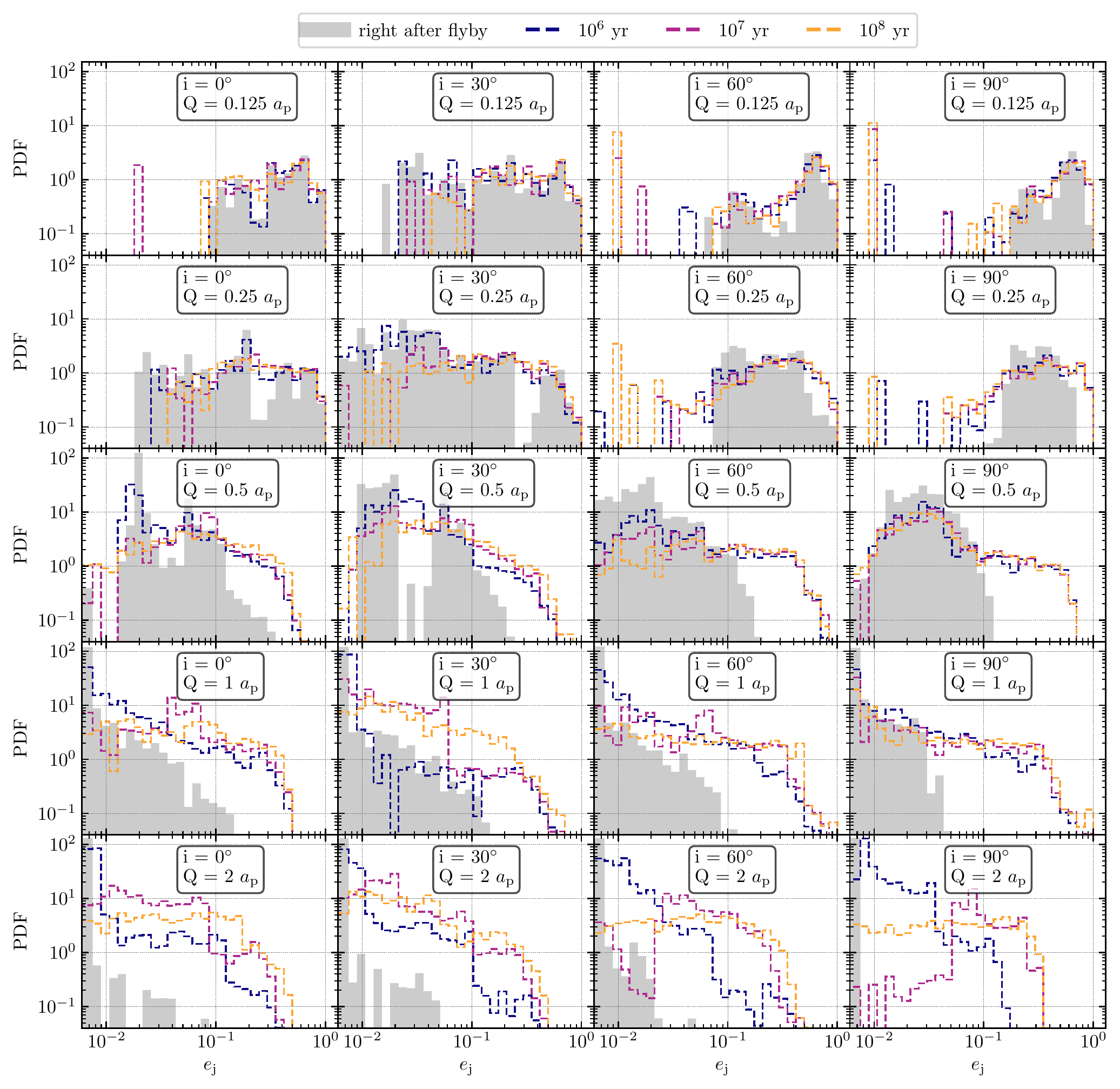}
    \caption{Snapshots of the distribution of the eccentricity of the Jupiter right after the flyby, after $10^6$, $10^7$ and $10^8$ years of evolution. The velocity dispersion of the cluster is 1 km s$^{-1}$, the initial SMA of the Jupiter is 5 AU and the initial SMA of the Saturn is 40 AU. The initial eccentricities are zero.}
    \label{fig:hj-ej-s1-r8}
\end{figure*}

\fgr{fig:hj-ej-s1-r2} to \fgr{fig:hj-ej-s1-r8} show the distribution of the eccentricity of the Jupiter right after the flyby (grey), 10$^6$ years after the flyby (blue), 10$^7$ years after the flyby (purple) and 10$^8$ years after the flyby (orange). In the close flyby regime where $Q<2a_{\rm j}$, the general trend of the Jupiter eccentricity after long interaction times is to decrease, while in the perturbative regime where $Q>2a_{\rm j}$ the general trend of the Jupiter eccentricity is to increase. Indeed, in the close flyby regime, the interloper increases the eccentricity of the Jupiter by extracting angular momentum from the planet. There is a large probability of creating a Jupiter orbit with a small pericenter. However, in the perturbative regime, the flyby interloper barely creates a small pericenter Jupiter orbit. Therefore, in the close flyby regime, the tidal dissipation will dominate in the long-time integration while in the perturbative regime, planet-planet interactions become the dominant effect. The tidal dissipation process will circularize the Jupiter orbit but planet-planet interactions will increase the eccentricity of the Jupiter orbit. Therefore, we see opposite general trends for the eccentricity evolution in the close flyby and perturbative regimes.

\begin{figure*}
    \centering
    \includegraphics[width=\textwidth]{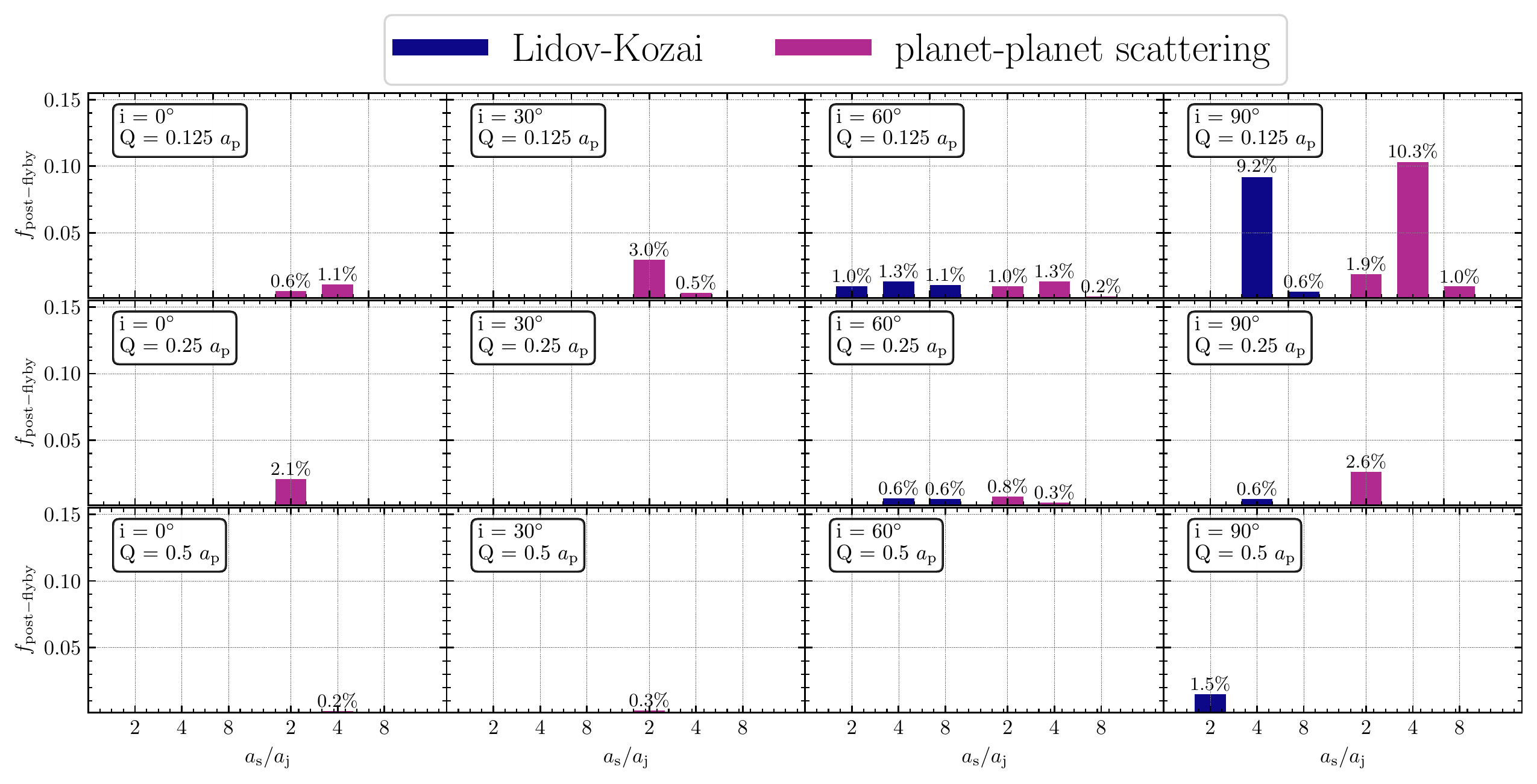}
    \caption{Hot Jupiter formation fraction of different channels with different initial semi-major axis ratios at $10^8$ years after the flyby. The velocity dispersion of the cluster is 1 km s$^{-1}$, the initial SMA of the Jupiter is 5 AU. For $Q>a_{\rm p}$, the fraction is smaller than 0.1\% and no hot Jupiter formed from the coplanar secular channel.}
    \label{fig:hj-regime-s1-r2}
\end{figure*}

\begin{figure*}
    \centering
    \includegraphics[width=\textwidth]{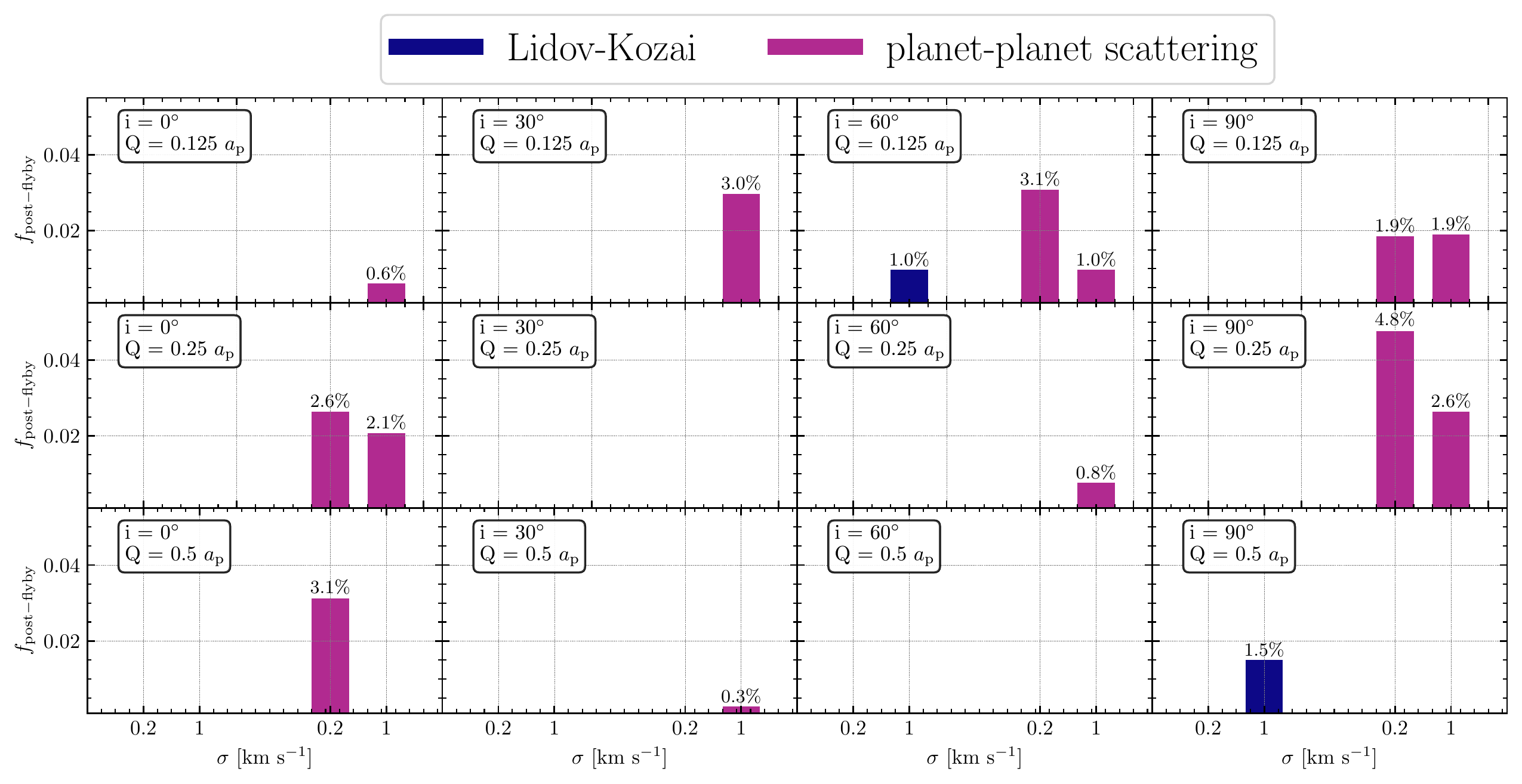}
    \caption{Hot Jupiter formation fraction for different channels with different velocity dispersions at $10^8$ years after the flyby. The initial semi-major axis of the Jupiter is 5 AU and the initial semi-major axis of the Saturn is 10 AU. For $Q>a_{\rm p}$, the fraction is smaller than 0.1\% and no hot Jupiter formed from the coplanar secular channel.}
    \label{fig:hj-regime-s02-r2}
\end{figure*}
\fgr{fig:hj-regime-s1-r2} shows the hot Jupiter formation fraction of all `Both-remain' systems after the flyby. Different colors represent the fractions from different channels and the sum of the fractions indicates the total formation fraction. We find that there is no hot Jupiter formed from the coplanar secular effect due to its very narrow parameter space, such that flybys cannot create the ideal initial conditions in this regime. For $a_{\rm s}/a_{\rm j}=2$, most of the hot Jupiters formed from planet-planet interactions. For $a_{\rm s}/a_{\rm j}=4$, the LK channel contributes most of the hot Jupiters, especially the perpendicular flybys with a formation fraction one order of magnitude higher. For $a_{\rm s}/a_{\rm j}=8$, we see that most of the hot Jupiters formed from the LK channel. Indeed, with $a_{\rm s}/a_{\rm j}=2$, the flybys more easily create closer planet orbits to activate planet-planet scatterings. However, with $a_{\rm s}/a_{\rm j}=8$, where the two orbits are distant from each other, it is harder for the flybys to bring the two orbits close enough together. \textit{Therefore, we see that hot Jupiter formation from planet-planet scattering dominates in the small SMA ratio regime while hot Jupiter formation from secular LK effects dominates in the large SMA ratio regime.} \fgr{fig:hj-regime-s02-r2} shows the formation fraction with lower velocity dispersion, where $\sigma=$0.2 ~km~s$^{-1}$. The formation fraction is slightly higher than for $\sigma=$1~km~s$^{-1}$.

\subsection{Hot Jupiter formation cross section and rate}
To estimate the hot Jupiter formation rate from different channels in different environments, we need the hot Jupiter formation fractions of each channel from isotropic scatterings. Then, the hot Jupiter formation rate in each channel  can be estimated by 
\begin{equation}
    \Gamma_{\rm HJ} = \Gamma_{\rm strong} f = \sigma_{\rm a_p} n v_{\rm rms} f
\end{equation}
where $\sigma_{\rm a_p}$ is the corresponding cross section for $Q=1a_{\rm p}$ for which a hot Jupiter can be formed ($Q<a_{\rm p}$), $n$ is the number density of the environment, $v_{\rm rms}$ is the root mean square velocity of the environment and $f$ is the hot Jupiter formation fraction of each channel after a close flyby. The cross section is given by
\begin{equation}
    \sigma_{\rm a_p} = \pi b_{\rm a_p}^2
\end{equation}
where
\begin{equation}
    b_{\rm a_p} =a_{\rm p}\sqrt{1 + 2\frac{G(M_0+M_1)}{v_\infty^2a_{\rm p}}}.
\end{equation}
The number density of a virialized cluster is \citep[e.g.][]{spitzer87,binney87}
\begin{equation}
    n\sim\frac{M_{\rm c}/\bar{m}}{4\pi R_{\rm c}^3/3}\sim\frac{6v_{\rm rms}^6}{\pi G^3M_{\rm c}^2\bar{m}}\,,
\end{equation}
where $M_{\rm c}$ and $\bar{m}$ are the total stellar mass and mean stellar mass of the cluster, respectively.
Thus, the hot Jupiter formation rate can be estimated from
\begin{eqnarray}
\Gamma_{\rm HJ} &=& \pi \bigg(a_{\rm p}^2 + 2\frac{G(M_0+M_1)}{v_\infty^2}a_{\rm p}\bigg)\frac{6v_{\rm rms}^7}{\pi G^3M_{\rm c}^2\bar{m}}f\nonumber\\
&=&\bigg(\frac{6v_{\rm rms}^7a_{\rm p}^2}{G^3M_{\rm c}^2\bar{m}} +  \frac{12v_{\rm rms}^5a_{\rm p}}{G^2M_{\rm c}^2} \bigg)f\nonumber\\
&=&\frac{6v_{\rm rms}^5a_{\rm p}}{G^2M_{\rm c}^2} \bigg( \frac{v_{\rm rms}^2}{v_{\rm p}^2} +  2 \bigg)f.
\end{eqnarray}
with $M_0\sim M_1\sim \bar{m}$, $v_{\rm rms}^2=2v_\infty^2$ and $G\bar{m}/a_{\rm p}\sim v_{\rm p}^2$. For most clusters $v_{\rm rms}\ll v_{\rm p}$, therefore,
\begin{eqnarray}
    \Gamma_{\rm HJ} &=& \frac{12v_{\rm rms}^5a_{\rm p}}{G^2M_{\rm c}^2} f.\\
&=&1.6\times10^{-4}\bigg(\frac{\sigma}{\rm 1kms^{-1}}\bigg)^5\bigg(\frac{a_{\rm p}}{\rm 20 AU}\bigg)\bigg(\frac{M_{\rm c}}{\rm 1000M_\odot}\bigg)^{-2} {\rm Gyr}^{-1}\nonumber
\end{eqnarray}
Since the hot Jupiter formation timescale is typically much shorter than the age of its host star cluster, the hot Jupiter percentage per star in a cluster is
\begin{eqnarray*}
    F_{\rm HJ}&=&
    \Gamma_{\rm HJ}N_*\tau_{\rm c}/N_*=\frac{12v_{\rm rms}^5a_{\rm p}}{G^2M_{\rm c}^2} f \tau_{\rm c}\,,\\
\end{eqnarray*}
where $\tau_{\rm c}$ is the age of the cluster.

\begin{figure}
    \includegraphics[width=0.5\textwidth]{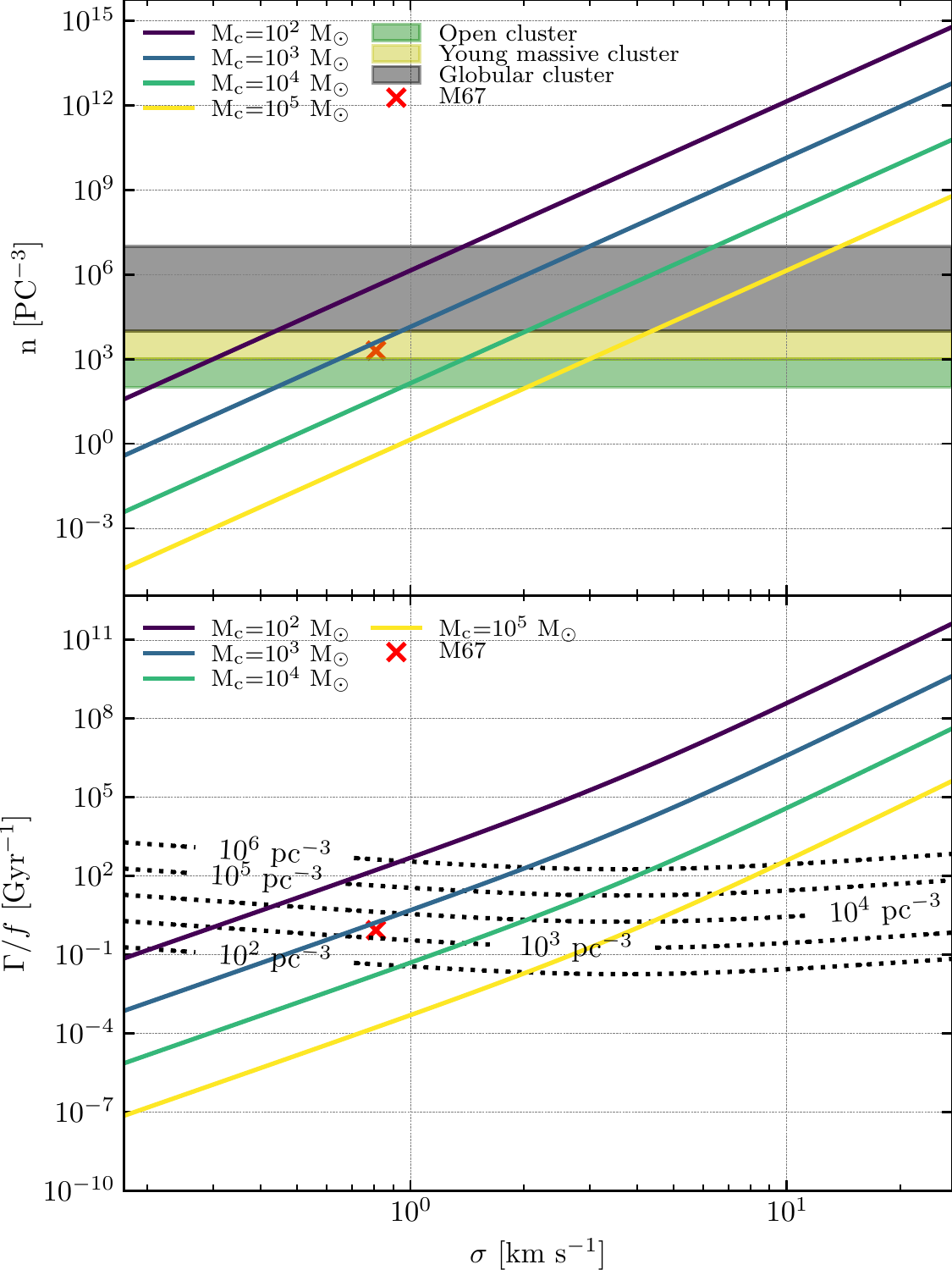}
    \caption{Rate of close flybys as a function of the velocity dispersion in the host virialized cluster, and for different values of the cluster mass. The dashed lines indicate the cluster number densities.}
    \label{fig:closeflyby}
\end{figure}

We performed isotropic scatterings with cluster velocity dispersions $\sigma = $ 0.2 and 1 km s$^{-1}$. The impact parameter $b$ of each scattering was randomly generated from the probability distribution function $f(b)\propto b$ within [$0, b_{\rm max}$], where $b_{\rm max}$ is chosen to be large enough that the system is not significantly perturbed. 
Based on our scattering experiments described in the last section, the planetary systems after the flyby perturbation will be almost unchanged if $Q>a_{\rm p}$. Therefore, we set the corresponding $b$ with $Q=a_{\rm p}$ as the $b_{\rm max}$ in isotropic scatterings. All other setups are the same as described in \sectn{method}.

Finally, it is worth emphasizing that the properties of star clusters change over time, primarily due to two-body relaxation and the evaporation of the cluster in a Galactic tidal field.  We have not accounted for these effects in any systematic way in this paper, and instead defer properly accounting for these effects to future work.  Briefly, the rate of stellar escape is inversely proportional to stellar mass, causing older clusters to become more depleted in low-mass stars and tending toward a top-heavy stellar mass function (relative to standard initial mass functions measured in young star-forming regions). In future work, a Monte Carlo approach will be needed for a given host galaxy to generate a realistic initial star cluster mass function, a realistic distribution of 3D star cluster galactocentric distances, and so on.

\section{Summary and conclusions}

In this work we have studied hot Jupiter and {{red} ultra-cold} Saturn formation in an environment of strong stellar interactions. Close flybys in stellar clusters can significantly change the configurations of the planetary systems that host the progenitors of hot Jupiters, significantly affecting the rate of hot Jupiter formation.  While primarily focused on the hot Jupiters, our simulations yield the full reconfiguration of the planetary system, hence allowing us to track also the distribution of orbital parameters of the outer planet, the Saturn, after the flyby. 

By performing high precision few-body scatterings in the flyby process and long-term integrations with tidal dissipation and general relativity corrections up to 1 Gyr, our main conclusions can be summarized as follows:

\begin{itemize}
    \item In stellar clusters where the velocity dispersion is relatively low, close encounters (where closest approach $\sim$ the size of the planetary system) that involve strong interactions give similar post-scattered configurations. In low velocity dispersion environments, the relative velocity at closest approach that determines the imparted impulse are mostly contributed by gravitational focusing, resulting in a weak dependence on the velocity dispersion.
    
    \item Without additional secular effects via additional degrees of freedom, it is improbable to create a hot Jupiter directly from a single close flyby via the extraction of a large amount of angular momentum from the planetary system. Hot Jupiter formation in dense clusters demands long-term interactions including perturbations from other bodies and subsequent tidal dissipation.
    
    \item Close flybys can activate high eccentricity tidal migration of a Jupiter by changing the semi-major axis ratio of the planets and the eccentricity of the outer planet. Close flybys in dense clusters thus significantly increase the hot Jupiter formation rate.
    
    \item Distant flybys, where the closest approach is much larger than the size of the planetary system (Q $\gg$ a$_{\rm p}$), do not by themselves activate high eccentricity tidal migration of Jupiters. They thus contribute little to the hot Jupiter formation rate in dense clusters.
    
    \item About 10\% of close flybys can activate high eccentricity tidal migration by significantly changing the configuration of the planetary system. The remaining 90\% of the outcomes correspond to planet ejections, making star-planet/star-star/planet-planet collisions rare.
    
    \item The most common formation channels that are activated by flybys are the Lidov-Kozai mechanism (i.e., the outer giant planet acts as the perturber) and planet-planet scatterings. With small initial planetary semi-major axis ratios, planet-planet scattering is the main mechanism operating to create hot Jupiters.  For larger initial ratios,  Lidov-Kozai activation is the main mechanism creating hot Jupiters.
    
    \item It is extremely rare to create hot Jupiters from coplanar secular effects in dense clusters. Thus, this naively predicts that retrograde hot Jupiters born in dense clusters should be rare.
    
    \item Perpendicular close flybys more frequently activate hot Jupiter formation from Lidov-Kozai oscillations for moderately small initial semi-major axis ratios (2 $\lesssim$ a$_{\rm s}/a_{\rm j}$ $\lesssim$4). Therefore, for even larger initial semi-major axis ratios (i.e., a$_{\rm s}/a_{\rm j}$ $\gtrsim$ 4), perpendicular close flybys become even less efficient at activating hot Jupiter formation from Lidov-Kozai oscillations.
    
    \item Most hot Jupiters in dense clusters are generated within $10^8$~years of the encounter that triggered them. We find that the distribution of hot Jupiters converges within $10^9$~years of the triggering encounter.
    
    \item 
     The dynamical reconfiguration produced by the flybys leads to frequent occurrences in which the Saturn moves onto a very wide orbit, larger than $\sim 100$~AU in many systems, but frequently exceeding $\sim 1000$~AU. Hence 
    flybys can readily explain the observed presence of giant planets on very wide orbits ({{red}ultra-cold} Saturns) \citep{Lafreniere2011}, which are not easily explained by conventional theories of planet formation.
    
    \item The properties of star clusters change over time, primarily due to two-body relaxation and the evaporation of the cluster in a Galactic tidal field.  We have not accounted for these effects in any systematic way in this paper, and instead defer properly accounting for these effects to future work, in which we perform a more sophisticated population synthesis-based study accounting for a realistic star cluster mass function, distribution of galactocentric distances, and so on.
\end{itemize}

\section*{Acknowledgements}
Professors Ward Wheeler and Cheryl Hayashi of the Richard Guilder Graduate school at the American Museum of Natural History made generous grants of CPU cycles on their computation clusters available to us, which we gratefully acknowledge.  NWCL gratefully acknowledges the support of a Fondecyt Iniciacion grant \#11180005.

\bibliography{refs}{}
\bibliographystyle{aasjournal}

\end{document}